\documentclass[aps,prd,nofootinbib,amssymb,eqsecnum,notitlepage]{revtex4-1}

\usepackage{bm}
\usepackage{amsmath,amsthm,amssymb}
\usepackage[dvipdfmx]{graphicx}
\usepackage{color}

\usepackage{amsfonts}
\usepackage{comment}

\begin{document}

\newcommand{\newc}{\newcommand}
\def\be{\begin{equation}}
\def\ee{\end{equation}}
\def\ba{\begin{eqnarray}}
\def\ea{\end{eqnarray}}
\def\Mpl{M_{\rm pl}}
\newc{\rhon}{\rho_{m,n}}
\newc{\rhonn}{\rho_{m,nn}}
\newc{\delj}{\delta j}
\newc{\delrho}{\delta \rho_m}
\newc{\pa}{\partial}
\newc{\dphi}{\delta\phi}
\newc{\tp}{\dot{\phi}}
\newc{\E}{{\cal E}}
\newc{\aM}{\alpha_{\rm M}}
\newc{\aB}{\alpha_{\rm B}}
\newc{\aT}{\alpha_{\rm T}}
\newc{\aK}{\alpha_{\rm K}}
\newc{\dphiv}{\delta\varphi}
\newc{\Om}{\tilde{\Omega}_m}
\newc{\qsn}{q_s^{\rm (N)}}
\newc{\qsu}{q_s^{\rm (u)}}
\newc{\dphiN}{\delta\phi_{\rm N}}
\newc{\vphiN}{\delta\varphi_{\rm N}}
\newc{\ep}{\epsilon_{\phi}}
\newc{\K}{{\cal K}}
\newc{\rk}[1]{\textcolor{red}{#1}}

\makeatother

\allowdisplaybreaks[1]

\title{Dark energy in Horndeski theories 
after GW170817: A review}

\author{Ryotaro Kase}

\affiliation{Department of Physics, Faculty of Science,
Tokyo University of Science, 1-3,
Kagurazaka, Shinjuku, Tokyo 162-8601, Japan}

\author{Shinji Tsujikawa}

\affiliation{Department of Physics, Faculty of Science, 
Tokyo University of Science, 1-3,
Kagurazaka, Shinjuku, Tokyo 162-8601, Japan}
\date{\today}

\begin{abstract}

The gravitational-wave event GW170817 from a binary neutron star 
merger together with the electromagnetic counterpart showed that 
the speed of gravitational waves $c_t$ is very close to that of light 
for the redshift $z<0.009$.
This places tight constraints on dark energy models constructed 
in the framework of modified gravitational theories. 
We review models of the late-time cosmic 
acceleration in scalar-tensor theories with 
second-order equations of motion (dubbed Horndeski theories) 
by paying particular attention to the evolution of dark energy 
equation of state and observables relevant to the cosmic 
growth history. We provide a gauge-ready formulation of scalar 
perturbations in full Horndeski theories and estimate observables  
associated with the evolution of large-scale structures, 
cosmic microwave background, and weak lensing by employing 
a so-called quasi-static approximation for the modes deep 
inside the sound horizon. 

In light of the recent observational bound of $c_t$, we 
also classify surviving dark energy models into four classes 
depending on different structure-formation patterns and 
discuss how they can be observationally distinguished from 
each other. In particular, the nonminimally coupled theories
in which the scalar field $\phi$ has a coupling with the Ricci 
scalar $R$ of the form $G_4(\phi) R$, including $f(R)$ gravity, 
can be tightly constrained not only from the cosmic expansion and 
growth histories but also from the variation of screened
gravitational couplings.
The cross correlation of integrated Sachs-Wolfe signal with 
galaxy distributions can be a key observable for placing bounds 
on the relative ratio of cubic Galileon density to total 
dark energy density. The dawn of gravitational-wave astronomy 
will open up a new window to constrain nonminimally coupled 
theories further by the modified luminosity distance of 
tensor perturbations.

\end{abstract}

\maketitle

\tableofcontents{}

\section{Introduction}
\label{sec1} 

The late-time cosmic acceleration was first discovered 
in 1998 from the observations of distant supernovae 
type Ia (SN Ia) \cite{SN1,SN2}.
The source for this phenomenon, which was dubbed 
dark energy \cite{Huterer}, occupies about 
70~\% of today's energy density of the Universe. 
The existence of dark energy has been independently 
confirmed from the observation data of 
Cosmic Microwave Background (CMB) \cite{WMAP,WMAP9,Planck,Planck2015,Planck2018} 
and baryon acoustic oscillations (BAO) \cite{BAO}. 
Despite the tremendous observational and theoretical 
progress over the past two 
decades \cite{dreview1,dreview2,dreview2a,dreview3,dreview4,dreview5,Tsuji10,dreview6,dreview7,deRham12,dreview8,Joyce16,dreview9,dreview10,dreview11,dreview12}, the origin of dark energy 
has not been identified yet.

A key quantity describing the property of dark energy is 
the equation of state (EOS) 
$w_{\rm DE}=P_{\rm DE}/\rho_{\rm DE}$, 
where $\rho_{\rm DE}$ and $P_{\rm DE}$ are the dark energy density and pressure respectively. 
If we use the EOS parametrization of the form  
$w_{\rm DE}(a)=w_0+(1-a)w_a$ \cite{CP,Linder}, 
where $w_0, w_a$ are constants and $a$ is a scale 
factor (with today's value $a=1$), the joint data analysis of 
Planck 2018 combined with the SN Ia and BAO data placed the bounds 
$w_0=-0.961 \pm 0.077$ and $w_a=-0.28^{+0.31}_{-0.27}$ 
at 68\,\% confidence level (CL) \cite{Planck2018}. 
This shows the overall consistency with the EOS 
$w_{\rm DE}=-1$, but the time variation of $w_{\rm DE}$ 
around $-1$ is also allowed from the data.
We also note that the parametrization $w_{\rm DE}(a)=w_0+(1-a)w_a$ is not necessarily 
versatile to accommodate  models with 
a fast varying dark energy EOS \cite{Bruce1,Cora03,Bruce2,LH05} or models 
in which $w_{\rm DE}$ has an extremum \cite{Savvas12,Lasenby}.

In the theoretical side,
the simplest candidate for dark energy is the 
cosmological constant $\Lambda$ characterized 
by the EOS $w_{\rm DE}=-1$. 
If the cosmological constant originates from the 
vacuum energy in particle physics, however, 
the theoretically predicted value is 
enormously larger than the observed dark energy 
scale \cite{Weinberg,MartinLa,Tony}.
There is a local theory of the vacuum energy sequestering 
in which quantum vacuum energy is cancelled by an auxiliary 
four-form field \cite{Kaloper2} 
(see also Ref.~\cite{Kaloper1} 
for a nonlocal version). 
In this formulation, what is left after the vacuum energy sequestering is 
a radiatively stable residual cosmological 
constant $\Delta \Lambda$. 
The value of $\Delta \Lambda$ is not uniquely 
fixed by the underlying theory, but it should be measured to match with observations. 
To explain today's cosmic acceleration, it is fixed as $\Delta \Lambda \simeq
10^{-12}$ eV$^4$.

If the cosmological constant problem is solved in such 
a way that the residual vacuum energy completely vanishes or it is much smaller than today's energy density of the Universe, 
we need to find an alternative mechanism for explaining the 
origin of dark energy.
There are dynamical models of dark energy 
in which $w_{\rm DE}$ changes in time.
The representative example  
is a minimally coupled scalar field $\phi$ 
with a potential energy 
$V(\phi)$, which is dubbed 
quintessence \cite{quin1,quin2,Ratra,quin3,quin4,quin4b,quin5,Joyce,quin6,quin7}. 
The quintessence EOS varies
in the region $w_{\rm DE}>-1$. 
While the ``freezing'' quintessence models \cite{Caldwell} in 
which the deviation of $w_{\rm DE}$ from $-1$ has been large by today are in strong tension with observations,  
the ``thawing'' models in which $w_{\rm DE}$ starts to 
deviate from $-1$ at late times have been consistent 
with the data for today's EOS 
$w_{\rm DE}(a=1) \lesssim -0.7$ \cite{CDT}.

We also have other minimally coupled scalar-field theories
dubbed k-essence in which the Lagrangian $G_2$ is 
a general function of $\phi$ and its derivative 
$X=-\partial^{\mu} \phi \partial_{\mu} 
\phi/2$ \cite{kinf,kes1,kes2,kes3}.
One of the examples of k-essence is the dilatonic 
ghost condensate model 
$G_2=-X+ce^{\lambda \phi}X^2$ \cite{gcon1,gcon2} 
($c$ and $\lambda$ are constants), in which case
the typical evolution of $w_{\rm DE}$ is similar to 
that in thawing quintessence. 
It is also possible to construct unified models of dark energy 
and dark matter in terms of a purely kinetic 
Lagrangian $G_2(X)$ \cite{unifiedkes,Ber1}.

There are also theories in which the scalar field is nonminimally 
coupled to the Ricci scalar $R$ in the form 
$G_4(\phi) R$ \cite{Fujiibook}.
One of the representative examples is Brans-Dicke (BD) 
theory \cite{Brans} with a scalar potential $V(\phi)$. 
This theory is given by the Lagrangian 
$L=\phi R/2+\omega_{\rm BD}X/\phi-V(\phi)$, where 
$\omega_{\rm BD}$ is a constant called the BD parameter.
The $f(R)$ gravity whose equations of motion are derived under 
the variation of metric tensor 
$g_{\mu \nu}$ \cite{Bergmann,Ruz,Staro} 
corresponds to the special case of BD theory with 
$\omega_{\rm BD}=0$ \cite{Ohanlon,Chiba03}.
The application of $f(R)$ gravity to the late-time 
cosmic acceleration has been extensively performed
in the literature \cite{fRearly1,fRearly1b,fRearly2,fRearly3,fR1,fR2,fR3,fR4,fR5}. 
In this case, it is possible to 
realize the dark energy EOS 
smaller than $-1$ without having ghosts \cite{fR1,fR4,Moto}. 
Moreover, the growth rate of matter perturbations is 
larger than that in General Relativity (GR) \cite{fR2,fR4,Tsuji07}. 
Hence it is possible to observationally distinguish 
$f(R)$ gravity models from quintessence and k-essence. 
The construction of dark energy models in BD theories with 
the scalar potential is also possible 
\cite{Brax04,BD1,BD2}, 
in which case observational signatures are different 
depending on the BD parameter.
In these models, the propagation of fifth forces can be 
suppressed in local regions of the 
Universe \cite{Bunn,Navarro,fR1,Capo07,Tamaki,Brax08} 
under the chameleon mechanism \cite{chame1,chame2}.

There are other modified gravity theories dubbed 
Galileons \cite{Nicolis,Galileons,Galileons1} containing scalar 
derivative self-interactions and nonminimal couplings to gravity. 
In the limit of Minkowski space, the equations of motion 
following from the Lagrangian of covariant Galileons 
are invariant under the Galilean shift 
$\partial_{\mu} \phi \to \partial_{\mu} \phi+b_{\mu}$ 
(see Ref.~\cite{brokengal} for the weakly broken version of Galilean shift symmetry).
The cubic Galileon of the form $X \square \phi$ arises 
in the Dvali-Gabadadze-Porrati braneworld model  
due to the mixture between longitudinal and transverse 
gravitons \cite{DGP} and also in the Dirac-Born-Infeld decoupling theory 
with bulk Lovelock invariants \cite{DTolley}. 
This derivative self-interaction can suppress the propagation 
of fifth forces in local regions of the Universe \cite{Cede,Luty,Ratta,Babichev,Babichev1,Burrage,Tana,Brax,DKT12,KKY12,deRham,Hira,Koyama13,Andrews,Babi13,Kase13}
under the Vainshtein mechanism \cite{Vain}, 
while modifying the gravitational interaction
at cosmological distances \cite{Kase10}. 
For covariant Galileons including quartic and quintic 
Lagrangians, there exist self-accelerating de Sitter attractors responsible 
for the late-time cosmic acceleration \cite{Sami,DT10,DT10b}. 
The self-accelerating solution of full covariant Galileons preceded by 
a tracker solution with $w_{\rm DE}=-2$ is in tension with observational 
data of SN Ia, CMB, and BAO \cite{NDT10,AppleLin,Neveu,Barreira1,Barreira2,Renk,Peirone2}.
However, this is not necessarily the case for covariant Galileons with a linear 
potential \cite{KTD15} or the model in which an additional term $X^2$ is present to 
the Galileon Lagrangian \cite{Kase18}.

The aforementioned theories belong to a subclass of more 
general scalar-tensor theories--dubbed Horndeski 
theories \cite{Horndeski}.
In 2011, Deffayet {\it et al.} \cite{Def11} derived the action 
of most general scalar-tensor theories with second-order 
equations of motion after the generalizations of  
covariant Galileons.
Kobayashi {\it et al.} \cite{KYY} showed that the corresponding 
action is equivalent to that derived by Horndeski 
in 1974 \cite{Horndeski} (see also Ref.~\cite{Char11}). 
The application of Horndeski theories to the late-time cosmic 
acceleration was extensively performed 
in the literature \cite{DKT11,DT12,Char12,Zhao,Cope12,Amen12,Zuma13,Sil13,Raveri14,Hu13,OTT13,HKi14,Bellini,Zuma17,Ken,Peirone,Ken}.
Since different modified gravity theories predict different background 
expansion and cosmic growth histories, it is possible to distinguish between 
dark energy models in Horndeski theories from 
the observations of 
SN Ia, CMB, BAO, large-scale structures, and weak lensing. 

The recent detection of gravitational waves (GWs) by 
GW170817 \cite{GW170817} from a binary neutron star 
merger together with the gamma-ray burst GRB 
170817A \cite{Goldstein} constrained the propagation 
speed $c_t$ of GWs, as \cite{Abbott}
\be
-3 \times 10^{-15} \le c_t-1 \le 
7 \times 10^{-16}\,,
\label{ctbound}
\ee
for the redshift $z \le 0.009$. 
If we strictly demand that $c_t=1$ and do not allow any tuning 
among functions in Horndeski theories, the Lagrangian 
needs to be of the form 
$L=G_2(\phi,X)+G_{3}(\phi,X)\square\phi+G_4(\phi) R$ 
\cite{Lon15,GWcon1,GWcon2,GWcon3,GWcon4,GWcon5,GWcon6}\footnote{
Even in the presence of quartic-order derivative and quintic-order couplings there are some specific models 
consistent with the bound (\ref{ctbound}), see, e.g., 
Ref.~\cite{Amendola18} for interacting dark energy and dark matter models in which $c_t$ approaches 1 by today.}.
This includes the theories such as quintessence, k-essence, 
$f(R)$ gravity, BD theories, and cubic Galileons, but the theories 
with quartic-order derivative and quintic-order couplings  
do not belong to this class\footnote{
This assumes that the effective field theory of dark energy 
with derivative interactions is valid up to the energy scale 
corresponding to the
gravitational-wave frequency observed by LIGO/Virgo 
($f \sim 100$~Hz). If the effective field theory is modified 
above a strong coupling scale (which is typically of the 
same order as the LIGO/Virgo frequency) in such a way that 
$c_t$ approaches 1, we may evade the 
observational bound (\ref{ctbound}) \cite{deRham18}. 
}.
Now, we are entering the era in which 
dark energy models in Horndeski theories 
can be tightly constrained from observations.

In this article, we review the application of Horndeski theories 
to cosmology and discuss observational signatures of surviving 
dark energy models. After reviewing Horndeski theories in 
Sec.~\ref{Hornsec}, we devote Sec.~\ref{backsec} for deriving 
the background equations of motion 
and the second-order action of tensor perturbations 
on the flat Friedmann-Lema\^{i}tre-Robertson-Walker 
(FLRW) spacetime in the presence of matter.
In Sec.~\ref{scalarsec}, we obtain the second-order action of 
scalar perturbations and their equations of motion 
in full Horndeski theories without fixing any gauge conditions.  
We note that this gauge-ready formulation was carried out in a subclass 
of Horndeski theories \cite{Hwang} and 
in scalar-vector-tensor theories \cite{HKT18}. 
The formalism developed in Ref.~\cite{HKT18} encompasses both 
Horndeski gravity and generalized Proca 
theories \cite{Heisenberg14,Tasinato,Allys,Jimenez} as specific cases.

In Sec.~\ref{stasec}, we derive conditions for the absence of ghost and 
Laplacian instabilities of scalar perturbations in the small-scale limit 
by choosing three different gauges and show that these conditions are
independent of the gauge choices. 
In Sec.~\ref{obsersec}, we also apply the results in Sec.~\ref{scalarsec}  
to the computation of observables relevant to the linear growth of matter 
density perturbations and the evolution of gravitational potentials in the gauge-invariant way. 
We also classify surviving dark energy models into four classes 
depending on their observational signatures. 
In Secs.~\ref{quinsec}, \ref{fRBDsec}, \ref{classCsec}, and \ref{mostsec}, 
we review the behavior of dark energy observables in each class of theories, i.e., 
(A) quintessence and k-essence, (B) $f(R)$ gravity and BD theories, 
(C) kinetic braidings, and (D) nonminimally coupled scalar with 
cubic derivative interactions. This is convenient to distinguish between 
surviving dark energy models in current and future observational data. 
We conclude in Sec.~\ref{concludesec}.

Throughout the review, we adopt the metric signature 
$(-,+,+,+)$. We also use the natural unit in which the 
speed of light $c$, the reduced Planck constant $\hbar$, 
and the Boltzmann constant $k_B$ are equivalent to 1. 
The reduced Planck mass $M_{\rm pl}$ is related to the 
Newton gravitational constant $G$, as 
$M_{\rm pl}=1/\sqrt{8 \pi G}$.

\section{Horndeski theories}
\label{Hornsec} 

The theories containing a scalar field $\phi$ coupled to gravity 
are generally called scalar-tensor theories \cite{Fujiibook}. 
This reflects the fact that the gravity sector has two tensor polarized degrees of freedom.
Horndeski theories \cite{Horndeski} are the most general 
scalar-tensor theories with second-order equations of motion.
Due to the second-order property, there is no Ostrogradski 
instability \cite{Ostro,Woodard} associated with 
the Hamiltonian unbounded from below. 

Horndeski theories are given by the action \cite{KYY}
\be
{\cal S}_{\rm H}=\int d^4 x \sqrt{-g}\,L\,, 
\label{action}
\ee
where $g$ is a determinant of the metric tensor $g_{\mu \nu}$, and 
\ba
\hspace{-0.8cm}
L &=& G_2(\phi,X)+G_{3}(\phi,X)\square\phi 
+G_{4}(\phi,X)\, R +G_{4,X}(\phi,X)\left[ (\square \phi)^{2}
-(\nabla_{\mu}\nabla_{\nu} \phi)
(\nabla^{\mu}\nabla^{\nu} \phi) \right] \nonumber \\
\hspace{-0.8cm}
& &
+G_{5}(\phi,X)G_{\mu \nu} \nabla^{\mu}\nabla^{\nu} \phi
-\frac{1}{6}G_{5,X}(\phi,X)
\left[ (\square \phi )^{3}-3(\square \phi)\,
(\nabla_{\mu}\nabla_{\nu} \phi)
(\nabla^{\mu}\nabla^{\nu} \phi)
+2(\nabla^{\mu}\nabla_{\alpha} \phi)
(\nabla^{\alpha}\nabla_{\beta} \phi)
(\nabla^{\beta}\nabla_{\mu} \phi) \right].
\label{LH}
\ea
Here, the symbol $\nabla_{\mu}$ stands for the covariant derivative operator with $\square \equiv \nabla^{\mu}\nabla_{\mu}$, 
$R$ is the Ricci scalar, $G_{\mu \nu}$ is the Einstein tensor, and 
\be
X \equiv -\frac12 \nabla^{\mu}\phi \nabla_{\mu}\phi\,.
\label{defX}
\ee
The functions $G_{2,3,4,5}$ depend on $\phi$ and $X$, with 
$G_{i,\phi} \equiv \partial G_i/\partial \phi$ and 
$G_{i,X} \equiv \partial G_i/\partial X$. 
Originally, Horndeski derived the Lagrangian of 
scalar-tensor theories with second-order equations 
of motion in a form different from Eq.~(\ref{LH}) \cite{Horndeski}, 
but their equivalence was explicitly shown 
in Ref.~\cite{KYY}. 
Depending on the papers \cite{KYY,DT12,Bellini,Gleyzes14}, 
different signs and notations were used for 
the quantities $G_3(\phi,X)$ and $X$, so we summarize them 
in Appendix~\ref{notation} to avoid confusion.

Below, we list the theories within the framework of 
the action (\ref{action}).
\begin{itemize}
\item (1) Quintessence and k-essence
\vspace{0.1cm}

K-essence \cite{kinf,kes1,kes2,kes3} is given by the functions
\be
G_2=G_2(\phi,X)\,,\qquad G_3=0\,,\qquad
G_4=\frac{M_{\rm pl}^2}{2}\,,\qquad G_5=0\,.
\ee
Quintessence \cite{quin1,quin2,Ratra,quin3,quin4,quin4b,quin5,Joyce,quin6,quin7} 
corresponds to the particular choice:
\be
G_2=X-V(\phi)\,,
\ee
where $V(\phi)$ is the potential of $\phi$.
\item (2) BD theory
\vspace{0.1cm}

In BD theory \cite{Brans} with the scalar potential 
$V(\phi)$, we have
\be
G_2=\frac{M_{\rm pl}\omega_{\rm BD}}{\phi}X-V(\phi)\,,
\qquad G_3=0\,,\qquad
G_4=\frac{1}{2}M_{\rm pl}\phi\,,\qquad G_5=0\,.
\label{BDaction}
\ee
In the limit that $\omega_{\rm BD} \to \infty$, 
we recover GR with a quintessence scalar field. 
We note that there are more general nonminimally coupled 
theories given by the couplings 
$G_2=\omega (\phi)X-V(\phi)$, $G_3=0$, $G_4=F(\phi)$, 
$G_5=0$ \cite{Gas92,Damour1,Damour2,Amen99,Uzan,Chiba99,Bartolo99,Perrotta,Boi00,Gille}. 
Since the basic structure of such theories is similar 
to that in BD theories, we do not discuss them in this review.

\item (3) $f(R)$ gravity
\vspace{0.1cm}

The action of $f(R)$ gravity \cite{Bergmann,Ruz,Staro} 
is given by  
\be
{\cal S}_{\rm H}=\int d^4x \sqrt{-g}\,\frac{M_{\rm pl}^2}{2}f(R)\,,
\label{fRaction}
\ee
where $f(R)$ is an arbitrary function of $R$.
The metric $f(R)$ gravity, which corresponds to the variation 
of (\ref{fRaction}) with respect to $g_{\mu \nu}$,
can be accommodated by the Lagrangian (\ref{LH}) 
for the choice
\be
G_2=-\frac{M_{\rm pl}^2}{2} (RF-f),\qquad
G_3=0\,,\qquad G_4=\frac{M_{\rm pl}^2}{2}F\,,\qquad G_5=0\,,
\label{fRcase}
\ee
where $F(R) \equiv \partial f/\partial R$. 
In this case, the scalar degree of freedom 
$\phi=M_{\rm pl}F(R)$ arises from the gravity sector.
Comparing Eq.~(\ref{BDaction}) with Eq.~(\ref{fRcase}), 
it follows that metric $f(R)$ gravity is equivalent to BD theory with 
$\omega_{\rm BD}=0$ and the scalar potential 
$V=(M_{\rm pl}^2/2) (RF-f)$ \cite{Ohanlon,Chiba03}.

\item (4) Covariant Galileons
\vspace{0.1cm}

In original Galileons \cite{Nicolis}, the field
equations of motion are invariant under the shift 
$\partial_{\mu} \phi \to \partial_{\mu} \phi+b_{\mu}$ 
in Minkowski spacetime \cite{Nicolis}.
In curved spacetime, the Lagrangian of covariant 
Galileons \cite{Galileons} is constructed 
to keep the equations of motion up to second order, while recovering 
the Galilean shift symmetry in the Minkowski limit.
Covariant Galileons are characterized by the functions
\be
G_2=\beta_1 X-m^3 \phi\,,\qquad G_3=\beta_3 X\,,\qquad
G_4=\frac{M_{\rm pl}^2}{2}+\beta_4 X^2\,,\qquad
G_5=\beta_5 X^2\,,
\ee
where $\beta_{1,3,4,5}$ and $m$ are constants. 
In absence of the linear potential $V(\phi)=m^3 \phi$, 
there exists a self-accelerating de Sitter solution 
satisfying $X={\rm constant}$ \cite{DT10,DT10b,Ginf1,Ginf2}.

\item (5) Derivative couplings
\vspace{0.1cm}

There is also a derivative coupling theory in which 
the scalar field couples to the Einstein tensor of the form 
$G_{\mu \nu}\nabla^{\mu}\phi \nabla^{\nu} \phi$ \cite{Amen93,Germani}.
This corresponds to the choice
\be
G_2=X-V(\phi)\,,\qquad G_3=0\,,\qquad
G_4=\frac{M_{\rm pl}^2}{2}\,,\qquad
G_5=c \phi\,,
\ee
where $V(\phi)$ is a scalar potential and 
$c$ is a constant.
Integrating the term $c\phi\,G_{\mu \nu} \nabla^{\mu} \nabla^{\nu}\phi$ by parts, it is equivalent to 
$-c\,G_{\mu \nu}\nabla^{\mu} \phi \nabla^{\nu} \phi$ 
up to a boundary term.

\item (6) Gauss-Bonnet couplings 
\vspace{0.1cm}

One can consider a coupling of the form 
$\xi(\phi){\cal G}$ \cite{Lovelock}, where 
$\xi(\phi)$ is a function of $\phi$ and ${\cal G}$ is  
the Gauss-Bonnet curvature invariant defined by 
\be
{\cal G}=R^2-4R_{\alpha \beta}R^{\alpha \beta}
+R_{\alpha \beta \gamma \delta}R^{\alpha \beta \gamma \delta}\,.
\ee
The theories given by the action \cite{NOS,Mota,Mota2,TS06}
\be
{\cal S}_{\rm H}=\int d^4 x \sqrt{-g} 
\left[ \frac{M_{\rm pl}^2}{2}R+X-V(\phi)+\xi(\phi){\cal G} \right]
\ee
can be accommodated in the framework of Horndeski theories 
for the choice \cite{KYY}
\ba
&&
G_2=X-V(\phi)+8\xi^{(4)} (\phi) X^2 \left( 3-\ln X  \right)\,,\qquad
G_3=-4\xi^{(3)} (\phi) X \left( 7-3\ln X \right)\,,\nonumber \\
& &
G_4=\frac{M_{\rm pl}^2}{2}+
4\xi^{(2)}(\phi)X \left( 2-\ln X \right)\,,\qquad
G_5=-4\xi^{(1)} (\phi) \ln X\,,
\ea
where $\xi^{(n)}(\phi) \equiv \partial^n \xi (\phi)/\partial \phi^n$.

\item (7) $f({\cal G})$ gravity 
\vspace{0.1cm}

There is also modified gravitational theories
given by the action \cite{NO05,DT08}
\be
{\cal S}_{\rm H}=\int d^4 x \sqrt{-g} 
\left[ \frac{M_{\rm pl}^2}{2}R+f({\cal G}) \right]\,,
\label{SfG}
\ee
where $f$ is a function of ${\cal G}$.
Since the Lagrangian $f({\cal G})$ is equivalent to 
$f(\phi)+f_{,\phi} ({\cal G}-\phi)$, the action (\ref{SfG}) belongs 
to a subclass of Horndeski theories with the couplings
\ba
&&
G_2=f(\phi)-f_{,\phi}\phi+8f^{(5)} (\phi) X^2 \left( 3-\ln X  \right)\,,\qquad
G_3=-4f^{(4)} (\phi) X \left( 7-3\ln X \right)\,,\nonumber \\
& &
G_4=\frac{M_{\rm pl}^2}{2}+
4 f^{(3)}(\phi)X \left( 2-\ln X \right)\,,\qquad
G_5=-4 f^{(2)} (\phi) \ln X\,.
\ea

\item (8) Kinetic braidings and its extensions
\vspace{0.1cm}

There are theories given by the Lagrangian 
\be
L=G_2(\phi,X)+G_3(\phi,X) \square \phi+G_4(\phi)R\,.
\label{Lagkin}
\ee
As we will show later, the Lagrangian (\ref{Lagkin}) corresponds 
to most general Horndeski theories with the tensor propagation 
speed $c_t$ equivalent to 1.
The kinetic braiding scenario \cite{braiding1,braiding2} 
corresponds to the minimally coupled case, i.e., $G_4=M_{\rm pl}^2/2$. 
The cubic Galileon given with  
$L=\beta_1 X-m^3 \phi+\beta_3 X \square \phi+(M_{\rm pl}^2/2)R$ 
belongs to a subclass of kinetic braidings.
The dark energy scenario given by 
$L=\beta_1 X+\beta_2 X^2+\beta_3 X \square \phi+(M_{\rm pl}^2/2)R$ \cite{Kase18} 
is also in the framework of kinetic braidings.
In presence of the nonminimal coupling $G_4(\phi)R$, 
it is known that a self-accelerating solution characterized by 
$\dot{\phi}/\phi={\rm constant}$ exists for the model 
$L=\omega (\phi/M_{\rm pl})^{1-n}X+(\lambda/\mu^3)
(\phi/M_{\rm pl})^{-n} X \square \phi+(M_{\rm pl}^2/2)
(\phi/M_{\rm pl})^{3-n}R$ with 
$2 \le n \le 3$ \cite{Silva,Koba1,Koba2,DTge}. 
There is also a nonminimally coupled model 
given by 
$L=\beta_1 \left( 1-6Q^2 \right) e^{-2Q \phi/M_{\rm pl}}X
-m^3 \phi+\beta_3 X \square \phi+(M_{\rm pl}^2/2)
e^{-2Q \phi/M_{\rm pl}}R$ \cite{Babi11,KTD15}, which recovers the minimally 
coupled cubic Galileon in the limit $Q \to 0$.
\end{itemize}

Thus, Horndeski theories can accommodate a wide variety 
of scalar-tensor theories with second-order equations of motion. 
Except for quintessence and k-essence, the scalar field 
has derivative self-interactions and nonminimal/derivative 
couplings to gravity. 
In such cases, we need to confirm whether 
there are neither ghost nor Laplacian instabilities.
In Secs.~\ref{tensec} and \ref{stasec}, we will address this issue 
after deriving the second-order actions of tensor and scalar perturbations 
on the flat FLRW background in full Horndeski theories. 
The same analysis also allows one to identify the speed of 
gravitational waves on the isotropic cosmological background.
Under the observational bound (\ref{ctbound}), 
this result can be used to pin down viable Horndeski theories 
relevant to  the late-time cosmic acceleration. 
In Sec.~\ref{classsec}, we classify surviving dark energy models into 
four classes depending on their observational signatures. 

To discuss the dynamics of cosmic acceleration preceded 
by the matter-dominated epoch, we take into account a matter 
perfect fluid minimally coupled to gravity.
The vector perturbations are nondynamical in Horndeski theories, 
so we only need to consider the evolution of scalar and tensor perturbations.
The perfect fluid without the vector degree of freedom can be 
described by the Schutz-Sorkin action \cite{Sorkin,DGS}: 
\be
{\cal S}_m=-\int d^{4}x \left[ \sqrt{-g}\,\rho_m(n)
+J^{\mu} \partial_{\mu}\ell \right]\,,
\label{Spf}
\ee
where $\rho_m$ is a fluid density, $\ell$ is 
a scalar quantity, and $J^{\mu}$ is a four vector associated with 
the fluid number density $n$, as
\be
n=\sqrt{\frac{J^{\mu}J^{\nu}g_{\mu \nu}}{g}}\,.
\label{ndef}
\ee
Varying the action (\ref{Spf}) with respect to $J^{\mu}$, 
we obtain 
\be
u_\mu\equiv\frac{J_\mu}{n\sqrt{-g}}
=\frac{\partial_{\mu}\ell}{\rho_{m,n}}\,,
\label{umudef}
\ee
where $u_{\mu}$ corresponds to a normalized four velocity, and  
$\rho_{m,n} \equiv \partial\rho_m/\partial n$.

In this review, we focus on the theories given by the sum 
of (\ref{action}) and (\ref{Spf}), i.e., 
\be
{\cal S}={\cal S}_{\rm H}+{\cal S}_m\,.
\label{actionfull}
\ee
After deriving the background and perturbation equations of motion, 
they can be applied to subclasses of Horndeski theories listed above.

\section{FLRW background and tensor perturbations}
\label{backsec} 

%
\subsection{Background equations of motion}
\label{backsec2}

Let us consider the flat FLRW spacetime given by the line element
\be
ds^2=-N^2(t)dt^2+a^2(t) \delta_{ij}dx^i dx^j\,,
\label{FLRW}
\ee
where $N(t)$ is a lapse and $a(t)$ is a scale factor.  
The lapse is introduced here for deriving the Friedmann equation, 
but we finally set $N=1$ after the variation of ${\cal S}$. 
The scalar field $\phi$ depends on the cosmic time $t$ alone on 
the background (\ref{FLRW}).
For the matter sector, the temporal component of $J^{\mu}$ 
in Eq.~(\ref{ndef}) is given by 
\be
J^{0}=n_0a^3\,, 
\label{Jback}
\ee
where $n_0$ is the background value of $n$. 
Then, the Schutz-Sorkin action (\ref{Spf}) reduces to 
\be
{\cal S}_m= -\int d^4 x\,a^3 
\left( N \rho_m+n_0 \dot{\ell}\right)\,.
\ee
Varying this matter action with respect to $\ell$, 
it follows that 
\be
{\cal N}_0 \equiv n_0 a^3={\rm constant}\,,
\label{N0con}
\ee
where ${\cal N}_0$ corresponds to the total 
conserved fluid number.

Now, we compute the action (\ref{actionfull}) on the spacetime 
metric (\ref{FLRW}) and vary it with respect to 
$N$, $a$ and $\phi$. Setting $N=1$ in the end, we obtain the 
background equations of motion:
\ba
& &
6 G_4H^2 +G_2 -\dot{\phi}^2 G_{2,X}
+\dot{\phi}^2 \left( 3H \dot{\phi}G_{3,X} -G_{3,\phi} \right)
+6 H \dot{\phi} \left( G_{4,\phi}+\dot{\phi}^2 G_{4,X \phi}
-2H \dot{\phi}G_{4,X}-H \dot{\phi}^3 G_{4,X X} \right)
\nonumber \\
& &
+H^2 \dot{\phi}^2 \left( 9 G_{5,\phi}+3\dot{\phi}^2 G_{5,X \phi} 
-5H \dot{\phi}G_{5,X}-H \dot{\phi}^3 G_{5,X X} \right)=\rho_m\,,
 \label{back1}\\
& &
2q_t \dot{H}-D_6 \ddot{\phi}+D_7 \dot{\phi}
=-\rho_m-P_m\,,
 \label{back2}\\
& & 
3D_6 \dot{H}+2D_1 \ddot{\phi}+3D_7 H-D_5=0\,,
 \label{back3}
 \ea
where $H \equiv \dot{a}/a$ is the Hubble expansion rate, 
and $P_m$ is the matter pressure defined by 
$P_m \equiv -n_0 \dot{\ell}-\rho_m$. 
Substituting $J_0=-n_0a^3$ into Eq.~(\ref{umudef}), the temporal 
component $u_{0}$ of four velocity is equivalent to $-1$ 
and hence 
$\dot{\ell}=-\rho_{m,n}$. Thus, the matter pressure can be 
written as 
\be
P_m=n_0 \rho_{m,n}-\rho_m\,.
\label{Pm}
\ee
{}From Eq.~(\ref{N0con}), the fluid number density 
obeys the differential equation $\dot{n}_0+3Hn_0=0$. 
On using the property $\dot{\rho}_{m}=\rho_{m,n} 
\dot{n}_0$ and the relation (\ref{Pm}), the conservation 
of total fluid number translates to 
\be
\dot{\rho}_m+3H \left( \rho_m+P_m \right)=0\,.
\label{coneq}
\ee
We note that this continuity equation also follows from 
Eqs.~(\ref{back1})-(\ref{back3}).

The quantity $q_t$ in Eq.~(\ref{back2}) is defined by 
\be
q_t=2G_4-2\dot{\phi}^2 G_{4,X}
+\dot{\phi}^2 G_{5,\phi}-H \dot{\phi}^3 G_{5,X}\,, 
\label{qt}
\ee
which is associated with the no-ghost condition of 
tensor perturbations discussed later in Sec.~\ref{tensec}.
The definitions of coefficients $D_{1,5,6,7}$ appearing in 
Eqs.~(\ref{back2}) and (\ref{back3}) are presented in Appendix~\ref{coeff}. 
As we will show in Sec.~\ref{scalarsec}, they also appear as coefficients 
in the second-order action of scalar perturbations. 
We note that Eq.~(\ref{back2}) has been derived by varying 
the action with respect to $a$ and then eliminating the term 
$G_2$ by using Eq.~(\ref{back1}). 

Solving Eqs.~(\ref{back2})-(\ref{back3}) for $\dot{H}$ 
and $\ddot{\phi}$, it follows that 
\ba
\dot{H} 
&=& -\frac{1}{q_s} 
\left[ 2D_1 D_7 \dot{\phi}-D_6 (D_5-3D_7 H)
+2D_1 (\rho_m+P_m) \right]\,,\label{dH} \\
\ddot{\phi}
&=& \frac{1}{q_s} \left[ 
3D_6 D_7 \dot{\phi}+2 (D_5-3D_7H)q_t
+3D_6 (\rho_m+P_m) \right]\,,
\label{ddphi}
\ea
where 
\be
q_s \equiv 4D_1 q_t+3D_6^2\,.
\label{calD}
\ee
To avoid the divergences on the right hand sides of 
Eqs.~(\ref{dH}) and (\ref{ddphi}), we require that 
\be
q_s \neq 0\,. 
\ee
As we will show later, the determinant $q_s$ is related to the no-ghost condition of scalar perturbations. 
The no-ghost condition corresponds to $q_s>0$, 
in which case $\dot{H}$ and $\ddot{\phi}$ can 
remain finite.

In the presence of matter, the condition for the cosmic 
acceleration is given by 
\be
w_{\rm eff} \equiv -1-\frac{2\dot{H}}{3H^2}<-\frac{1}{3}\,.
\label{weff}
\ee
We can express Eqs.~(\ref{back1}) and (\ref{back2}) in the forms
\ba
& &
3M_{\rm pl}^2 H^2=\rho_{\rm DE}+\rho_m\,,
\label{Frieq}\\
& &
2M_{\rm pl}^2 \dot{H}=-\rho_{\rm DE}-P_{\rm DE}
-\rho_m-P_m\,,
\label{Frieq2}
\ea
where the density $\rho_{\rm DE}$ and pressure $P_{\rm DE}$ 
of the ``dark'' component are given, respectively, by 
\ba
\rho_{\rm DE} 
&=& 3H^2 \left( M_{\rm pl}^2 -2G_4 \right)
-G_2+\dot{\phi}^2 G_{2,X}
-\dot{\phi}^2 \left( 3H \dot{\phi}G_{3,X} -G_{3,\phi} \right) 
-6 H \dot{\phi} \left( G_{4,\phi}+\dot{\phi}^2 G_{4,X \phi} 
\right. \nonumber \\
& &
\left.
-2H \dot{\phi}G_{4,X}-H \dot{\phi}^3 G_{4,X X} \right) 
-H^2 \dot{\phi}^2 \left( 9 G_{5,\phi}+3\dot{\phi}^2 G_{5,X \phi} 
-5H \dot{\phi}G_{5,X}-H \dot{\phi}^3 G_{5,X X} \right)\,,\\
P_{\rm DE} 
&=& 2 \left( q_t-M_{\rm pl}^2 \right) \dot{H}-D_6 \ddot{\phi}
+D_7 \dot{\phi}-\rho_{\rm DE}\,.
\ea
We define the dark energy EOS, as 
\be
w_{\rm DE} \equiv \frac{P_{\rm DE}}{\rho_{\rm DE}}
=-1+\frac{2 ( q_t-M_{\rm pl}^2 ) \dot{H}-D_6 \ddot{\phi}
+D_7 \dot{\phi}}{\rho_{\rm DE}}\,,
\label{wdedef}
\ee
which is different from the effective EOS (\ref{weff}) 
due to the presence of additional matter (dark matter, baryons, radiation). 
The necessary condition for the late-time cosmic acceleration is 
given by $w_{\rm DE}<-1/3$, but we caution that 
this is not a sufficient condition.
In quintessence and k-essence we have $q_t=M_{\rm pl}^2$, 
so the time variation of $\phi$ leads to the deviation 
of $w_{\rm DE}$ from $-1$. 
In other theories listed in Sec.~\ref{Hornsec}, the quantity $q_t$ 
generally differs from $M_{\rm pl}^2$, so the term 
$2 (q_t-M_{\rm pl}^2 ) \dot{H}$ in Eq.~(\ref{wdedef}) 
also contributes to the additional deviation 
of $w_{\rm DE}$ from $-1$. 
Since the evolution of $w_{\rm DE}$ is different depending on 
dark energy models, it is possible to distinguish between 
them from the observations of SN Ia, CMB, and BAO. 

\subsection{Tensor perturbations}
\label{tensec}

The speed of gravitational waves on the flat FLRW background 
can be derived by expanding 
the action (\ref{actionfull}) up to second order in tensor 
perturbations $h_{ij}$. As a by-product, we can identify
conditions for the absence of ghost and Laplacian instabilities 
in the tensor sector.  
The perturbed line element, which contains 
traceless and divergence-free tensor perturbations obeying 
the conditions $h _{ii}=0$ and $\partial _{i}h_{ij}=0$, 
is given by 
\be
ds^2=-dt^2+a^2(t)(\delta_{ij}+h_{ij})dx^i dx^j\,.
\ee
Without loss of generality, we can choose nonvanishing components 
of $h_{ij}$ in the forms 
\be
h_{11}=h_1(t,z)\,,\qquad 
h_{22}=-h_1(t,z)\,,\qquad
h_{12}=h_{21}=h_2(t,z)\,,
\ee
where $h_1$ and $h_2$ characterize the
two polarization states.

Expanding the Horndeski action (\ref{action}) up to second order 
in perturbations and integrating it by parts, the quadratic action 
$({\cal S}_{\rm H}^{(2)})_t$
contains the time derivative $\dot{h}_i^2$, the spatial derivative 
$(\partial h_i)^2$, and the mass term $h_i^2$ (where $i=1,2$). 
The second-order action of tensor perturbations arising from the 
matter action (\ref{Spf}) can be written in the form 
\be
({\cal S}_m^{(2)})_t=-\int dtd^3x \left[
(\sqrt{-g})^{(2)} \rho_m
+\sqrt{-\bar{g}} \rho_{m,n}\delta n \right]\,,
\label{Sm2}
\ee
where $(\sqrt{-g})^{(2)}=-a^3(h_1^2+h_2^2)/2$ 
and $\delta n=n_0 (h_1^2+h_2^2)/2$ 
with $\sqrt{-\bar{g}}=a^3$. 
Then, Eq.~(\ref{Sm2}) reduces to
\be
({\cal S}_m^{(2)})_t=-\int dt d^3 x \sum_{i=1}^{2} 
\frac{1}{2}a^3 P_m h_i^2\,,
\ee
where $P_m$ is the matter pressure given by Eq.~(\ref{Pm}).
Using the background Eqs.~(\ref{back1}) and (\ref{back2}), 
the terms proportional to $h_i^2$ identically vanish from 
the total second-order action 
${\cal S}_t^{(2)}=({\cal S}_{\rm H}^{(2)})_t+({\cal S}_{m}^{(2)})_t$.
Finally, we can express ${\cal S}_t^{(2)}$ in a compact form:
\be
{\cal S}_t^{(2)}=\int dt d^3x \sum_{i=1}^{2}
\frac{a^3}{4}q_t \left[ \dot{h}_i^2-\frac{c_t^2}{a^2} 
(\partial h_i)^2 \right]\,,
\label{actionSt}
\ee
where $q_t$ was already introduced in Eq.~(\ref{qt}), 
and $c_t^2$ is the tensor propagation speed squared given by 
\be
c_t^2=\frac{1}{q_t} 
\left( 2G_4-\dot{\phi}^2G_{5,\phi} -\dot{\phi}^2 \ddot{\phi} G_{5,X} \right)\,.
\label{ct}
\ee

To avoid the ghost and Laplacian instabilities, 
we require that 
\ba
& &
q_t>0\,,\label{Qtcon}\\
& &
c_t^2>0\,.
\label{ctcon}
\ea
Varying the action (\ref{actionSt}) with respect to $h_i$, 
we obtain the tensor perturbation equation of motion 
in Fourier space, as
\be
\ddot{h}_i+\left(3H +\frac{\dot{q}_t}{q_t} \right) \dot{h}_i
+c_t^2 \frac{k^2}{a^2} h_i=0\,,
\label{heq}
\ee
where $k$ is a coming wavenumber. 
The time variation of $q_t$ and the deviation of $c_t^2$ 
from 1 
lead to the modified evolution of $h_i$ compared to that in GR.

The observational bound (\ref{ctbound}) of $c_t$ places tight 
constraints on surviving dark energy models.
To realize the exact value $c_t^2=1$, we require the following condition 
\be
2G_{4,X}-2G_{5,\phi}+\left( H \dot{\phi} -\ddot{\phi} 
\right) G_{5,X}=0\,.
\ee
If we do not allow any tuning among functions, 
the dependence of $G_4$ on $X$ and $G_5$ on $\phi, X$ 
is forbidden. Then, the Horndeski Lagrangian is restricted 
to be of the 
form \cite{Lon15,GWcon1,GWcon2,GWcon3,GWcon4,GWcon5,GWcon6}
\be
L=G_2(\phi,X)+G_3(\phi,X) \square \phi
+G_4(\phi)R\,.
\label{lagcon}
\ee
Among the theories listed in Sec.~\ref{Hornsec}, the theories 
(5), (6), (7) lead to $c_t^2$ different from 1. 
In particular, as long as the derivative and Gauss-Bonnet couplings 
contribute to the late-time cosmological dynamics, the deviation of 
$c_t^2$ from 1 is large and hence such couplings are excluded from 
the bound (\ref{ctbound}) as a source for dark 
energy \cite{Gong}\footnote{Besides this problem, 
the matter perturbation 
in $f({\cal G})$ gravity is subject to strong instabilities 
attributed to the negative sound speed squared \cite{FMT10},  
so $f({\cal G})$ cosmological models are not 
cosmologically viable unless the deviation from the 
$\Lambda$CDM model is very tiny.}.

Quintessence and k-essence, BD theory, and $f(R)$ gravity give rise to 
the exact value $c_t^2=1$, so they automatically satisfy the bound 
(\ref{ctbound}). The quartic and quintic Galileon couplings 
$G_4=\beta_4 X^2$ and $G_5=\beta_5 X^2$ lead to the deviation of 
$c_t^2$ from 1, but the Galileon Lagrangian up to 
the cubic interaction $G_3=\beta_3 X$ is allowed 
(which includes the model with the additional term
$G_2=\beta_2 X^2$ to the cubic Galileon Lagrangian \cite{Kase18}). 
The kinetic braidings \cite{braiding1,braiding2} and its extensions \cite{Silva,Koba1,Koba2,DTge,Babi11,KTD15} are also consistent 
with the bound (\ref{ctbound}). 

Even if $c_t^2$ is constrained to be close to 1, there is another modification 
to Eq.~(\ref{heq}) arising from the time variation of $q_t$. 
This leads to the modified luminosity distance of GWs 
relative to that of electromagnetic 
signals \cite{Maggiore1,Amen17,Maggiore2}.
Since $q_t=2G_4(\phi)$ for the Lagrangian (\ref{lagcon}), 
it is possible to distinguish between nonminimally and 
minimally coupled theories from the luminosity 
distance measurements of GWs 
together with the electromagnetic counterpart. 
After the accumulation of GW events in future, the 
luminosity distance can be a key observable to probe 
the existence of nonminimal couplings.

For the Lagrangian (\ref{lagcon}), we have 
$D_6=-\dot{\phi}^2 G_{3,X}-2G_{4,\phi}$ and
$D_7=\dot{\phi}(G_{2,X}+2G_{3,\phi}+2G_{4,\phi \phi})
-H(2G_{4,\phi}+3\dot{\phi}^2 G_{3,X})$, so 
the dark energy EOS (\ref{wdedef}) reduces to 
\be
w_{\rm DE}
=-1+\frac{2(2G_4-M_{\rm pl}^2) \dot{H}
+(\dot{\phi}^2 G_{3,X}+2G_{4,\phi}) \ddot{\phi}
+[\dot{\phi}(G_{2,X}+2G_{3,\phi}+2G_{4,\phi \phi})-H(2G_{4,\phi}+3\dot{\phi}^2 G_{3,X})] \dot{\phi}}
{3H^2 ( M_{\rm pl}^2 -2G_4 )
-G_2+\dot{\phi}^2 G_{2,X}
-\dot{\phi}^2 ( 3H \dot{\phi}G_{3,X} -G_{3,\phi}) 
-6 H \dot{\phi}G_{4,\phi}}\,.
\label{wde}
\ee
In Secs.~\ref{quinsec}-\ref{mostsec}, we will study 
the evolution of 
$w_{\rm DE}$ for concrete models 
of the late-time cosmic acceleration in the framework of Horndeski theories.

\section{Gauge-ready formulation of scalar perturbations}
\label{scalarsec} 

In this section, we derive the second-order action of scalar 
perturbations in full Horndeski theories without fixing 
gauge conditions. The resulting linear perturbation equations 
of motion are written in a gauge-ready form \cite{Hwang,HKT18}, 
so that one can choose convenient gauges depending on the problems at hand. 
For generality, we do not restrict the Horndeski Lagrangian to  
the form (\ref{lagcon}).

We begin with the linearly perturbed line-element
given by \cite{Bardeen,Kodama,MFB92,BTW05,Malik}
\be
ds^2=-(1+2\alpha)dt^2+2 \partial_i \chi dt dx^i
+a^2(t) \left[ (1+2\zeta) \delta_{ij}
+2\partial_i \partial_j E \right]dx^i dx^j\,,
\label{permet}
\ee
where $\alpha, \chi, \zeta, E$ are scalar metric perturbations, and 
the symbol $\partial_i$ stands for the partial derivative 
$\partial/\partial x^i$. 
We do not take into account vector perturbations, 
as they are nondynamical in scalar-tensor theories. 
The scalar field is decomposed into the form
\be
\phi=\bar{\phi}(t)+\delta \phi (t, x^i)\,,
\ee
where $\bar{\phi}(t)$ and $\delta \phi(t, x^i)$ are 
the background and perturbed values, respectively. 
In the following, we omit the bar from the background quantities.

\subsection{Second-order matter action}

We first expand the Schutz-Sorkin action (\ref{Spf}) up to 
second order in perturbations.
We decompose the temporal and spatial components of $J^{\mu}$ 
into the background and perturbed parts, as \cite{GPcosmo,GPGeff}
\be
J^{0} = \mathcal{N}_{0}+\delta J\,,\qquad
J^{i} =\frac{1}{a^2(t)}\,\delta^{ik}\partial_{k}\delta j\,,
\label{elldef}
\ee
where $\mathcal{N}_{0}=n_0a^3$ is the conserved background 
fluid number defined by Eq.~(\ref{N0con}), 
and $\delta J, \delta j$ are scalar perturbations. 
The spatial component of four velocity can be expressed as
$u_i=-\partial_i v$, where $v$ is the velocity potential. 
On using Eq.~(\ref{umudef}), the scalar quantity $\ell$ 
contains the perturbation $-\rho_{m,n}v$. Then,  
$\ell$ can be decomposed as
\be
\ell=-\int^{t} \rho_{m,n} 
(\tilde{t})d\tilde{t} 
-\rho_{m,n}v\,,
\label{ells}
\ee
where the first contribution to the right hand side corresponds 
the background quantity. 
Defining the matter density perturbation 
\be
\delta \rho_m \equiv \frac{\rho_{m,n}}{a^3} \left[\delta J
-\mathcal{N}_{0}(3\zeta+\partial^2E)\right]\,,
\label{drhom}
\ee
where $\partial^2E \equiv \partial_i \partial_i E$ 
(the same latin subscripts are summed over),  
the perturbation of fluid number density $n$, up to 
second order, is expressed as 
\be
\delta n= \frac{\delta \rho_m}{\rho_{m,n}}
-\frac{({\cal N}_0 \partial \chi+\partial \delta j)^2}{2{\cal N}_0a^5}
-\frac{(3\zeta+\partial^2E)\delta \rho_m}{\rho_{m,n}}
-\frac{{\cal N}_0(\zeta+\partial^2E)(3\zeta-\partial^2E)}{2a^3}\,.
\ee
Since $\delta n$ reduces to $\delta \rho_m/\rho_{m,n}$ at linear order, 
this confirms the consistency of the definition (\ref{drhom}).

Now, we are ready for expanding the Schutz-Sorkin action (\ref{Spf}) 
up to quadratic order in scalar perturbations. 
This manipulation gives the second-order matter 
action \cite{HKT18}:
\ba
({\cal S}_m^{(2)})_s
&=&\int dt d^3 x\, a^3
\Bigg[
\frac{\rhon}{2a^8n_0}(\pa \delj)^2+\frac{\rhon}{a^5}(\pa\chi+\pa v) (\pa\delj)
+\left(\dot{v}-3Hc_m^2v-\alpha\right)\delrho
-\frac{c_m^2}{2n_0\rhon}\delrho^2+\frac{\rho_m}{2}\alpha^2
\notag\\
&&
+\frac{n_0 \rhon-\rho_m}{2}\left\{\frac{(\pa\chi)^2}{a^2}
+(\zeta+\pa^2E)(3\zeta-\pa^2E)\right\}
+(3\zeta+\pa^2E)\left\{ n_0\rhon(\dot{v}-3Hc_m^2 v)-\rho_m\alpha\right\}
\Bigg]\,,
\label{SMS}
\ea
where $c_m^2$ is the matter sound speed squared given by 
\be
c_m^2=\frac{P_{m,n}}{\rho_{m,n}}
=\frac{n_0 \rho_{m,nn}}{\rho_{m,n}}\,.
\ee
Variation of the action (\ref{SMS}) with respect to $\delj$ 
leads to
\be
\partial \delta j=-a^3 n_0 
\left( \partial v+\partial \chi \right)\,.
\label{deljre}
\ee
Substituting the relation (\ref{deljre}) into 
Eq.~(\ref{SMS}), we obtain
\ba
\hspace{-0.3cm}
({\cal S}_m^{(2)})_s
&=&\int dt d^3x\,a^3 
\Bigg[ \left( \dot{v}-3H c_m^2v-\alpha 
\right) \delta \rho_m-\frac{c_m^2}{2(\rho_m+P_m)}\delrho^2
-\frac{\rho_m+P_m}{2a^2} 
\left\{ (\partial v)^2+2\partial v \partial \chi \right\}
-\frac{\rho_m}{2a^2} (\partial \chi)^2
+\frac{\rho_m}{2}\alpha^2
\notag\\
\hspace{-0.3cm}
&&
+\frac{P_m}{2}(\zeta+\pa^2E)(3\zeta-\pa^2E)
+(3\zeta+\pa^2E)\left\{(\rho_m+P_m)(\dot{v}-3Hc_m^2 v)
-\rho_m\alpha\right\} \Bigg]\,,
\label{SMS2}
\ea
where we used Eq.~(\ref{Pm}). 

\subsection{Full second-order scalar 
action and linear perturbation equations of motion}
\label{actsec}

We also expand the Horndeski action (\ref{action}) up to second order 
in scalar perturbations and take the sum with Eq.~(\ref{SMS2}).
Using the background Eq.~(\ref{back1}), the term 
$\rho_m\alpha^2/2$ in Eq.~(\ref{SMS2}) is cancelled
by a part of contributions proportional to $\alpha^2$ arising from 
the Horndeski action. 
After the integration by parts, we can write the full second-order action 
of ${\cal S}$ in the gauge-ready form 
\be
{\cal S}_s^{(2)}=\int dt d^3x 
\left( L_{\rm flat}+L_{\zeta}+L_{E}\right)\,, 
\label{Ss}
\ee
where 
\ba
\hspace{-1.0cm}
L_{\rm flat}
&=& a^3\left[
D_1\dot{\dphi}^2+D_2\frac{(\partial\dphi)^2}{a^2}+D_3\dphi^2
+\left(D_4\dot{\dphi}+D_5\dphi+D_6\frac{\partial^2\dphi}{a^2}\right) \alpha
-\left(D_6\dot{\dphi}-D_7\dphi\right)\frac{\partial^2\chi}{a^2}
\right.\notag\\
\hspace{-1.0cm}
&&\left.~~~
+\left(\dot{\phi} D_6-2Hq_t\right)\alpha\frac{\partial^2\chi}{a^2}
+\left(\dot{\phi}^2 D_1+3H \dot{\phi} D_6-3H^2 q_t\right)\alpha^2
\right. \nonumber \\
\hspace{-1.0cm}
& &
\left.~~~+\left( \rho_m+P_m \right)v\frac{\partial^2 \chi}
{a^2}-v\dot{\delta \rho}_m-3H (1+c_m^2) v\delta \rho_m 
-\frac{1}{2} (\rho_m+P_m) \frac{(\partial v)^2}{a^2}
-\frac{c_m^2}{2 (\rho_m+P_m)} \delta \rho_m^2 
-\alpha \delta \rho_m \right]\,,
\label{LF}\\
\hspace{-1.0cm}
L_{\zeta}
&=& a^3\left[
\left\{
3D_6\dot{\dphi}-3D_7\dphi-3\left(\dot{\phi} D_6-2Hq_t\right)\alpha-3(\rho_m+P_m)v
+2q_t\frac{\pa^2\chi}{a^2}\right\}\dot{\zeta}-3q_t\dot{\zeta}^2
\right.\notag\\
\hspace{-1.0cm}
&&\left.~~~
-\left(B_1\dphi+2q_t\alpha\right)\frac{\pa^2\zeta}{a^2}
+q_t c_t^2\frac{(\pa\zeta)^2}{a^2}
\right]\,,
\label{Lze}\\
\hspace{-1.0cm}
L_{E}
&=& a^3\Bigg[
2q_t\ddot{\zeta}+2B_2\dot{\zeta}
-D_6\ddot{\dphi}-B_3\dot{\dphi}+B_4\dphi
+\frac{1}{a^3}\frac{d}{dt}\left\{a^3 \left(\dot{\phi} D_6-2Hq_t 
\right) \alpha\right\}
+(\rho_m+P_m)(\dot{v}-3Hc_m^2v)
\Bigg]\,\pa^2 E,
\label{LE}
\ea
where the explicit forms of coefficients $D_{1,...,7}$ are 
presented in Appendix~\ref{coeff}.  
The quantities $q_t$ and $c_t^2$ are given by 
Eqs.~(\ref{qt}) and (\ref{ct}), respectively.
The coefficients $B_{1,2,3,4}$ in Eqs.~(\ref{Lze}) and (\ref{LE}) 
can be expressed by using other quantities, as
\be
B_1=\frac{2}{\tp}\left[\dot{q_t}+(1-c_t^2)Hq_t\right]\,,
\qquad 
B_2=\dot{q_t}+3Hq_t\,,
\qquad 
B_3=\dot{D_6}+3HD_6-D_7\,,
\qquad
B_4=\dot{D_7}+3HD_7\,.
\label{Bi}
\ee
The Lagrangian $L_{\rm flat}$ is present for the flat 
gauge $\zeta=0=E$ \cite{KTdark18}, 
while the other two Lagrangians 
$L_{\zeta}$ and $L_{E}$ arise in the presence of metric perturbations 
$\zeta$ and $E$. 
We note that, in scalar-vector-tensor theories, the generalized version 
of the action (\ref{Ss}) was derived in Ref.~\cite{HKT18}.

Since there are no kinetic terms for the variables $\alpha,\chi,v,E$ 
in Eqs.~(\ref{LF})-(\ref{LE}), they correspond to nondynamical perturbations. 
Varying the second-order action (\ref{Ss}) with respect to $\alpha,\chi,v,E$, 
their constraint equations in Fourier space are given, respectively, by 
\ba
\E_{\alpha} &\equiv&
D_4\dot{\dphi}-3\left(\dot{\phi} D_6-2Hq_t\right)\dot{\zeta}
+D_5\dphi+2\left(\dot{\phi}^2 D_1+3H \dot{\phi} D_6-3H^2 q_t\right)\alpha
\notag\\
&&
+\frac{k^2}{a^2}\left[ 2q_t\zeta-\left(\dot{\phi} D_6-2Hq_t\right)\left(\chi-a^2\dot{E}\right)
-D_6\dphi\right]-\delta \rho_m=0\,,
\label{eqalpha}\\
\E_{\chi} &\equiv&
D_6\dot{\dphi}-2q_t\dot{\zeta}-D_7\dphi-\left(\dot{\phi} D_6-2Hq_t\right)\alpha
-\left( \rho_m+P_m \right)v=0\,,
\label{eqchi}\\
\E_{v} &\equiv&
\dot{\delta \rho}_m+3H \left( 1+c_m^2 \right)
\delta \rho_m
+3(\rho_m+P_m)\dot{\zeta}
+\frac{k^2}{a^2} \left( \rho_m+P_m \right) 
\left( v+\chi-a^2\dot{E} \right)=0\,,
\label{eqdrho} \\
\E_{E} &\equiv&
2q_t\ddot{\zeta}+2B_2\dot{\zeta}
-D_6\ddot{\dphi}-B_3\dot{\dphi}+B_4\dphi
+\frac{1}{a^3}\frac{d}{dt}\left\{a^3(\dot{\phi} D_6-2Hq_t)\alpha\right\}
+(\rho_m+P_m)(\dot{v}-3Hc_m^2v)=0\,,
\label{eqE}
\ea
where $k$ is a comoving wavenumber. 
Variations of the action (\ref{Ss}) with respect to the remaining 
variables $\dphi,\delrho,\zeta$ lead to
\ba
\E_{\delta \phi} &\equiv&
\dot{\cal Z}+3H {\cal Z}+3D_7\dot{\zeta}-2D_3 \delta \phi-D_5 \alpha
-\frac{k^2}{a^2} \left( 2D_2 \delta \phi -D_6 \alpha 
-D_7 \chi+B_1\zeta-a^2B_4E
 \right)=0\,,\label{calZeq}\\
\E_{\delta \rho_m} &\equiv&
\dot{v}-3Hc_m^2 v-\frac{c_m^2}{\rho_m+P_m} \delta \rho_m 
-\alpha=0\,,
\label{veq}\\
\E_{\zeta} &\equiv&
\dot{\cal W} +3H{\cal W}+(\rho_m+P_m)(\dot{v}-3Hc_m^2v)
+\frac{k^2}{3a^2}\left(2q_t\alpha+2q_tc_t^2\zeta
+B_1\dphi\right)=0\,,
\label{calWeq}
\ea
where 
\ba
{\cal Z} &\equiv& 2D_1 \dot{\delta \phi}+3D_6\dot{\zeta}
+D_4 \alpha+\frac{k^2}{a^2}\left[D_6\chi 
-a^2(D_6\dot{E}+D_7E)
\right]\,, \label{Zdef}\\
{\cal W} &\equiv& 2q_t\dot{\zeta}-D_6\dot{\dphi}+D_7\dphi
+\left(\dot{\phi} D_6-2Hq_t\right)\alpha
+\frac{2k^2}{3a^2} q_t (\chi-a^2 \dot{E})\,. 
\label{Wdef}
\ea
By combining Eq.~(\ref{eqE}) with Eq.~(\ref{calWeq}), we can eliminate 
the time derivatives $\ddot{\zeta}$ and $\ddot{\dphi}$. 
This manipulation leads to 
\be
q_t \left( \alpha+\dot{\chi}
+c_t^2 \zeta+H\chi-a^2 \ddot{E}-3a^2 H \dot{E}  
\right)+  \dot{q}_t \left( \chi-a^2 \dot{E} \right)
+\frac{B_1}{2} \dphi=0\,. 
\label{eqE2}
\ee
The perturbation equations of motion (\ref{eqalpha})-(\ref{calWeq}) 
and (\ref{eqE2}) are written in the gauge-ready form, such that they can 
be used for  arbitrary gauge choices. 

The consistency of the above perturbation equations can be 
confirmed in terms of Bianchi identities. 
For this purpose, we resort to the following properties: 
\ba
&&
2\tp^2 D_{2}=-2H\left[\dot{q_t}+H\left(1-c_t^2\right)q_t\right]
-\tp\left(\dot{D_6}+HD_6+D_7\right)\,,\label{conD2}\\
&&
2\tp D_{3}=\frac{1}{a^3}\frac{d}{dt}(a^3D_5)-\frac{3H}{a^3}\frac{d}{dt}(a^3D_7)\,,\\
&&
D_4=-2\tp D_1-3HD_6\,.\label{conD4}
\ea
Using also the background Eqs.~(\ref{back2}), (\ref{back3}) and (\ref{coneq}), 
we find that the perturbation equations (\ref{eqalpha})-(\ref{calWeq}) 
satisfy the two particular relations: 
\ba
&&
\frac{1}{a^3}\frac{d}{dt}\left(a^3\E_{\alpha}\right)
-3H\E_{\zeta}-\tp\,\E_{\dphi}
-\frac{k^2}{a^2}\E_{\chi}
+3H(\rho_m+P_m) \E_{\delta \rho_m}+\E_v=0\,,
\label{Bianchi1}\\
&&
\E_{E}-\frac{1}{a^3}\frac{d}{dt}\left(a^3\E_{\chi}\right)
=0\,, 
\label{Bianchi2}
\ea
which correspond to the temporal and spatial components 
of Bianchi identities, respectively. 
Thus, we have confirmed the consistency of 
Eqs.~(\ref{eqalpha})-(\ref{calWeq}) with Bianchi identities.

\subsection{Gauge issues}
\label{gaugesec}

Since there are residual gauge degrees of freedom, we can 
fix the gauges in the perturbation equations derived 
in Sec.~\ref{actsec}.
We discuss the issues of gauge transformations, 
gauge-invariant 
variables, and gauge choices.
Let us consider a scalar infinitesimal gauge transformation from one coordinate 
$x^{\mu}=(t, x^i)$ to another coordinate $\tilde{x}^{\mu}
=(\tilde{t}, \tilde{x}^i)$, as
\be
\tilde{t}=t+\xi^{0}\,,\qquad 
\tilde{x}^{i}=x^{i}+\delta^{ij} \partial_{j} \xi\,,
\label{gaugetra}
\ee
where $\xi^{0}$ and $\xi$ are scalar quantities.
Then, the metric perturbations are subject to 
the transformations:
\be
\tilde{\alpha}=\alpha-\dot{\xi}^{0}\,,\qquad 
\tilde{\chi}=\chi+\xi^{0}-a^2 \dot{\xi}\,,\qquad 
\tilde{\zeta}=\zeta-H \xi^{0}\,,\qquad 
\tilde{E}=E-\xi\,, 
\label{gaugetra1}
\ee
while the perturbations associated with the scalar field and matter  
transform as 
\be
\widetilde{\delta \phi}=\delta \phi-\dot{\phi}\,\xi^{0}\,,
\qquad 
\widetilde{\delta \rho_m}=\delta \rho_m-\dot{\rho}_m\,\xi^{0}\,,
\qquad 
\tilde{v}=v-\xi^{0}\,.
\label{gaugetra2}
\ee

One can construct a family of gauge-invariant variables whose 
forms are unchanged under the transformation (\ref{gaugetra}). 
{}From Eq.~(\ref{gaugetra1}), the Bardeen gravitational 
potentials defined by \cite{Bardeen}
\be
\Psi=\alpha+\frac{d}{dt} 
\left( \chi - a^2 \dot{E} \right)\,,\qquad 
\Phi=\zeta+H \left( \chi - a^2 \dot{E} \right) 
\label{PsiPhi}
\ee
are gauge-invariant. 
On using the properties (\ref{gaugetra2}) as well, 
there are the following gauge-invariant combinations: 
\ba
& &
\delta \phi_{\rm f}=\delta \phi
-\frac{\dot{\phi}}{H}\zeta\,,
\qquad 
\delta \phi_{\rm N}=\delta \phi+\dot{\phi}\left(\chi-a^2 \dot{E}\right)\,,
\label{delphif}\nonumber \\
& &
\delta \rho_{\rm f}=\delta \rho_m -\frac{\dot{\rho}_m}{H}\zeta\,,\qquad 
\delta \rho_{\rm u}=\delta \rho_m -\frac{\dot{\rho}_m}
{\dot{\phi}}\delta \phi\,,\qquad  
\delta \rho_{\rm N}=\delta \rho_m+\dot{\rho}_m \left(\chi-a^2 \dot{E}\right)\,,
\label{rhof}\nonumber \\
& &
\delta_m=\frac{\delta \rho_m}{\rho_m}
+3H \left( 1+\frac{P_m}{\rho_m} \right)v\,,
\qquad 
{\cal R}=\zeta-\frac{H}{\dot{\phi}} \delta \phi\,,
\qquad
{\cal B}=Hv-\zeta\,.
\label{delmB}
\ea
The Mukhanov-Sasaki variable $\delta \phi_{\rm f}$ \cite{Mukhanov,Sasaki} 
is related to the gauge-invariant curvature perturbation 
${\cal R}$ \cite{Lukash,Lyth}, as
\be
{\cal R}=-\frac{H}{\dot{\phi}}\delta \phi_{\rm f}\,.
\label{R}
\ee

The residual gauge degrees of freedom can be removed by 
fixing $\xi^0$ and $\xi$ in the transformation (\ref{gaugetra2}).
There are several gauge conditions commonly 
used in the literature:
\ba
& &
\delta \phi=0\,,\quad ~\,E=0\,,\qquad\, \mathrm{(Unitary~gauge)}\,, 
\label{Ugauge}\\
& &
\zeta=0\,,\qquad E=0\,,\qquad\, \mathrm{(Flat~gauge)}\,,
\label{Fgauge}\\
& &
\chi=0\,,\qquad E=0\,,\qquad 
\mathrm{(Newtonian~gauge)}\,.
\label{Ngauge}
\ea
In the so-called synchronous gauge characterized by $\alpha=0$ and $\chi=0$, 
the transformation scalar $\xi^0$ is not unambiguously fixed and hence 
this gauge is not chosen in the following.
In Sec.~\ref{stasec}, we derive no-ghost conditions and the speed of scalar 
perturbations in full Horndeski theories 
by taking the three gauge conditions 
(\ref{Ugauge})-(\ref{Ngauge}). 
Independent of the gauge choices, we are dealing 
with the same physics.

\subsection{Perturbation equations expressed in terms of 
gauge-invariant variables}
\label{gauginsec}

We rewrite the perturbation equations of motion by using the gauge-invariant variables 
$\Psi, \Phi, \dphiN, {\cal B}$ without fixing gauge conditions. 
The perturbations $\alpha, \zeta, \dphi$ can be 
expressed in terms of $\Psi, \Phi, \dphiN$ as well 
as $E$, $\chi$ and their time derivatives.
On using the background Eqs.~(\ref{back2}), (\ref{back3}), (\ref{coneq}) with Eqs.~(\ref{conD2})-(\ref{conD4}), 
the terms containing $E$, $\chi$ and their time derivatives
identically vanish. 

First of all, the perturbation Eqs.~(\ref{calZeq}) and (\ref{calWeq}), 
which correspond to the equations of motion for $\dphi$ and $\Phi$ 
respectively, can be expressed as
\ba
&&
2D_1\ddot{\dphiN}+2(\dot{D_1}+3HD_1)\dot{\dphiN}
+\left(M_{\phi}^2-2D_2\frac{k^2}{a^2}\right)\dphiN
+3D_6\ddot{\Phi}+3(\dot{D_6}+3HD_6+D_7)\dot{\Phi}
-(2\tp D_1+3HD_6)\dot{\Psi}\notag\\
&&
-B_1\frac{k^2}{a^2}\Phi
-\left[2\tp(\dot{D_1}+3HD_1)+2D_5
+3H(\dot{D_6}+3HD_6-D_7)
-D_6\frac{k^2}{a^2}\right]\Psi=0\,,
\label{dphiNeq}\\
&&
-2q_t\ddot{\Phi}-\left(2\dot{q_t}+6Hq_t+\frac{\rho_m+P_m}{H}\right)\dot{\Phi}
+(2Hq_t-\tp D_6)\dot{\Psi}+\left[(\rho_m+P_m)\left(\frac{\dot{H}}{H^2}
+3c_m^2\right)-\frac{2q_tc_t^2}{3}\frac{k^2}{a^2}\right]\Phi
\notag\\
&&
-\left[\tp(\dot{D_6}+3HD_6+D_7)-2H(\dot{q_t}+3Hq_t)
+\rho_m+P_m+\frac{2q_t}{3}\frac{k^2}{a^2}\right]\Psi
-(\rho_m+P_m)\left[\frac{\dot{\cal B}}{H}
-\left(\frac{\dot{H}}{H^2}+3c_m^2\right){\cal B}\right] 
\nonumber \\
& &
+D_6\ddot{\dphiN}+(\dot{D_6}+3HD_6-D_7)\dot{\dphiN}
-\left( \dot{D_7}+3HD_7+\frac{B_1}{3}
\frac{k^2}{a^2}\right) \dphiN=0\,,
\label{ddPhi}
\ea
where 
\be
M_{\phi}^2 \equiv -2D_3\,.
\ee
The quantity $M_{\phi}^2$ corresponds to the mass 
squared of scalar-field perturbation $\delta \phi$. 
In quintessence with the Lagrangian $G_2=X-V(\phi)$, 
the term $G_{2,\phi \phi}/2$ in $D_3$ gives rise to 
the contribution $V_{,\phi \phi}\dphi_{\rm N}$ 
to Eq.~(\ref{dphiNeq}). 
For the models in which $M_{\phi}^2$ exceeds 
the order of $H^2$ in the past (like $f(R)$ models of the late-time 
cosmic acceleration \cite{fR1,fR2,fR3,fR4,fR5}), 
the condition $M_{\phi}^2>0$ is required for 
avoiding the tachyonic instability\footnote{The model 
$f(R)=R-\beta/R^n$ ($n>0$) proposed in 
Refs.~\cite{fRearly1,fRearly1b,fRearly2,fRearly3} leads to 
$M_{\phi}^2<0$, so it is ruled out by the tachyonic instability 
of scalar perturbations.} \cite{fR1,fR3,fRreview0,fRreview}. 
Equations (\ref{dphiNeq}) and (\ref{ddPhi}) can be 
closed by solving them for $\ddot{\dphiN}$ and $\ddot{\Phi}$.

The remaining perturbation equations of motion can be 
compactly expressed by using the following 
dimensionless variables:
\be
\aM \equiv \frac{\dot{q}_t}{Hq_t}\,,
\qquad 
\aB\equiv-\frac{\tp D_6}{2Hq_t}\,,
\qquad 
\aK\equiv\frac{2\tp^2D_1}{H^2q_t}\,, 
\label{nodim}
\ee
and 
\be
\dphiv_{\rm N} \equiv\frac{H}{\tp}\dphi_{\rm N}\,, 
\qquad 
h\equiv\frac{\dot{H}}{H^2}\,,
\qquad 
\tilde{\Omega}_m \equiv \frac{\rho_m}{3H^2q_t}\,, 
\qquad 
w_m\equiv\frac{P_m}{\rho_m}\,,
\qquad 
\K\equiv\frac{k}{aH}\,,
\qquad 
{\cal N} \equiv \ln a\,.
\label{nodim2}
\ee
The quantity $\aB$ introduced by Bellini and 
Sawicki \cite{Bellini}, which is denoted as 
$\aB^{\rm (BS)}$, is related to our $\aB$ according to 
\be
\aB=-\frac{1}{2}\aB^{\rm (BS)}\,.
\label{abre}
\ee
Our notations of $\alpha_{\rm M}$, $\alpha_{\rm B}$, and 
$\alpha_{\rm K}$ match with those used in Ref.~\cite{Gleyzes14} 
in the context of effective field theory of 
dark energy \cite{EFT1,EFT2,EFT3,EFT4,Piazza,EFT5,Gergely,Gao14,Kase15}.

{}From Eq.~(\ref{conD4}), the coefficient $D_4$ is written as 
$D_4=H^2 q_t (6 \alpha_{\rm B}-\alpha_{\rm K})/\dot{\phi}$. 
We also solve Eqs.~(\ref{back2}) and  (\ref{back3}) for $D_5$ and $D_7$ 
to express them in terms of $\alpha_{\rm B}$, $\alpha_{\rm K}$, 
and other dimensionless variables given in Eq.~(\ref{nodim2}). 
When we take the time derivative of 
$\delta_m$ in Eq.~(\ref{delmB}), we use Eq.~(\ref{coneq}) and 
the relation $c_m^2=\dot{P}_m/\dot{\rho}_m$.
Then, Eqs.~(\ref{eqalpha})-(\ref{eqdrho}), 
(\ref{veq}) and (\ref{eqE2}) reduce, respectively, to
\ba
&&
(6\aB-\aK)\vphiN'+6(1+\aB)\Phi'
+\left[h(\aK-12\aB-6)-9(1+w_m)\tilde{\Omega}_m+2\aB\K^2\right]\vphiN\notag\\
&&
+\left[9(1+w_m)\tilde{\Omega}_m+2\K^2\right]\Phi+(\aK-12\aB-6)\Psi
-3\tilde{\Omega}_m\delta_m+9(1+w_m)\tilde{\Omega}_m{\cal B}=0\,,
\label{eqalpha2}\\
&&
\aB\vphiN'+\Phi' 
-\left[h(1+\aB)+\frac32(1+w_m)\tilde{\Omega}_m\right]\vphiN
-(1+\aB)\Psi+\frac32(1+w_m)\tilde{\Omega}_m(\Phi+{\cal B})=0\,,
\label{eqchi2}\\
&&
\delta_m'-3(1+w_m){\cal B}'
+(1+w_m)\K^2(\Phi+{\cal B})
+3(c_m^2-w_m)\delta_m=0\,,
\label{eqdrho2}\\
&&
\Phi'+{\cal B}'-h(\Phi+{\cal B})-\Psi
-\frac{c_m^2}{1+w_m}\delta_m=0\,,
\label{veq2}\\
&&
\Psi+c_t^2 \Phi+\left( 1-c_t^2+\alpha_{\rm M} 
\right) \delta \varphi_{\rm N}=0\,,
\label{aniso}
\ea
where a prime represents a derivative 
with respect to ${\cal N}$.
Since ${\cal R}=\Phi-\delta \varphi_{\rm N}$, Eq.~(\ref{aniso}) can be also expressed as  
\be
\Psi+\left( 1+\alpha_{\rm M} \right) \Phi
+\left( c_t^2-1-\alpha_{\rm M} \right){\cal R}
=0\,.
\label{anire}
\ee
The relation (\ref{anire}) shows that the time-dependence of $q_t$ 
and the deviation of $c_t^2$ from 1 lead to the difference 
between $-\Psi$ and $\Phi$.
Under the bound (\ref{ctbound}), the gravitational slip 
($-\Psi \neq \Phi$) mostly arises from the time variation 
of $q_t$. 

We solve Eqs.~(\ref{eqalpha2})-(\ref{veq2}) for 
$\Phi', \vphiN',  \delta_m', {\cal B}'$ 
by using Eq.~(\ref{aniso}) to eliminate the perturbation $\Psi$. 
Then, we obtain
\ba
\hspace{-.5cm}&&
\Phi'=
-(c_t^2-b_2+4\aB b_1\K^2)\Phi-(1-c_t^2+\aM-h+b_2+4\aB^2b_1\K^2)\vphiN
+6\aB b_1\tilde{\Omega}_m\delta_m+b_2{\cal B}\,,
\label{Phieq}\\
\hspace{-.5cm}&&
\vphiN'=
-(c_t^2+b_3-4b_1\K^2)\Phi-(1-c_t^2+\aM-h-b_3-4\aB b_1\K^2)\vphiN
-6b_1\tilde{\Omega}_m\delta_m-b_3{\cal B}\,,\label{vphieq}\\
\hspace{-.5cm}&&
\frac{\delta_m'}{3(1+w_m)}=
\notag\\
\hspace{-.5cm}&&
\left[h-b_2+\left(4\aB b_1-\frac13\right)\K^2\right]\Phi
-(h-b_2-4\aB^2b_1\K^2)\vphiN
+\left(\frac{w_m}{1+w_m}-6\aB b_1\tilde{\Omega}_m\right)\delta_m
+\left(h-b_2-\frac13\K^2\right){\cal B}\,,\label{dmeq} \\
\hspace{-.5cm}&&
{\cal B}'=
(h-b_2+4\aB b_1\K^2)\Phi-(h-b_2-4\aB^2b_1\K^2)\vphiN
+\left(\frac{c_m^2}{1+w_m}-6\aB b_1\tilde{\Omega}_m\right)\delta_m+(h-b_2){\cal B}\,,
\label{Beq}
\ea
where $b_1, b_2, b_3$ are dimensionless quantities 
defined by 
\be
b_1 \equiv \frac{1}{2(\alpha_{\rm K}+6 \alpha_{\rm B}^2)}\,,
\qquad 
b_2 \equiv \frac32(1+w_m)(12\aB^2b_1-1)\tilde{\Omega}_m\,,
\qquad 
b_3 \equiv 18(1+w_m)\aB b_1\tilde{\Omega}_m\,.
\ee

The evolution of $\Phi$, $\vphiN$, $\delta_m$, ${\cal B}$, and 
$\Psi$ is known by integrating Eqs.~(\ref{Phieq})-(\ref{Beq}) 
with (\ref{aniso}) together with the background 
Eqs.~(\ref{back1}), (\ref{coneq}), (\ref{dH}), (\ref{ddphi}). 
In GR without the scalar field $\phi$, the gravitational 
potentials and the matter density contrast evolve as 
$-\Psi \simeq \Phi={\rm constant}$ and 
$\delta_m' \simeq \delta_m \propto a$ in the 
deep matter era. In modified gravity theories, the viable 
models of late-time cosmic acceleration are usually 
constructed to recover the behavior close to GR in the 
asymptotic past. In such cases, the initial conditions of perturbations 
can be chosen to satisfy $\Phi' \simeq 0$, $\vphiN' \simeq 0$, 
and $\delta_m' \simeq \delta_m$ in the early matter era.
Hence $\Phi, \vphiN, {\cal B}$ and $\Psi$ can be expressed 
in terms of $\delta_m$ by using Eqs.~(\ref{Phieq})-(\ref{dmeq}) 
and (\ref{aniso}). 

The choice of these initial conditions amounts to neglecting 
the oscillating mode of $\delta \phi_{\rm N}$ 
induced by a heavy mass term $M_{\phi}$ larger than $H$.
The large-field mass arises for $f(R)$ dark energy models 
in the asymptotic past, so in such cases, there is a 
fine-tuning problem of initial conditions for avoiding the 
dominance of the oscillating mode over the mode induced 
by $\delta_m$ \cite{fR2,fR4}. 
This problem does not arise for dark energy models 
in which $M_{\phi}$ does not exceed the order 
$H$ in the past (like k-essence and Galileons).

\section{Stability conditions in the small-scale limit}
\label{stasec}

By using the scalar perturbation equations of motion obtained in Sec.~\ref{scalarsec}, 
we derive conditions for the absence of 
ghost and Laplacian instabilities in the small-scale limit. 
We choose the three different gauges (\ref{Ugauge})-(\ref{Ngauge}) 
and show that these stability conditions are independent of the choice of gauges.

\subsection{Unitary gauge}

We begin with the unitary gauge (\ref{Ugauge}), under which the dynamical perturbations 
are given by  ${\cal R}=\zeta$ and $\delta \rho_{\rm u}=\delta \rho_m$. 
We solve Eqs.~(\ref{eqalpha}), (\ref{eqchi}), (\ref{eqdrho}) for $\alpha$, $\chi$, $v$, and substitute them
into Eq.~(\ref{Ss}).
After the integration by parts, the quadratic action in Fourier space is 
expressed in the form 
\be
{\cal S}_s^{(2)}=\int dt d^3x\,a^{3}\left( 
\dot{\vec{\mathcal{X}}}^{t}{\bm K}\dot{\vec{\mathcal{X}}}
-\frac{k^2}{a^2}\vec{\mathcal{X}}^{t}{\bm G}\vec{\mathcal{X}}
-\vec{\mathcal{X}}^{t}{\bm M}\vec{\mathcal{X}}
-\vec{\mathcal{X}}^{t}{\bm B}\dot{\vec{\mathcal{X}}}
\right)\,,
\label{Ss2}
\ee
where ${\bm K}$, ${\bm G}$, ${\bm M}$, ${\bm B}$ 
are $2 \times 2$ matrices, and 
\be
\vec{\mathcal{X}}^{t}=\left({\cal R}, \delta \rho_{\rm u}/k \right) \,.
\ee
The leading-order contributions to 
${\bm M}$ and ${\bm B}$ are of order $k^0$. 

In the small-scale limit ($k \to \infty$), 
the non-vanishing components of 
matrices ${\bm K}$ and ${\bm G}$ are give by 
\ba
&&
K_{11}^{(\rm u)}=\frac{\tp^2 q_t q_s}{(2Hq_t-\tp D_6)^2}\,,
\qquad 
K_{22}^{(\rm u)}=\frac{a^2}{2(\rho_m+P_m)}\,,
\notag\\
&&
G_{11}^{(\rm u)}=-q_tc_t^2-\frac{\rho_m+P_m}{2Hq_t-\tp D_6}{\cal F}_1
+\frac{1}{a}\frac{d}{dt}\left(a{\cal F}_1\right)\,,
\qquad 
G_{22}^{(\rm u)}=\frac{a^2c_m^2}{2(\rho_m+P_m)}\,,
\label{KGu}
\ea
where $q_s$ is defined by Eq.~(\ref{calD}), and 
\be
{\cal F}_1=\frac{2q_t^2}{2Hq_t-\tp D_6}\,.
\ee
Since there are no off-diagonal components in 
${\bm K}$ and ${\bm G}$, the matter perturbation 
$\delta \rho_{\rm u}$ is decoupled from the other 
field ${\cal R}$. For the matter sector, the conditions 
for the absence of ghost and Laplacian instabilities 
correspond to $\rho_m+P_m>0$ and $c_m^2>0$. 
For the perturbation ${\cal R}$, the ghost does not arise for
\be
q_s^{(\rm u)} \equiv K_{11}^{(\rm u)}
=\frac{\tp^2 q_t q_s}{(2Hq_t-\tp D_6)^2}>0\,. 
\label{qsue}
\ee
Since the absence of tensor ghosts requires that $q_t>0$, 
the condition (\ref{qsue}) translates to
\be
q_s>0\,,
\label{noghost}
\ee
under which the denominators of Eqs.~(\ref{dH}) and 
(\ref{ddphi}) do not cross 0.

Taking the small-scale limit in Eq.~(\ref{Ss2}), we obtain 
the dispersion relation of the form 
\be
{\rm det} \left( c_{\rm s}^2 {\bm K}-{\bm G} \right)=0\,,
\ee
where $c_{\rm s}$ is the propagation speed of scalar perturbations. 
One of the solutions is the matter propagation speed 
squared $c_m^2$, while the other solution is 
$c_s^2={G_{11}^{(\rm u)}}/{q_s^{(\rm u)}}$. 
On using Eqs.~(\ref{back2}), (\ref{Bi}), and (\ref{conD2}) 
to solve for $\ddot{\phi}$, $\dot{q}_t$, and $\dot{D_6}$, 
the latter solution can be simply expressed as  
\be
c_s^2=\frac{G_{11}^{(\rm u)}}{q_s^{(\rm u)}}
=-\frac{c_t^2D_6^2+2B_1D_6+4q_t D_2}{q_s}\,.
\label{cs}
\ee
The small-scale Laplacian instability can be avoided for 
\be
c_s^2>0\,.
\ee
In sections after \ref{quinsec}, we will compute $c_s^2$ 
for concrete dark energy models in 
the framework of Horndeski theories.

\subsection{Flat gauge}

In the flat gauge (\ref{Fgauge}), the dynamical 
scalar perturbations are given by the matrix 
$\vec{\mathcal{X}}^{t}=
\left(\dphi_{\rm f}, \delta \rho_{\rm f}/k \right)$. 
Solving Eqs.~(\ref{eqalpha}), (\ref{eqchi}), (\ref{eqdrho}) for $\alpha$, $\chi$, $v$  
and substituting them into Eq.~(\ref{Ss}), the second-order action reduces to 
the same form as Eq.~(\ref{Ss2}) after the integration by parts.
In the small-scale limit, the matrix components
$K_{22}^{(\rm f)}$ and $G_{22}^{(\rm f)}$ 
are identical to those in the unitary gauge. 
The other nonvanishing matrix components 
are given by 
\ba
&&
q_s^{(\rm f)} \equiv K_{11}^{(\rm f)}=\frac{H^2 q_t q_s}
{(2Hq_t-\tp D_6)^2}\,,
\qquad 
G_{11}^{(\rm f)}=-D_2+\frac{D_6D_7-(\rho_m+P_m){\cal F}_2}{2Hq_t-\tp D_6}
+\frac{1}{a}\frac{d}{dt}\left(a{\cal F}_2\right)\,,
\label{qsf}
\ea
where 
\be
{\cal F}_2=\frac{D_6^2}{2(2Hq_t-\tp D_6)}\,.
\ee
Note that we eliminated the term $D_4$ by using the relation (\ref{conD4}).
As long as there is no tensor ghost, the condition for avoiding 
the scalar ghost again corresponds to $q_s>0$.  

On using Eqs.~(\ref{back2}) and (\ref{conD2}), 
it follows that 
\be
\frac{q_s^{(\rm f)}}{q_s^{(\rm u)}}=
\frac{G_{11}^{(\rm f)}}{G_{11}^{(\rm u)}}=
\frac{H^2}{\tp^2}\,.
\ee
Hence the scalar propagation speed squared 
$c_s^2=G_{11}^{(\rm f)}/q_s^{(\rm f)}$ 
in the flat gauge is equivalent to that in the unitary gauge.

\subsection{Newtonian gauge}
\label{Newtonsec}

Let us finally consider the Newtonian gauge (\ref{Ngauge}), under 
which $\Phi=\zeta$, $\delta \phi_{\rm N}=\delta \phi$, and 
$\delta \rho_{\rm N}=\delta \rho_m$. 
We first solve Eqs.~(\ref{eqalpha}) and (\ref{eqdrho}) for nondynamical perturbations 
$\alpha$ and $v$, and substitute them into Eq.~(\ref{Ss}). 
Then, the terms proportional to $\dot{\Phi}^2$,  
$\dot{\dphi}_{\rm N}^2$, and $\dot{\Phi}\dot{\dphi}_{\rm N}$ appear in the 
second-order action. Apparently, this looks as if the two fields 
$\Phi$ and $\dphi_{\rm N}$ were dynamical, but the second-order action 
in the small-scale limit is factorized as
\be
{\cal S}_s^{(2)}=\int dt d^3x\,a^{3}\left[ 
\frac{3q_s}{4\{ 3H^2 q_t-\dot{\phi} (D_1 \dot{\phi}+3H D_6)\}} 
\left( \dot{\phi} \dot{\Phi}-H \dot{\delta \phi}_{\rm N} \right)^2
+\frac{a^2}{2(\rho_m+P_m)} \frac{ \dot{\delta \rho}_{\rm N}^2}{k^2}
+\cdots
\right]\,,
\label{SsN}
\ee
where the abbreviation corresponds to terms without containing the product 
of first-order time derivatives of perturbed quantities.
If we consider the combination
\be
{\cal R}=\Phi-\frac{H}{\dot{\phi}}\delta \phi_{\rm N}\,,\qquad {\rm or} \qquad
\delta \phi_{\rm f}=\delta \phi_{\rm N}-\frac{\dot{\phi}}{H}\Phi\,,
\ee
then the action (\ref{SsN}) can be expressed in terms of $\dot{{\cal R}}^2$ or 
$\dot{\delta \phi}_{\rm f}^2$ in addition to $\dot{\delta \rho}_{\rm N}^2$, 
without any other products of first-order time derivatives of perturbations. 
In other words, ${\cal R}$ or $\dphi_{\rm f}$ corresponds to 
the dynamical perturbation besides the matter perturbation $\delta \rho_{\rm N}$. 

If we choose ${\cal R}$ as a dynamical perturbation, the terms in the abbreviation 
of Eq.~(\ref{SsN}) contain nondynamical variables like $\dot{\delta \phi}_{\rm N}$ and 
$\delta \phi_{\rm N}$, which can be eliminated by using Eqs.~(\ref{eqchi}) and (\ref{eqE2}). 
After the integration by parts, the second-order action (\ref{SsN}) reduces to 
the same form as Eq.~(\ref{Ss2}) with dynamical perturbations
$\vec{\mathcal{X}}^{t}=
\left({\cal R}, \delta \rho_{\rm N}/k \right)$. 
In the small-scale limit, the components of matrices ${\bm K}$ and ${\bm G}$ 
are identical to those in the unitary gauge given in Eq.~(\ref{KGu}).

If the combination $\delta \phi_{\rm f}$ is chosen as a dynamical perturbation besides 
$\delta \rho_{\rm N}$,
the nondynamical variables like $\dot{\Phi}$ and $\Phi$ can be eliminated from
the action (\ref{SsN}) by using Eqs.~(\ref{eqchi}) and (\ref{eqE2}). 
In the small-scale limit, the resulting second-order action is expressed 
in the form (\ref{Ss2}) with the same matrix components of 
${\bm K}$ and ${\bm G}$ as those in the flat gauge.

Thus, the conditions for the absence of ghost and Laplacian instabilities in the Newtonian gauge
are equivalent to those in the unitary and flat gauges.

\subsection{Summary}

We have shown that the quantities 
$q_s^{(\rm u)}$ and $q_s^{(\rm f)}$
contain the common term $q_s$.
The positivity of this term, i.e., 
\be
q_s=4D_1 q_t+3D_6^2>0\,,
\label{qscon}
\ee
is the condition for the absence of scalar ghosts 
for any gauge choices. 
The small-scale Laplacian instability can be 
avoided as long as the gauge-invariant sound speed 
squared is positive, i.e., 
\be
c_s^2 
=\frac{G_{11}^{(\rm u)}}{q_s^{(\rm u)}}
=\frac{G_{11}^{(\rm f)}}{q_s^{(\rm f)}}
=-\frac{c_t^2D_6^2+2B_1D_6+4q_t D_2}{q_s}>0\,.
\label{csge}
\ee
Depending on the problems at hand, we can choose most 
convenient gauges. 

We caution that the perturbations $\delta \phi_{\rm f}$ 
and $\delta \rho_{\rm f}$ contain the term $H$ 
in their denominators, so they are not well defined 
at $H=0$. In this case, it apparently 
looks possible to regulate the combination $\zeta/H$ 
by choosing the flat gauge ($\zeta=0$). 
However, the problem arises for the gravitational 
potential $\Phi$ and the curvature perturbation ${\cal R}$, 
both of which exactly vanish at $H=0$ for the gauge 
choice $\zeta=0$.
This suggests that, for the cosmological evolution crossing 
$H=0$ (such as the  bouncing cosmology),  
it is not appropriate to choose the flat gauge \cite{HKT18}.
In the bouncing Universe, the quantity 
$q_s^{(\rm f)}$ vanishes at the bounce ($H=0$). Apparently, 
this looks as if a strong coupling problem arises, but 
it simply comes from the inappropriate gauge choice 
for this problem. 
The real strong coupling problem arises when the 
quantity $q_s$ in Eq.~(\ref{qscon}) approaches 0, 
in which case the background Eqs.~(\ref{dH}) 
and (\ref{ddphi}) exhibit divergences.
In the unitary gauge, ${\cal R}$ and $\delta \rho_{\rm u}$ corresponds to dynamical perturbations, but 
they are not well defined at $\dot{\phi}=0$. 
For this gauge choice, the problems similar to those mentioned 
above arise for the case in which $\dot{\phi}$ crosses 0. 
If neither $H$ nor $\dot{\phi}$ vanishes during the cosmological evolution, 
we can choose any gauge among 
(\ref{Ugauge})-(\ref{Ngauge}). 

\section{Growth of large-scale structures}
\label{obsersec} 

The dark energy EOS $w_{\rm DE}$ introduced 
in Sec.~\ref{backsec2} is a key quantity to distinguish 
between different dark energy models at the background level. 
In modified gravity theories, the gravitational coupling with 
the matter sector is different from that in GR. 
In this case, the growth rate of matter perturbations and 
the evolution of gravitational potentials are subject to 
modifications. Then, we can distinguish between different
models of the late-time cosmic acceleration from 
the observations of large-scale structures, weak lensing, 
and CMB. 

In this section, we study observables associated 
with the evolution of linear cosmological perturbations in full 
Horndeski theories given by the action (\ref{actionfull}). 
By using the linear perturbation equations of motion derived 
in Sec.~\ref{scalarsec}, it is possible to estimate the effective gravitational coupling 
with matter perturbations on scales 
relevant to the growth of large-scale structures in the gauge-invariant way. 
We also discuss the gravitational coupling around local objects 
screened by nonlinear interactions.  

\subsection{Observable quantities}

Since we are interested in the evolution of perturbations 
after the end of the radiation-dominated epoch, 
we consider nonrelativistic matter satisfying 
\be
P_m=0\,,\qquad c_m^2=0\,,
\ee
for the action ${\cal S}_m$. 
Taking the time derivative of Eq.~(\ref{eqdrho}) and 
using Eq.~(\ref{veq}), the gauge-invariant matter 
density contrast $\delta_m=\delta \rho_m/\rho_m+3Hv$ 
obeys
\be
\ddot{\delta}_m+2H\dot{\delta}_m+\frac{k^2}{a^2}\Psi
=3 \left( \ddot{\cal B}+2H \dot{\cal B} \right)\,,
\label{delmeq}
\ee
where $\Psi$ and ${\cal B}$ are gauge-invariant 
perturbations defined in Eqs.~(\ref{PsiPhi}) and (\ref{delmB}). 
We relate the gravitational potential $\Psi$ with  
the density contrast $\delta_m$ through the modified 
Poisson equation 
\be
\frac{k^2}{a^2}\Psi
=-4\pi G\mu \rho_m \delta_m\,,
\label{Poi}
\ee
where
\be
\mu=\frac{G_{\rm eff}}{G}\,.
\label{mudef}
\ee
The quantity $\mu$ characterizes the ratio between the 
effective gravitational coupling $G_{\rm eff}$ and 
the Newton constant $G=1/(8\pi M_{\rm pl}^2)$. 
In Sec.~\ref{quasisec}, we will derive an explicit form of 
$\mu$ in full Horndeski theories by using a so-called 
quasi-static approximation for perturbations deep 
inside the sound horizon.

We also define the gravitational slip parameter $\eta$ and 
the effective gravitational potential $\psi_{\rm eff}$ 
relevant to the light bending in weak lensing and CMB 
observations, as \cite{weak1,weak2} 
\be
\eta=-\frac{\Phi}{\Psi}\,,\qquad 
\psi_{\rm eff}=\Phi-\Psi\,.
\label{etadef}
\ee
{}From Eqs.~(\ref{Poi}) and (\ref{etadef}), it follows that 
\be
\frac{k^2}{a^2}\psi_{\rm eff}=
8\pi G\,\Sigma\rho_m \delta_m\,, 
\label{psieff}
\ee
where 
\be
\Sigma=\frac{1+\eta}{2}\frac{G_{\rm eff}}{G}\,.
\label{Sigmadef}
\ee
The dimensional ratios $\mu$ and $\Sigma$ are 
two key quantities characterizing the linear growth of  
$\delta_m$ as well as $\Psi, \psi_{\rm eff}$.

\subsection{Quasi-static approximation deep inside 
the sound horizon}
\label{quasisec}

To confront dark energy models in the framework of Horndeski 
theories with the observations of large-scale structures and 
weak lensing, we would like to derive analytic expressions of 
$\mu$ and $\Sigma$ for the perturbations 
deep inside the sound horizon. 
In doing so, we exploit the scalar perturbation equations of motion 
expressed in terms of gauge-invariant variables.
As we will see below, we do not need to fix gauges for the derivation 
of $\mu$ and $\Sigma$.

In Sec.~\ref{stasec}, we showed that the curvature perturbation 
${\cal R}$ (or equivalently $\dphi_{\rm f}=-\tp {\cal R}/H$) and 
the matter perturbation are the dynamical degrees of freedom. 
Let us derive the closed-form equation of motion for ${\cal R}$ 
in the small-scale limit. 
First of all, we replace $\Phi$ with ${\cal R}$ in the perturbation equations 
(\ref{dphiNeq}), (\ref{ddPhi}) and (\ref{eqalpha2})-(\ref{aniso}) by using 
the relation ${\cal R}=\Phi-\dphiv_{\rm N}$. 
Secondly, we solve Eqs.~(\ref{eqalpha2})-(\ref{aniso}) and the 
time derivative of Eq.~(\ref{aniso}) for $\dot{\cal B}$, ${\cal B}$, 
$\dot{\delta \varphi_{\rm N}}$, $\delta \varphi_{\rm N}$, $\dot{\Psi}$ and $\Psi$ 
in order to express them in terms of $\dot{\cal R}$, ${\cal R}$, 
$\dot{\delta_m}$ and $\delta_m$. 
Finally, we substitute these solutions into the equation derived after 
eliminating $\ddot{\dphi_{\rm N}}$ from Eqs.~(\ref{dphiNeq}) 
and (\ref{ddPhi}). Taking the $k\to\infty$ limit, 
this equation reduces to 
\be
\ddot{\cal R}+\left(3H+\frac{\dot{q_s}^{(\rm u)}}{q_s^{(\rm u)}}\right) \dot{\cal R}
+c_s^2\frac{k^2}{a^2}{\cal R}\simeq
\frac{1}{2q_t q_s^{(\rm u)}}\left(q_t c_t^2+q_s^{(\rm u)} c_s^2-\frac{1+\aM}{1+\aB}q_t\right)
\rho_m\delta_m\,, 
\label{ddotR}
\ee
where we used Eqs.~(\ref{Bi}), (\ref{conD2}), (\ref{nodim}), (\ref{qsue}) and (\ref{cs}) 
to remove $D_1$, $D_2$, $D_6$, $D_7$ and $\dot{q_t}$. 
The general solution to Eq.~(\ref{ddotR}) can be expressed in the form 
\be
{\cal R}={\cal R}_{\rm ho}+{\cal R}_{\rm ind}\,,
\ee
where ${\cal R}_{\rm ho}$ is a homogenous solution obtained by setting 
the right hand side of Eq.~(\ref{ddotR}) to be 0, and ${\cal R}_{\rm ind}$ 
is the special solution induced by the matter perturbation $\delta_m$. 
The solution ${\cal R}_{\rm ho}$ contains an oscillating mode induced 
by the second time derivative $\ddot{\cal R}$.
Provided that the variation of ${\cal R}$ over the Hubble time scale is not large such that 
the conditions $|\ddot{{\cal R}}| \lesssim |H^2 {\cal R}|$ and  $|\dot{{\cal R}}| \lesssim |H {\cal R}|$ 
are satisfied with 
$|\dot{q_s}^{(\rm u)}/q_s^{(\rm u)}| \lesssim H$, the dominant contribution to 
the left hand side of Eq.~(\ref{ddotR}) is the term $c_s^2k^2 {\cal R}/a^2$ 
for the modes deep inside the sound horizon ($c_s^2k^2/a^2 \gg H^2$).
This term balances with the matter perturbation $\delta_m$, in which case 
the curvature perturbation ${\cal R}$ is dominated by the matter 
induced mode ${\cal R}_{\rm ind}$.
 
In the so-called quasi-static approximation, the dominant contributions to 
the perturbation equations are regarded as those containing the terms 
$\delta_m$ and $k^2/a^2$, without any time derivatives 
of metric perturbations \cite{Boi00,dreview4,Tsuji07,DKT12,Nesseris,Song}. 
Provided that $c_s^2$ is not very close to 0, the quasi-static approximation 
is sufficiently accurate for sub-horizon perturbations in dark energy models 
where the mass $M_{\phi}$ 
associated with the field perturbation $\delta \phi_{\rm N}$ is at most of 
order $H$. This is generally the case for a nearly massless scalar field 
like quintessence, k-essence, and Galileons \cite{Kase10}. 
One exception is $f(R)$ models of the late-time cosmic acceleration, in which case 
the scalar mass $M_{\phi}$ becomes larger than $H$ as we go back to the past. 
Then, the oscillating mode of field perturbations cannot be ignored 
relative to the matter-induced mode in the asymptotic past \cite{fR2,fR4}. 
In other words, in $f(R)$ gravity, we need a fine tuning for initial conditions 
of perturbations such that $|{\cal R}_{\rm ind}| \gg |{\cal R}_{\rm ho}|$.

In the following, we apply the quasi-static approximation 
to the scalar perturbation equations of motion.
First of all, we remind that there is the relation (\ref{aniso}) among two 
gravitational potentials $\Psi$ and $\Phi$. 
Applying the quasi-static approximation 
to Eq.~(\ref{eqalpha2}), it follows that 
\be
2q_t \frac{k^2}{a^2} \left( \Phi + \aB \dphiv_{\rm N} \right)
\simeq \rho_m \delta_m\,.
\label{quasi2}
\ee

In Eq.~(\ref{dphiNeq}), there exists the mass term 
$M_{\phi}^2\dphi_{\rm N}$ of the scalar-field perturbation. 
In viable models of the late-time cosmic acceleration 
based on $f(R)$ gravity \cite{fR1,fR2,fR3,fR4,fR5} 
and BD theories with the field potential \cite{BD1,BD2}, 
the mass of $\dphi_{\rm N}$ tends to be large 
in the early cosmological epoch. 
Taking into account its contribution 
and using the quasi-static approximation 
in Eq.~(\ref{dphiNeq}), we obtain 
\be
M_{\phi}^2 \delta \phi_{\rm N}
+\frac{k^2}{a^2} \left( D_6 \Psi-B_1 \Phi-2D_2 
\delta \phi_{\rm N} \right) \simeq 0\,. 
\label{quasi3d}
\ee
{}From Eq.~(\ref{cs}), the coefficient $D_2$ can be expressed 
as $D_2=-(q_s c_s^2+c_t^2 D_6^2+2B_1 D_6)/(4q_t)$. 
We also note that $B_1$ defined in Eq.~(\ref{Bi})
is written as $B_1=2Hq_t(1-c_t^2+\alpha_{\rm M})/\dot{\phi}$.
Then, Eq.~(\ref{quasi3d}) can be expressed as
\be
q_t\frac{k^2}{a^2}\left[\aB\Psi+(1-c_t^2+\aM)\Phi\right]
-\left[ q_t \frac{k^2}{a^2} \alpha_{\rm B} 
\left\{ c_t^2 (2+\alpha_{\rm B})-2(1+\alpha_{\rm M}) 
\right\}+\frac{\dot{\phi}^2 (c_s^2 k^2 q_s
+2M_{\phi}^2a^2 q_t)}{4a^2H^2 q_t}
\right] \delta \varphi_{\rm N} \simeq 0\,, 
\label{quasi3}
\ee
where we used $D_6=-2Hq_t \alpha_{\rm B}/\dot{\phi}$.

Solving Eqs.~(\ref{aniso}), (\ref{quasi2}), and (\ref{quasi3}) 
for $\Psi, \Phi$, and $\dphiv_{\rm N}$, we obtain
\be
\Psi=-\frac{1}{2\Delta_2}
\left(\Delta_1^2+\frac{c_t^2\Delta_2}{q_t}\right)
\frac{a^2}{k^2}\rho_m\delta_m\,,
\qquad
\Phi=\frac{1}{2\Delta_2}
\left(\aB\Delta_1+\frac{\Delta_2}{q_t}\right)
\frac{a^2}{k^2}\rho_m\delta_m\,,
\qquad
\dphiv_{\rm N}=-\frac{\Delta_1}{2\Delta_2}
\frac{a^2}{k^2}\rho_m\delta_m\,,
\ee
where
\be
\Delta_1 \equiv c_t^2(1+\aB)-1-\aM\,, 
\qquad
\Delta_2 \equiv \frac{\tp^2q_sc_s^2}{4H^2q_t}
\left( 1+\frac{2a^2 M_{\phi}^2 q_t}{c_s^2 k^2 q_s} 
\right)\,.
\label{Delta2}
\ee

Then the quantities $\mu$, $\eta$, and $\Sigma$ defined 
by Eqs.~(\ref{mudef}), (\ref{etadef}), and 
(\ref{Sigmadef}) reduce to
\ba
\mu
&=& \frac{c_t^2}{8\pi Gq_t} 
\left(1+\frac{q_t \Delta_1^2}{c_t^2\Delta_2} \right)\,,
\label{Gefff}\\
\eta &=&
\frac{q_t \aB\Delta_1+\Delta_2}
{q_t \Delta_1^2+c_t^2\Delta_2}\,,
\label{etaf}\\
\Sigma &=& 
\frac{1+c_t^2}{16\pi Gq_t} 
\left[ 1+\frac{q_t (\aB+\Delta_1)\Delta_1}
{(1+c_t^2)\Delta_2} \right]\,.
\label{Sigmaf}
\ea
The effective gravitational coupling $\mu=G_{\rm eff}/G$ is composed of the following two contributions :
\be
\mu_t=\frac{c_t^2}{8\pi G q_t}\,,\qquad 
\xi_s=\frac{q_t \Delta_1^2}{c_t^2\Delta_2}\,.
\ee
The term $\mu_t$ arises from the modification of gravity in the tensor sector. Under the no-ghost and stability conditions $q_t>0$ and $c_t^2>0$, we have $\mu_t>0$. 
The term $\xi_s$ quantifies the interaction between 
the scalar field $\phi$ and matter.  
For the field mass squared $M_{\phi}^2>0$, the quantity 
$\xi_s$ is positive under the no-ghost conditions 
$q_t>0, q_s>0$ and the stability conditions $c_t^2>0, c_s^2>0$. 
Thus, the scalar-matter interaction in full Horndeski theories 
is attractive under theoretically consistent 
conditions \cite{Tsuji15,Kase18}. 
Since $\mu, \eta, \Sigma$ given in 
Eqs.~(\ref{Gefff})-(\ref{Sigmaf}) have been derived without 
fixing any gauge conditions, they are gauge-invariant 
quantities for the modes deep inside the sound horizon.

The modification to the effective gravitational coupling 
manifests itself in the ``massless'' regime in which the condition 
$M_{\phi}^2 \lesssim (c_s^2 k^2/a^2)(q_s/q_t)$ holds. 
For $M_{\phi}^2 \lesssim H^2$ with $q_s/q_t={\cal O}(1)$, 
this condition is satisfied for perturbations inside 
the sound horizon ($H^2 \lesssim c_s^2 k^2/a^2$). 
Taking the massless limit $M_{\phi}^2 \to 0$ in $\Delta_2$, 
the $k$ dependence disappears in the expressions of 
$\mu, \eta, \Sigma$. 
In Ref.~\cite{Kase18}, the present authors derived these quantities 
by choosing the unitary gauge. 
Since there is the relation 
$q_s^{(\rm u)}/q_s=\tp^2/[4H^2q_t(1+\aB)^2]$ from Eq.~(\ref{qsue}), 
the massless limits of Eqs.~(\ref{Gefff}) 
and (\ref{etaf}) yield
\be
\mu=\frac{c_t^2}{8\pi G q_t} \left[ 1+
\frac{q_t \Delta_1^2}{q_s^{(\rm u)} 
c_s^2 c_t^2 (1+\alpha_{\rm B})^2} \right]\,,\qquad 
\eta=\frac{q_t \aB \Delta_1+q_s^{(\rm u)} 
c_s^2 (1+\alpha_{\rm B})^2}
{q_t \Delta_1^2+q_s^{(\rm u)} 
c_s^2 c_t^2 (1+\alpha_{\rm B})^2}\,,
\ee
which coincide with Eqs.~(3.33) and (3.34) of 
Ref.~\cite{Kase18}, respectively, after replacing $q_t$ 
with $2Q_t$. 

The GW170817 event \cite{GW170817} placed the 
tight bound (\ref{ctbound}) on the tensor propagation speed $c_t$.
Setting $c_t^2=1$ in Eqs.~(\ref{Gefff})-(\ref{Sigmaf}), 
it follows that 
\ba
\mu
&=& \frac{1}{8\pi Gq_t} 
\left[1+\frac{q_t (\aB-\aM)^2}{\Delta_2} \right]\,,
\label{Gefff2}\\
\eta &=&
1+\frac{q_t\aM(\aB-\aM)}{\Delta_2+q_t(\aB-\aM)^2}\,, 
\label{eta2}\\
\Sigma &=& \frac{1}{8\pi Gq_t} 
\left[ 1+\frac{q_t (2\aB-\aM)(\aB-\aM)}{2\Delta_2} \right]\,, 
\label{Sigmaf2}
\ea
where we used the property $\Delta_1=\aB-\aM$.
The nonvanishing values of $\alpha_{\rm B}$ and 
$\alpha_{\rm M}$ satisfying $\alpha_{\rm B} 
\neq \alpha_{\rm M}$ lead to the enhancement 
of the gravitational interaction with matter, such that 
$\mu>1/(8\pi G q_t)$. 
The gravitational slip ($\eta \neq 1$) arises for the theories 
with $\alpha_{\rm M} \neq 0$ and $\alpha_{\rm B} \neq \alpha_{\rm M}$.
Under the no-ghost and stability conditions, 
the quantity $\Sigma$ is larger
than $1/(8\pi Gq_t)$ for $(2\alpha_{\rm B}-\alpha_{\rm M})
(\alpha_{\rm B}-\alpha_{\rm M})>0$.

{}From Eq.~(\ref{veq2}), the perturbation ${\cal B}$ is at most 
of order $\Psi$. Then the terms on the right hand side 
of Eq.~(\ref{delmeq}) is at most of order $H^2 \Psi$, so 
they can be ignored compared to those on its left hand side 
for the perturbations deep inside the Hubble radius.
On using Eq.~(\ref{Poi}), the matter density contrast 
approximately obeys 
\be
\ddot{\delta}_m+2H \dot{\delta}_m 
-4\pi G \mu \rho_m \delta_m \simeq 0\,.
\label{delmeqf}
\ee
For a given model the quantity $\mu$ is known 
from Eq.~(\ref{Gefff2}), so we can integrate
Eq.~(\ref{delmeqf}) to solve for $\delta_m$. 
The matter power spectrum and the growth rate of 
matter perturbations can be constrained 
from the measurements of galaxy 
clusterings \cite{2df,Tegmark} and redshift-space 
distortions \cite{Kaiser,RSD1,RSD2,RSD3,RSD4,Okumura}, 
so it is possible to distinguish between different dark energy models.
The evolution of two gravitational potentials is also 
determined from Eqs.~(\ref{Poi}) and (\ref{psieff}).
This information can be used to place further constraints 
on dark energy models from the observations of 
CMB \cite{Planck2015,Planck2018}  
and weak lensing \cite{lensing1,lensing2}.

\subsection{Screened gravitational coupling}
\label{screensec}

We have shown that Horndeski theories generally give 
rise to modifications to the gravitational interaction 
for scales relevant to the growth of large-scale structures 
and weak lensing. 
In local regions of the Universe, the fifth force induced by the scalar-matter interaction needs to be small 
for the consistency with solar-system tests of gravity.
There are several mechanisms to suppress the propagation of fifth forces in regions of the high density: 
(i) chameleon mechanism \cite{chame1}, and 
(ii) Vainshtein mechanism \cite{Vain}.

The chameleon mechanism can be at work for a scalar 
potential whose mass is different depending on the matter 
densities in the surrounding environment.
If the effective scalar mass is sufficiently large in regions of the high density, the coupling between the field and matter can 
be suppressed by having a thin shell inside a spherically symmetric body (see Ref.~\cite{chame2} for detail). 
In $f(R)$ gravity or BD theories with 
a scalar potential, it is possible to design functional forms of $f(R)$ or the potential $V(\phi)$ to realize the large mass 
squared $M_{\phi}^2$ for increasing $R$, while realizing 
the late-time cosmic acceleration by the potential of a light 
scalar field \cite{fR1,fR2,fR3,fR4,BD1,BD2}. 
Cosmologically, $M_{\phi}^2$ decreases in time, 
so there is a transition from the ``massive'' regime 
$M_{\phi}^2 \gg (c_s^2 k^2/a^2)(q_s/q_t)$ to 
the ``massless'' regime 
$M_{\phi}^2 \ll (c_s^2 k^2/a^2)(q_s/q_t)$. 
In the massive limit $M_{\phi}^2 \to \infty$, we have 
$\xi_s \to 0$ and $\mu \to c_t^2/(8\pi G q_t)$, by reflecting 
the fact that the scalar degree of freedom does not propagate.
This is the regime in which the chameleon mechanism is 
at work in the local region whose matter density $\rho_m$ 
is much higher than today's critical cosmological 
density $\rho_c$.

The Vainshtein mechanism operates around local sources 
in the presence of nonlinear scalar derivative interactions. 
One of the representative examples is the cubic Galileon 
given by the Lagrangian $X \square \phi$.
This nonlinear interaction leads to the decoupling of 
the field $\phi$ from matter within a radius $r_V$ called 
the Vainshtein radius (see Refs.~\cite{DKT12,KKY12,Kase13} 
for detail). 
For the Sun, the Vainshtein radius 
can be of order $10^{20}$~cm, which is much larger 
than the solar-system scale. On scales relevant to 
the growth of large-scale structures ($\gtrsim 10^{24}$~cm), 
the cubic Galileon modifies the effective gravitational 
coupling $\mu$, while, in the solar system, 
the scalar-matter interaction term 
$\xi_s=q_t \Delta_1^2/(c_t^2 \Delta_2)$ in Eq.~(\ref{Gefff}) 
is much smaller than 1.

In the rest of this section, we focus the case in which 
$c_t^2$ is close to 1, so that $\mu, \eta, \Sigma$ are 
given by Eq.~(\ref{Gefff2})-(\ref{Sigmaf2}) on 
scales relevant to the growth of large-scale structures.
If the screening of fifth forces occurs efficiently around 
local sources, the screened gravitational coupling 
$G_{\rm sc}$ is given by \cite{KWY}
\be
G_{\rm sc}(t)=\frac{1}{8\pi q_t(t)}
=\frac{1}{16\pi} \left[ G_4-\tp^2G_{4,X}
+\frac12\tp^2G_{5,\phi}
-\frac12H \dot{\phi}^3G_{5,X} \right]^{-1}\,.
\label{Gsc}
\ee
If we strictly demand that $c_t^2=1$, we have
$G_{\rm sc}(t)=1/[16\pi G_4(\phi)]$. 
Since we live in a screened environment, today's value  
of Eq.~(\ref{Gsc}), i.e., 
$G_{\rm sc}(t_0)=1/(8\pi q_t(t_0))$ should be close 
to the Newton gravitational constant $G$, such that
\be
q_t(t_0) \simeq \frac{1}{8 \pi G}\,.
\ee
The quantity $\mu_t$ reduces to
\be
\mu_t (t) \simeq \frac{q_t(t_0)}{q_t(t)}\,.
\ee
If $q_t(t)<q_t(t_0)$ in the past, we have $\mu_t>1$ 
and hence $\mu>1$ under the conditions 
$q_s>0, q_t>0, c_s^2>0$, and $M_{\phi}^2>0$. 
In this case, the effective gravitational coupling 
$G_{\rm eff}$ is larger than $G$ for scales relevant 
to the linear growth of large-scale structures. 
In the opposite case, $q_t(t)>q_t(t_0)$, $\mu_t$ is 
smaller than 1. This is the necessary condition 
for realizing $G_{\rm eff}<G$, but it is not sufficient
due to the existence of the positive term 
$\xi_s=q_t (\aB-\aM)^2/\Delta_2$ in Eq.~(\ref{Gefff2}).
In other words, even if $q_t(t)>q_t(t_0)$ in the past, 
the scalar-matter interaction $\xi_s$ can lead to 
$G_{\rm eff}$ larger than $G$.

There are bounds on the variation of the gravitational coupling 
constrained from Lunar Laser Ranging experiments \cite{William,Babi11}. 
In the screened environment, the experimental bound 
corresponds to $|\dot{G}_{\rm sc}/G_{\rm sc}|<0.02H_0$, 
where $H_0$ is today's value of the Hubble parameter.
On using Eq.~(\ref{Gsc}), this bound translates to 
\be
|\alpha_{\rm M}(t_0)|<0.02\,.
\label{alpM}
\ee
Assuming that the quantity $\alpha_{\rm M}$ is nearly constant 
around today, we obtain
$q_t(t)=q_t(t_0) e^{(t-t_0)H_0 \alpha_{\rm M}(t_0)}$ and 
hence $\mu_t \simeq e^{-(t-t_0)H_0 \alpha_{\rm M}(t_0)}$. 
The difference of $\mu_t$ from 1 over the cosmological 
time scale ($t-t_0 \sim 1/H_0$) is of order 
$\alpha_{\rm M}(t_0)$, so the modification to $\mu=G_{\rm eff}/G$ 
arising from the tensor contribution $\mu_t$ is 
suppressed under the bound (\ref{alpM}). 
The scalar-matter contribution 
$\xi_s=q_t (\aB-\aM)^2/\Delta_2$ is the main source 
for modifying the gravitational interaction on scales 
relevant to the growth of large-scale structures. 
The two quantities $\alpha_{\rm B}$ and $\alpha_{\rm M}$ 
play important roles for the evolution of gravitational potentials 
$\Psi$ and $\Phi$.

\subsection{Classification of surviving Horndeski 
theories in terms of $\mu$ and $\Sigma$}
\label{classsec} 

In the following, we focus on Horndeski theories given by 
the Lagrangian (\ref{lagcon}), i.e., those satisfying  
the condition $c_t^2=1$.
Then, the quantities $\alpha_{\rm B}$ and $\alpha_{\rm M}$ 
reduce, respectively, to 
\be
\alpha_{\rm B}=\frac{2\dot{\phi}G_{4,\phi}+\dot{\phi}^3G_{3,X}}
{4H G_4}\,,\qquad 
\alpha_{\rm M}=\frac{\dot{\phi}G_{4,\phi}}{HG_4}\,,
\label{alBM}
\ee
with $q_t=2G_4>0$.
Depending on the values of $\alpha_{\rm B}$ and $\alpha_{\rm M}$, 
the surviving theories can be classified 
into the following four classes.
\begin{itemize}

\item (A) $G_2=G_2(\phi,X)$, $G_3=0$, $G_4=M_{\rm pl}^2/2$.

This class accommodates both quintessence and k-essence. Since 
$\alpha_{\rm B}=0=\alpha_{\rm M}$ in Eqs.~(\ref{Gefff2})-(\ref{Sigmaf2}), 
it follows that 
\be
\mu=1\,,\qquad \eta=1\,,\qquad \Sigma=1\,.
\ee
Hence $G_{\rm eff}$ is equivalent to $G$ without the gravitational slip.

\item (B) $G_2=G_2(\phi,X)$, $G_3=0$, $G_4=G_4(\phi)$.

This class includes metric $f(R)$ gravity and BD theories. 
Since there is the specific relation 
\be
\alpha_{\rm B}=\frac{\alpha_{\rm M}}{2}\,,
\ee
we have 
\be
\mu=\frac{1}{16\pi G\,G_4} \left( 1+\frac{G_4 \alpha_{\rm M}^2}
{2\Delta_2} \right)\,, \qquad
\eta=\frac{2\Delta_2-G_4\alpha_{\rm M}^2}
{2\Delta_2+G_4\alpha_{\rm M}^2}\,,\qquad
\Sigma=\frac{1}{16\pi G\,G_4}\,.
\ee
The nonminimal coupling $G_4(\phi)$ enhances the gravitational 
interaction with matter ($\xi_s=G_4 \alpha_{\rm M}^2/(2\Delta_2)>0$). 
There is the difference between $\mu$ and $\Sigma$, so the Newtonian 
gravitational potential $\Psi$ and the weak lensing potential $\psi_{\rm eff}$
evolve in different ways.

\item (C)  $G_2=G_2(\phi, X)$, $G_3=G_3(\phi, X)$, 
$G_4=M_{\rm pl}^2/2$.

The theories of this class are known as kinetic 
braidings, which accommodate cubic Galileons 
as a specific case. Since 
\be
\alpha_{\rm B}=\frac{\dot{\phi}^3 G_{3,X}}{2H M_{\rm pl}^2}\,,
\qquad \alpha_{\rm M}=0\,,
\ee
we obtain
\be
\mu=\Sigma=1+\frac{M_{\rm pl}^2 \alpha_{\rm B}^2}{\Delta_2}\,,
\qquad \eta=1\,.
\label{muC}
\ee
Unlike the case (B) there is no gravitational slip. 
The cubic derivative coupling $G_3(X)$ enhances the two gravitational 
potentials $-\Psi$ and $\Phi$ in the same manner.

\item (D)  $G_2=G_2(\phi, X)$, $G_3=G_3(\phi,X)$, $G_4=G_4(\phi)$.

This is the most general case including kinetic braidings and its extensions. 
The relation between $\alpha_{\rm B}$ and $\alpha_{\rm M}$ is 
\be
\alpha_{\rm B}-\frac{\alpha_{\rm M}}{2}
=\frac{\dot{\phi}^3 G_{3,X}}{4H G_4}\,.
\ee
The original kinetic braiding scenario \cite{braiding1} corresponds to 
$G_4=M_{\rm pl}^2/2$, in which case 
$\mu, \eta, \Sigma$ are of the same forms as those 
given in Eq.~(\ref{muC}).

{}From Eqs.~(\ref{Gefff2}) and (\ref{Sigmaf2}), the difference between 
$\mu$ and $\Sigma$ is 
\be
\mu-\Sigma=-\frac{\alpha_{\rm M}(\alpha_{\rm B}-\alpha_{\rm M})}
{16\pi G \Delta_2}\,.
\ee
For the theories with $\alpha_{\rm M}=0$ (including the class (C)) or 
$\alpha_{\rm B}=\alpha_{\rm M}$, $\mu$ is equivalent to $\Sigma$. 
The case $\alpha_{\rm B}=\alpha_{\rm M}$ is special in that 
Eqs.~(\ref{Gefff2})-(\ref{Sigmaf2}) reduce to 
$\mu=\Sigma=1/(16\pi G\,G_4)$ and $\eta=1$.
\end{itemize}

In subsequent sections, we will discuss observational signatures 
for concrete dark energy models which belong to the classes 
(A), (B), (C), (D) in more detail.

\section{Class (A): Quintessence and k-essence}
\label{quinsec}

Quintessence and k-essence belong to the class (A) 
given by the Lagrangian 
\be
L=G_2(\phi,X)+\frac{M_{\rm pl}^2}{2}R\,,
\ee
which is within the framework of GR. 
{}From Eq.~(\ref{wde}), the dark energy EOS 
of k-essence yields
\be
w_{\rm DE}=-\frac{G_2}{G_2-\dot{\phi}^2 G_{2,X}}\,.
\label{wdeke}
\ee

The quantities $\alpha_{\rm K}$, $\alpha_{\rm B}$, and 
$\alpha_{\rm M}$ are
\be
\alpha_{\rm K}=\frac{\dot{\phi}^2}{H^2q_t} 
\left( G_{2,X}+\dot{\phi}^2 G_{2,XX} \right)\,,
\qquad
\alpha_{\rm B}=0\,,\qquad 
\alpha_{\rm M}=0\,,
\label{alM2}
\ee
with $q_t=M_{\rm pl}^2>0$.
{}From Eqs.~(\ref{qscon}) and (\ref{csge}), we obtain 
\be
q_s=2M_{\rm pl}^2 \left( G_{2,X}+\dot{\phi}^2 G_{2,XX} 
\right)\,,\qquad 
c_s^2=\frac{G_{2,X}}{G_{2,X}+\dot{\phi}^2 G_{2,XX}}\,.
\label{qsku}
\ee
The ghost and Laplacian instabilities of scalar perturbations 
are absent for 
\be
G_{2,X}+\dot{\phi}^2\,G_{2,XX}>0\,, \quad 
{\rm and} \quad
G_{2,X} > 0\,.
\label{kescon}
\ee

The background Eqs.~(\ref{back1}), (\ref{dH}), and (\ref{ddphi}) 
reduce, respectively, to 
\ba
& &
3M_{\rm pl}^2 H^2=-G_2+\dot{\phi}^2 G_{2,X}+\rho_m\,,\label{bake1}\\
& &
2M_{\rm pl}^2 \dot{H}=-\dot{\phi}^2 G_{2,X}-\rho_m-P_m\,,\label{bake2}\\
& &
\ddot{\phi}=-\frac{3H\dot{\phi} G_{2,X}+\dot{\phi}^2 G_{2,X \phi}
-G_{2,\phi}}{G_{2,X}+\dot{\phi}^2 G_{2,XX}}\,.\label{bake3}
\ea
Provided that the field energy density 
$\rho_{\rm DE}=-G_2+\dot{\phi}^2 G_{2,X}$ is 
positive, the k-essence EOS (\ref{wdeke}) is in the range 
\be
w_{\rm DE}>-1\,,
\ee
under the second condition of (\ref{kescon}).
The evolution of $w_{\rm DE}$ is different depending on 
the potential of quintessence and the form of k-essence Lagrangian.

On using Eq.~(\ref{alM2}), the quantities (\ref{Gefff2}) 
and (\ref{Sigmaf2}) reduce, respectively, to  
\be
\mu=1\,,\qquad \Sigma=1\,,
\ee
independent of the models of quintessence and k-essence. 
Although this situation is degenerate, the evolution of 
matter perturbations and gravitational potentials is affected 
by the difference of the background cosmology \cite{TDA13}. 
Moreover, as we will see 
in the following, the scalar sound speed squared $c_s^2$ 
in k-essence differs from that in quintessence.
Hence there are still possibilities 
for distinguishing between models of 
quintessence and k-essence.

\subsection{Quintessence}

Quintessence is given by the function 
\be
G_2(\phi, X)=X-V(\phi)\,,
\ee
under which Eq.~(\ref{wdeke}) reduces to 
\be
w_{\rm DE}=\frac{\dot{\phi}^2/2-V(\phi)}
{\dot{\phi}^2/2+V(\phi)}> -1\,.
\ee
The cosmological constant $\Lambda$ corresponds to 
the limit $\dot{\phi}^2/2 \to 0$ and $V(\phi) \to \Lambda$, 
i.e., $w_{\rm DE} \to -1$. 
For quintessence, the quantities given in Eq.~(\ref{qsku}) reduce to 
$q_s=2M_{\rm pl}^2$ and $c_s^2=1$, so the conditions for the 
absence of ghost and Laplacian instabilities are trivially satisfied.

To study the dark energy dynamics, we introduce the 
following dimensionless quantities:
\be
\Omega_{\rm DE} \equiv \frac{\rho_{\rm DE}}{3M_{\rm pl}^2 H^2}\,,\qquad 
\Omega_{m} \equiv \frac{\rho_{m}}{3M_{\rm pl}^2 H^2}\,,\qquad 
w_m \equiv \frac{P_m}{\rho_m}\,, 
\ee
where $\rho_{\rm DE}=\dot{\phi}^2/2+V(\phi)$. 
For the matter sector, we consider nonrelativistic matter characterized 
by the constant equation of state $w_m$ close to $+0$.
The Hamiltonian constraint (\ref{bake1}) gives the relation 
$\Omega_{\rm DE}+\Omega_m=1$.
On using Eqs.~(\ref{bake2}) and (\ref{bake3}), we obtain the following 
differential equations \cite{Sch,CDT}:
\ba
w_{\rm DE}'
&=& (w_{\rm DE}-1) \sqrt{3(1+w_{\rm DE})}
\left[ \sqrt{3(1+w_{\rm DE})}
-\lambda \sqrt{\Omega_{\rm DE}}
\right]\,,
\label{quinw}\\
\Omega_{\rm DE}'
&=& 3(w_m-w_{\rm DE}) \Omega_{\rm DE}
\left( 1-\Omega_{\rm DE} \right)\,,
\label{quinOme} \\
\lambda' 
&=& -\sqrt{3(1+w_{\rm DE}) \Omega_{\rm DE}}\,
(\Gamma-1) \lambda^2\,,
\label{quinlam}
\ea
where a prime represents a derivative with respect to 
${\cal N}=\ln a$, and
\be
\lambda \equiv -\frac{M_{\rm pl} V_{,\phi}}{V}\,, \qquad 
\Gamma \equiv \frac{V V_{,\phi \phi}}{V_{,\phi}^2}\,.
\ee 

{}From Eq.~(\ref{quinw}), besides the trivial case 
$w_{\rm DE} \simeq 1$, there are two important 
situations in which 
$w_{\rm DE}$ stays nearly constant\footnote{The case $w_{\rm DE} \simeq 1$ is irrelevant to the late-time cosmic acceleration, as Eq.~(\ref{quinOme}) 
shows that $\Omega_{\rm DE}$ decreases  
for $0<\Omega_{\rm DE}<1$.}: 
(1) $w_{\rm DE} \simeq -1$, and 
(2) $\sqrt{3(1+w_{\rm DE})} \simeq \lambda \sqrt{\Omega_{\rm DE}}$. 
In the following, we will discuss these two cases separately.

\subsubsection{Thawing quintessence}
\label{thawsec}

In case (1),  $w_{\rm DE}$ is close to $-1$, but 
the deviation of $w_{\rm DE}$ from $-1$ still occurs at late times. 
{}From Eq.~(\ref{quinOme}), we find that 
$\Omega_{\rm DE}$ increases according to 
$\Omega_{\rm DE}' \simeq 3(1+w_m)\Omega_{\rm DE}
(1-\Omega_{\rm DE})>0$.
Eventually, the growth of $\Omega_{\rm DE}$ in Eq.~(\ref{quinw}) leads to 
the variation of $w_{\rm DE}$, such that 
$w_{\rm DE}' \simeq 2 \lambda \sqrt{3(1+w_{\rm DE}) \Omega_{\rm DE}}$. 
This belongs to the class of {\it thawing quintessence} \cite{Caldwell} in which 
the scalar field is nearly frozen by the Hubble friction in the early cosmological 
epoch and it starts to evolve only recently.
The representative potential of this class is the one 
arising from the pseudo-Nambu-Goldstone (PNGB)
boson \cite{Frieman}:
\be
V(\phi)=\mu^4 \left[ 1+\cos \left(\frac{\phi}{f} 
\right) \right]\,,
\label{pngbpo}
\ee
where $\mu$ and $f$ are constants characterizing the 
energy scale and the mass scale of spontaneous symmetry 
breaking, respectively. 
The axion can be the candidate for the ultra-light PNGB boson.
When a global $U(1)$ symmetry is spontaneously broken, the axion appears as an angular massless field $\phi$ with an expectation 
value $\phi=f_s e^{i \phi/f_s}$ of a complex scalar 
at a scale $f_s$ \cite{Quinn,Kim0,Shifman}.

If the field mass squared 
$|m_{\phi}^2|=|V_{,\phi \phi}| \simeq \mu^4/f^2$ around 
the potential maximum is smaller than $H^2$, 
the field is stuck there with the dark energy EOS $w_{\rm DE} \simeq -1$. 
After $H^2$ drops below $|m_{\phi}^2|$, $w_{\rm DE}$ 
starts to increase from $-1$. 
For $|m_{\phi}^2|$ of order $H_0^2$ 
(i.e., $\mu \approx \sqrt{H_0 M_{\rm pl}} \approx 10^{-3}$~eV with the mass scale  
$f \approx M_{\rm pl}$), the growth of 
$w_{\rm DE}$ occurs at low redshifts. 
The radiative correction, which is proportional to $\mu^4$, 
does not give rise to explicit symmetry breaking terms. 
Then, the small mass term $|m_{\phi}| \approx \mu^2/M_{\rm pl} 
\approx 10^{-33}$~eV relevant to the late-time cosmic acceleration 
can be protected against  radiative corrections.
There are several interesting attempts for explaining the small
mass scale $\mu$ of order $10^{-3}$~eV
in the context of supersymmetric 
theories \cite{Watari,Choi,Kim,Axiverse,Hall,Kim2}.

\begin{figure}[h]
\begin{center}
\includegraphics[height=3.4in,width=3.6in]{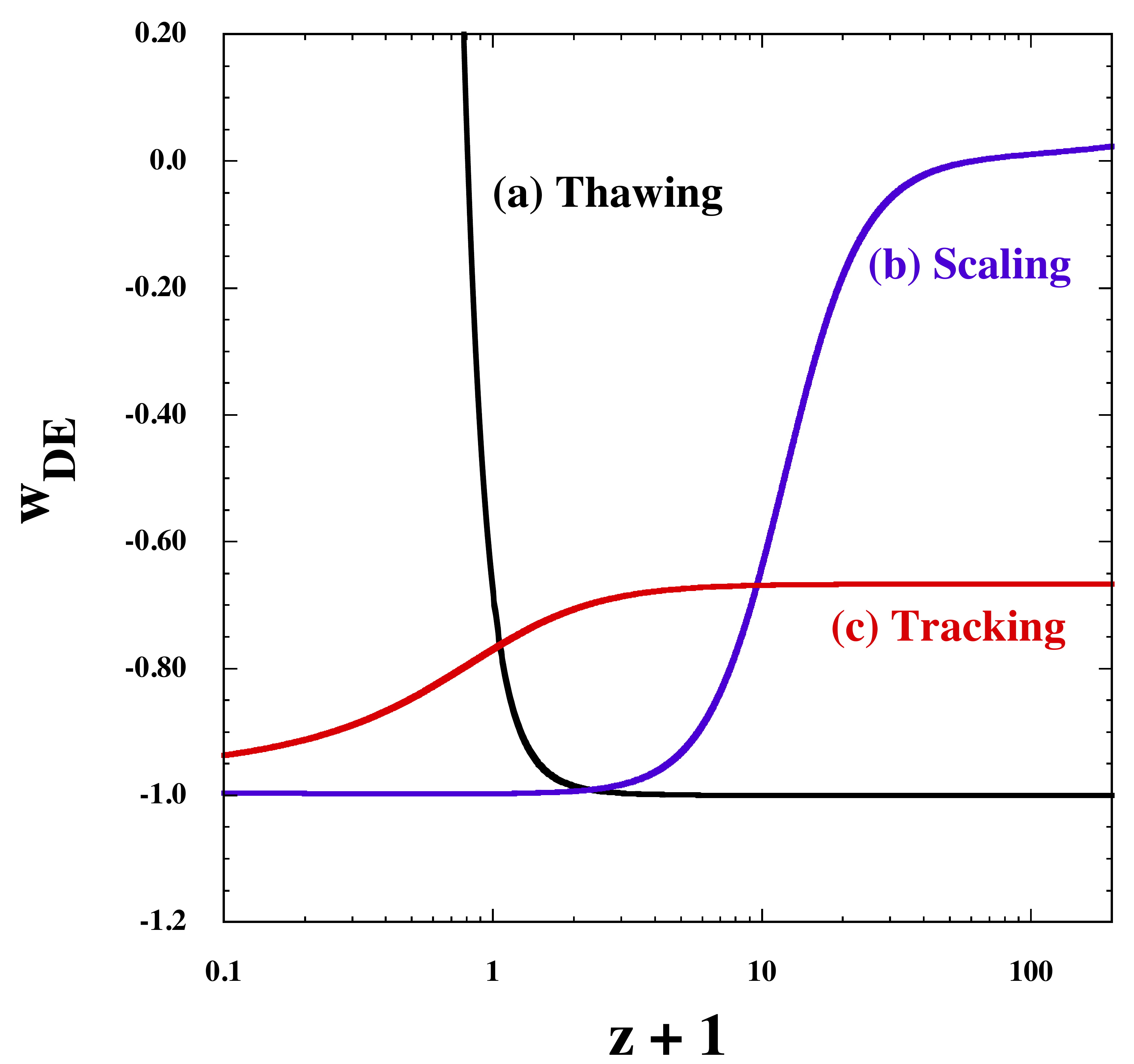}
\end{center}
\caption{\label{fig1}
Examples for the evolution of $w_{\rm DE}$ versus 
$z+1$ in (a) thawing quintessence, (b) scaling freezing quintessence, 
and (c) tracking freezing quintessence. 
The present epoch is identified by the condition 
$\Omega_{\rm DE}(z=0)=0.68$.}
\end{figure}

In case (a) of Fig.~\ref{fig1}, we plot one example for the evolution of 
$w_{\rm DE}$ in terms of $z+1=1/a$, where $z$ is 
the redshift (with today's value $z=0$).
In this case, $w_{\rm DE}$ starts to deviate from $-1$ 
around the redshift $z \lesssim 2$, with today's value 
$w_{\rm DE}(z=0) \simeq -0.7$. 
The likelihood analysis using CMB shift parameters measured 
by WMAP7 \cite{WMAP7} combined with SN Ia and BAO data 
placed the bound $w_{\rm DE}(z=0)<-0.695$ (95 \%\,CL) with 
the quintessence prior $w_{\rm DE}>-1$ \cite{CDT}. 
An updated data analysis based on the 
Planck 2015+SN Ia+BAO data provided the bound 
$w_{\rm DE}(z=0)<-0.473$ (95 \%\,CL) \cite{Ooba}.
This difference is mostly attributed to the fact that today's Hubble parameter 
$H_0$ constrained from the Planck data \cite{Planck2015,Planck2018} 
favors a lower value than that constrained from the WMAP7 data \cite{WMAP7}.
The bound of $H_0$ derived by the Planck team
has been in tension with direct measurements of 
$H_0$ at low redshifts \cite{RiessH0}. 
This is the main reason why the analysis of 
Ref.~\cite{Ooba} provided a more conservative bound on  
$w_{\rm DE}(z=0)$ than that obtained in Ref.~\cite{CDT}.

%
\subsubsection{Freezing quintessence}

In case (2), the field density parameter obeys the particular relation
\be
\Omega_{\rm DE}=\frac{3(1+w_{\rm DE})}{\lambda^2}\,.
\label{Ometra}
\ee
If $w_{\rm DE}=w_m$, then $\Omega_{\rm DE}=3(1+w_m)/\lambda^2=
{\rm constant}$ and hence $\lambda$ is constant. 
This is known as a scaling solution \cite{CLW,Joyce}, along which the ratio 
$\Omega_{\rm DE}/\Omega_{m}$ does not vary in time. 
This can be realized by the exponential potential 
$V(\phi)=V_0 e^{-\lambda \phi/M_{\rm pl}}$,
where $V_0$ is a constant. Since $\Gamma=1$ in this case, 
Eq.~(\ref{quinlam}) is trivially satisfied. 
The scaling solution has a nice feature in that the field  
density is not much smaller than the background density 
in the past, which can be compatibile with the energy scale 
relevant to particle physics. 
Hence it can alleviate the coincidence problem of 
dark energy (see 
Refs.~\cite{gcon2,Tsuji04,Quartin,Chiba14,Gomes,Amendola18,Fru18a,Fru18b} 
for more general models 
allowing for scaling solutions). 
However, since  $w_{\rm DE}=w_m \simeq 0$ 
for nonrelativistic matter,  the scaling solution is not appropriate 
to be used for the cosmic 
acceleration. If the single exponential potential is 
modified at late times to slow down the evolution 
of $\phi$, it is possible to realize 
a scaling matter-dominated epoch followed by the 
accelerated expansion \cite{Nelson,Scap1,Scap2,Scap3,Scap4}.
We call this model {\it scaling freezing quintessence} \cite{CDT}. 

The typical example of scaling freezing quintessence is
given by the potential \cite{Nelson}
\be
V(\phi)=V_1 e^{-\lambda_1 \phi/M_{\rm pl}}
+V_2 e^{-\lambda_2 \phi/M_{\rm pl}}\,,
\label{scaling}
\ee
where $\lambda_i$ and $V_i$ ($i=1, 2$) are constants. 
We consider the case in which the slopes 
$\lambda_1$ and $\lambda_2$ are in the ranges  
$\lambda_1> {\cal O}(1)$ and $\lambda_2 <{\cal O}(1)$.
In the early cosmological epoch, the steep exponential 
potential $V_1 e^{-\lambda_1 \phi/M_{\rm pl}}$ 
dominates over the other potential 
$V_2 e^{-\lambda_2 \phi/M_{\rm pl}}$. 
The solution enters the scaling radiation-dominated epoch
($w_{\rm DE}=1/3$ and $\Omega_{\rm DE}=4/\lambda_1^2$), 
which is followed by the scaling matter era
($w_{\rm DE}=0$ and $\Omega_{\rm DE}=3/\lambda_1^2$).
{}From the big bang nucleosynthesis there is the bound 
$\Omega_{\rm DE}<0.045$ in the radiation era \cite{Bean}, 
which translates to $\lambda_1>9.4$ \cite{Ohashi}. 
The Universe finally enters the epoch of cosmic acceleration 
after the second potential $V_2 e^{-\lambda_2 \phi/M_{\rm pl}}$ 
in Eq.~(\ref{scaling}) starts to contribute to the cosmological dynamics. 
The fixed point relevant to the late-time cosmic acceleration 
satisfies $\sqrt{3(1+w_{\rm DE})}=\lambda \sqrt{\Omega_{\rm DE}}$, 
$\Omega_{\rm DE}=1$, $\Gamma=1$, with $\lambda=\lambda_2$  
in Eqs.~(\ref{quinw})-(\ref{quinlam}), so that 
\be
w_{\rm DE}=-1+\frac{\lambda_2^2}{3}\,.
\label{wdefi}
\ee
The necessary condition for the cosmic acceleration is given by 
$\lambda_2^2<2$, under which the fixed point is a stable 
attractor \cite{CLW,dreview4}. 
In case (b) of Fig.~\ref{fig1}, we plot the evolution of  
$w_{\rm DE}$ for $\lambda_1=50$ and $\lambda_2=0.1$ with 
today's field density parameter $\Omega_{\rm DE}(z=0)=0.68$. 
We observe that $w_{\rm DE}$ is close to 0 in the scaling matter era 
and it approaches the asymptotic value (\ref{wdefi}).

In order to confront scaling freezing quintessence with observations, 
it is convenient to quantify the evolution of $w_{\rm DE}$ 
in terms of the transition scale factor $a_t$ and the thickness of 
transition to the freezing regime driven by the 
potential $V_2 e^{-\lambda_2 \phi/M_{\rm pl}}$. 
In Ref.~\cite{CDT}, it was shown that the change of $w_{\rm DE}$ 
from the scaling matter era to the accelerated attractor with 
$\lambda_2^2 \ll 1$ can be well approximate by 
the parametrization \cite{Linder05}:
\be
w_{\rm DE}=-1+\frac{1}{1+(a/a_t)^{1/\tau}}\,,
\label{wdetran}
\ee
with $\tau \simeq 0.33$. 
The joint likelihood analysis based on the Planck 2015 data  
combined with SN Ia and BAO data showed that the transition 
scale factor $a_t$ is constrained to be $a_t<0.11$ (95\,\%\,CL) \cite{Ooba} 
(which updated the bound $a_t<0.23$ (95\,\%\,CL) derived 
in Ref.~\cite{CDT}).
This translates to the transition redshift $z_t>8.1$, so the scaling 
matter era needs to end at quite early time. 
This is mostly attributed to the fact that, since the sound 
speed squared $c_s^2$ of quintessence is equivalent to 1, 
the perturbation of scaling quintessence hardly contributes 
to gravitational potentials. This slows down the growth of structures 
and leads to a large early Integrated-Sachs-Wolfe (ISW) 
effect \cite{Ooba}. 
For decreasing $z_t$ in the range $z_t<8$, the CMB angular 
power spectrum is subject to stronger modification.

There is the other class of freezing models dubbed 
{\it tracking freezing quintessence}. 
In this case, the slope $\lambda$ decreases in time with 
$w_{\rm DE}$ nearly constant, so that $\Omega_{\rm DE}$ 
in Eq.~(\ref{Ometra}) grows in time.
{}From Eq.~(\ref{quinlam}), the condition for decreasing 
$\lambda$ translates to
\be
\Gamma>1\,.
\label{Gamcon}
\ee
Since $\Omega_{\rm DE}$ increases, Eq.~(\ref{quinOme}) shows 
that $w_{\rm DE}<w_m$ for this tracker solution. 
This property is different from the scaling solution along which 
$w_{\rm DE}=w_m$ and $\Omega_{\rm DE}={\rm constant}$. 
Taking the ${\cal N}$ derivative of Eq.~(\ref{Ometra}), 
it follows that $\Omega_{\rm DE}'/\Omega_{\rm DE}=-2\lambda'/\lambda$. 
Substituting Eqs.~(\ref{quinOme}), (\ref{quinlam}), 
and (\ref{Ometra}) into 
this relation under the condition that $\Omega_{\rm DE} \ll 1$, 
the dark energy EOS in the matter era 
can be estimated as \cite{Paul98}
\be
w_{\rm DE} \simeq w_{(0)} \equiv 
-\frac{2(\Gamma-1)}{2\Gamma-1}\,,
\label{traeq}
\ee
where we used $w_m \simeq 0$.
As long as $\Gamma$ evolves slowly, $w_{\rm DE}$ stays 
nearly constant at high redshifts ($z \gg 1$). 
The tracker solution can be realized by the inverse power-law 
potential given by 
\be
V(\phi)=M^{4+p} \phi^{-p}\,,
\label{inpo}
\ee
where $M$ and $p$ are positive constants. 
This potential gives the value $\Gamma=1+1/p={\rm constant}>1$, 
so it satisfies the tracking condition (\ref{Gamcon}). 
The tracker EOS (\ref{traeq}) reduces to 
\be
w_{(0)}=-\frac{2}{p+2}\,.
\ee
For $p$ closer to 0, $w_{(0)}$ approaches the cosmological 
constant value $-1$. 

As $\Omega_{\rm DE}$ grows to the order of 0.1, 
$w_{\rm DE}$ starts to deviate from the value (\ref{traeq}), see 
case (c) in Fig.~\ref{fig1}.
It is possible to estimate the deviation $\delta w_{\rm DE}$ from  
$w_{(0)}$ by dealing with $\Omega_{\rm DE}$ as a perturbation to 
the 0-th order solution (\ref{traeq}). 
The leading-order solution to $\Omega_{\rm DE}$ can be derived by 
substituting $w_{\rm DE} \simeq w_{(0)}$ into Eq.~(\ref{quinOme}) 
and integrating it with respect to $a$, such that 
\be
\Omega_{\rm DE}(a)=\Omega_{\rm DE}^{(0)}
\left[ \Omega_{\rm DE}^{(0)}+a^{3w_{(0)}}
\left( 1-\Omega_{\rm DE}^{(0)} \right)
\right]^{-1}\,,
\label{Omedetra}
\ee
where $\Omega_{\rm DE}^{(0)}$ is today's value of $\Omega_{\rm DE}$. 
Plugging Eq.~(\ref{Omedetra}) into Eq.~(\ref{quinw}) 
and integrating the linear perturbation equation 
of $\delta w_{\rm DE}$ with  
respect to $a$, it follows that \cite{Chiba10}
\be
w_{\rm DE}(a)=w_{(0)}+\sum_{n=1}^{\infty}
\frac{(-1)^{n-1}w_{(0)}(1-w_{(0)}^2)}
{1-(n+1)w_{(0)}+2n(n+1)w_{(0)}^2}
\left( \frac{\Omega_{\rm DE}(a)}{1-\Omega_{\rm DE}(a)}
\right)^n\,.
\label{wtracker}
\ee
As shown in Refs.~\cite{Chiba10,CDT}, the iterative formula (\ref{wtracker})
up to the order $n=3$ exhibits fairly good agreement with the numerically 
integrated solution. 
Under the quintessence prior $w_{(0)}>-1$, 
the likelihood analysis using the iterative solution (\ref{wtracker})
put the bounds $w_{(0)}<-0.923$ 
and $0.675<\Omega_{\rm DE}^{(0)}<0.703$ (95 \%\,CL) 
from the observational data 
of Planck 2015, SN Ia, and CMB \cite{Ooba}.
For the potential (\ref{inpo}), this bound translates to $p<0.17$. 
For example, the tracker solution (c) plotted in Fig.~\ref{fig1} corresponds 
to $p=1$, which is strongly disfavored by the data. 
Thus, the tracker solutions arising from the inverse power-law potential with  
positive integer powers ($p \geq 1$) are observationally excluded.

\subsection{k-essence}

The Lagrangian of k-essence is a general function of $\phi$ and $X$. 
For example, the tachyon field given by the Lagrangian 
$G_2=-V(\phi) \sqrt{-{\rm det} \left(g_{\mu \nu}+
\nabla_{\mu} \phi \nabla_{\nu} \phi \right)}$, where 
$V(\phi)$ is a tachyon potential, arises in open string theory 
living on a non-BPS D3-brane \cite{tac1,tac2,tac3}.
There is also the so-called Dirac-Born-Infeld (DBI) scenario in which 
the movement of a probe D3-brane along the radial direction in
the AdS$_5$ spacetime is described by 
the action $P=-f(\phi)^{-1} \sqrt{1-2f(\phi)X}
+f(\phi)^{-1}-V(\phi)$ \cite{DBI1,DBI2}.
In the original setup of string theory, however, these two 
scenarios do not account for the late-time cosmic acceleration. 
Allowing for the freedom to modify the functions $V(\phi)$ and $f(\phi)$, 
there are possibilities for applying such theories to dark 
energy \cite{Paddy2002,Abramo,Lazkoz,CGST,Martin,Guo}.
In both tachyon and DBI theories, we require the existence of scalar potential $V(\phi)$ 
for driving the cosmic acceleration (as in quintessence).

There are k-essence scenarios in which nonlinear terms in $X$ play crucial roles 
for the late-time cosmological dynamics \cite{kes1,kes2,kes3,gcon1,gcon2,unifiedkes}. 
In the following, we review two typical k-essence theories of this type 
and discuss their observational signatures.

\subsubsection{Ghost condensate}

If the scalar field has a negative kinetic energy $-X$, 
this is typically a sign for the appearance of scalar ghosts.
However, the existence of an additional term $X^2$ to $-X$ can allow a possibility for avoiding the ghost \cite{gcon1}. 
The so-called dilatonic ghost condensate \cite{gcon2} 
given by the Lagrangian 
\be
G_2(\phi, X)=-X+e^{\lambda \phi/M_{\rm pl}}
\frac{X^2}{M^4}
\label{Lagghost}
\ee
belongs to such a class (where $\lambda$ and $M$ 
are positive constants). 
In the limit $\lambda \to 0$, there exists a de Sitter solution 
at $X=M^4/2$. As we will see below, the exponential term 
$e^{\lambda \phi/M_{\rm pl}}$ in Eq.~(\ref{Lagghost}) leads to 
the deviation from the de Sitter solution.
 
To discuss the cosmological dynamics of dilatonic ghost condensate, 
we introduce the dimensionless quantities:
\be
x_1 \equiv \frac{\dot{\phi}}{\sqrt{6}H M_{\rm pl}}\,,\qquad 
x_2 \equiv \frac{\dot{\phi}^2 e^{\lambda \phi/M_{\rm pl}}}{2M^4}\,.
\ee
From Eq.~(\ref{qsku}), the quantities $q_s$ 
and $c_s^2$ are
\be
q_s=2 \left( 6x_2-1 \right) M_{\rm pl}^2\,,\qquad 
c_s^2=\frac{2x_2-1}{6x_2-1}\,,
\ee
which are both positive for $x_2>1/2$.
The dark energy density parameter and its EOS 
are given, respectively, by 
\be
\Omega_{\rm DE}=x_1^2 \left( -1+3x_2 \right)\,,\qquad 
w_{\rm DE}=\frac{1-x_2}{1-3x_2}\,.
\ee
The necessary condition for the cosmic acceleration 
($-1<w_{\rm DE}<-1/3$) translates to $1/2<x_2<2/3$, 
in which regime $0<c_s^2<1/9$ \cite{gcon2}.
On using the background Eqs.~(\ref{dH}) and (\ref{ddphi}), 
we obtain the differential equations for
$w_{\rm DE}$ and $\Omega_{\rm DE}$, as
\ba
w_{\rm DE}' &=& 
\frac{(1-3w_{\rm DE})(1-w_{\rm DE})
[\sqrt{3(1-3w_{\rm DE})\Omega_{\rm DE}}\,\lambda-3(1+w_{\rm DE})]}{5-3w_{\rm DE}}\,,
\label{wdeke2}\\
\Omega_{\rm DE}' &=& 
3(w_m-w_{\rm DE}) \Omega_{\rm DE}
\left( 1-\Omega_{\rm DE} \right)\,,\label{Omedeke}
\ea
where Eq.~(\ref{Omedeke}) is the same as Eq.~(\ref{quinOme}) 
derived for quintessence.
{}From Eq.~(\ref{wdeke2}), there are cases in which 
$w_{\rm DE}$ stays nearly constant. 
Among them, the fixed point corresponding 
to $w_{\rm DE}<-1/3$ is characterized by 
\be
w_{\rm DE}=-1+\sqrt{\left( \frac{\lambda^2}{2}\Omega_{\rm DE} \right)^2
+\frac43 \lambda^2 \Omega_{\rm DE}}-\frac{\lambda^2}{2}\Omega_{\rm DE}\,.
\label{wdekes1}
\ee
{}From Eq.~(\ref{Omedeke}), we have three fixed points satisfying $w_{\rm DE}=w_m$, 
$\Omega_{\rm DE}=0$, and $\Omega_{\rm DE}=1$. 
The first case corresponds to the scaling solution, 
but this is not responsible for dark energy unless the form 
of $G_2(\phi,X)$ is appropriately modified at late times.
The second fixed point ($\Omega_{\rm DE}=0$) is relevant 
to the early matter era, during which $w_{\rm DE}$ 
is close to $-1$. 
The third fixed point ($\Omega_{\rm DE}=1$) 
is associated with 
the late-time cosmic acceleration with 
$w_{\rm DE}=-1+\sqrt{\lambda^4/4+4\lambda^2/3}-\lambda^2/2>-1$. 
Then, the evolution of $w_{\rm DE}$ in the
diatonic ghost condensate is similar to that 
in thawing quintessence discussed in Sec.~\ref{thawsec}.
Provided that $w_{\rm DE}$ does not significantly deviate from $-1$, 
the integrated solution to Eq.~(\ref{Omedeke}) in the intermediate 
regime $0<\Omega_{\rm DE}<1$ can be derived 
by setting $w_{m} \simeq 0$ and $w_{\rm DE} \simeq -1$, 
such that 
\be
\Omega_{\rm DE}(a)=a^3 \Omega_{\rm DE}^{(0)}
\left[ 1- \Omega_{\rm DE}^{(0)}+a^3  \Omega_{\rm DE}^{(0)} 
\right]^{-1}\,.
\label{Omedekes1}
\ee
Substituting Eq.~(\ref{Omedekes1}) into Eq.~(\ref{wdekes1}), 
we obtain the approximate solution to $w_{\rm DE}$ 
as a function of $a$. If we adopt the observational bound 
$w_{\rm DE}(a=1) \lesssim -0.7$ with $\Omega_{\rm DE}^{(0)} 
=0.68$, the parameter $\lambda$ is constrained to be 
$\lambda \lesssim 0.36$.

We note that, in the dilatonic ghost condensate, $c_s^2$ is 
initially close to $+0$ and then it starts to deviate from $+0$
only recently. In this case, the k-essence scalar with $c_s^2 \simeq +0$ 
can work as a part of dark matter because of additional gravitational clusterings.
This property is different from that in quintessence where 
$c_s^2$ is always equivalent to 1, so there is a possibility 
for distinguishing between thawing quintessence and 
diatonic ghost condensate at the level of perturbations.

\subsubsection{K-essence as unified dark energy and dark matter}

There is a unified k-essence model of 
dark energy and dark matter given by \cite{unifiedkes}
\be
G_2(X)=-b_0+b_2 \left( X-X_0 \right)^2\,,
\label{unilag}
\ee
where $b_0, b_2, X_0$ are positive constants.
In this case, the k-essence pressure and density are 
$P=-b_0+b_2 \left( X-X_0 \right)^2$ and 
$\rho=b_0+b_2(X-X_0)(3X+X_0)$, respectively, with 
the propagation speed squared: 
\be
c_s^2=\frac{X-X_0}{3X-X_0}\,.
\label{csuni}
\ee
As long as $X$ stays around $X_0$, $c_s^2$ is close to 0. 
Substituting $X=X_0 \left[ 1+ \epsilon(t) \right]$ with 
$0<\epsilon \ll 1$ into Eq.~(\ref{bake3}), we obtain the 
integrated solution $\epsilon(t)=\epsilon_1 (a_1/a)^3$, 
where $\epsilon_1$ and $a_1$ are positive constants. 
Then, the sound speed squared (\ref{csuni}) approximately 
reduces to 
\be
c_s^2 \simeq \frac{\epsilon(t)}{2}
=\frac{\epsilon_1}{2} \left( \frac{a_1}{a} \right)^3\,,
\ee
which always stays in the region $0<c_s^2 \ll 1$. 
Around $X=X_0$, we have $P \simeq -b_0$ and 
$\rho \simeq b_0+4b_2 X_0 (X-X_0)$, so the k-essence 
EOS, $w=P/\rho$, reads 
\be
w \simeq -\left[ 1+\frac{4b_2}{b_0}X_0^2 \epsilon_1 
\left( \frac{a_1}{a} \right)^3 \right]^{-1}\,.
\ee
In the early matter era, the k-essence field behaves as dark matter 
with $w \simeq 0$. Since $w$ approaches $-1$ 
at late times due to the dominance of the term $-b_0$ 
in Eq.~(\ref{unilag}), the same field behaves as dark energy. 
Thus, the k-essence Lagrangian (\ref{unilag}) provides the 
unified description of dark energy and dark matter 
with $0<c_s^2 \ll 1$. We note that the purely k-essence 
with the Lagrangian $G_2=G_2(X)$ is equivalent to a barotropic 
perfect fluid on the cosmological background \cite{Hu05,Arroja,DMT10,KT14b}.

In Refs.~\cite{Ber1,Ber2}, the authors proposed several unified models 
of dark matter and dark energy different from (\ref{unilag}). 
In particular, there is a model in which $c_s^2$ starts to evolve from a 
value close to 0 and approaches an asymptotic value $c_{\infty}^2$.
In such a case, the likelihood analysis based on the galaxy-ISW correlation 
data showed that $c_{\infty}^2$ is constrained to be smaller than 
$c_{\infty}^2 \lesssim 9 \times 10^{-3}$ \cite{Ber3}.

\section{Class (B): $f(R)$ gravity and Brans-Dicke theories}
\label{fRBDsec}

The theories of class (B) include BD theories and 
$f(R)$ gravity as specific cases.
The Lagrangian of BD theories with a scalar 
potential $V(\phi)$ is given by  
\be
L=\left(1-6Q^2 \right)e^{-2Q\phi/M_{\rm pl}}X
-V(\phi)+\frac{M_{\rm pl}^2}{2} 
e^{-2Q\phi/M_{\rm pl}}R\,,
\label{lagbra}
\ee
where $Q$ is a constant. 
In terms of the redefined dimensionless field 
\be
\chi=e^{-2Q\phi/M_{\rm pl}}\,,
\ee
the Lagrangian (\ref{lagbra}) is equivalent to 
\be
L=-\frac{M_{\rm pl}^2 \omega_{\rm BD}}{2\chi} 
g^{\mu \nu} \nabla_{\mu}\chi  \nabla_{\nu}\chi
-V(\phi(\chi))+\frac{M_{\rm pl}^2}{2} \chi R\,,
\label{lagbra2}
\ee
where the BD parameter $\omega_{\rm BD}$ 
is related to the constant $Q$, as 
\be
3+2\omega_{\rm BD}=\frac{1}{2Q^2}\,.
\ee
The original BD theory \cite{Brans} is written in the form 
(\ref{lagbra2}) without the scalar potential ($V=0$). 

The $f(R)$ gravity, which is given by the Lagrangian 
$L=M_{\rm pl}^2 f(R)/2$, is equivalent to the scalar-tensor 
theory with $L=M_{\rm pl}^2 
[f(\varphi)+f_{,\varphi}(R-\varphi)]/2$, 
where $\varphi$ is a scalar quantity. 
Varying the latter Lagrangian with respect to $\varphi$, 
we obtain $f_{,\varphi \varphi}(R-\varphi)=0$ and hence 
$\varphi=R$ for  $f_{,\varphi \varphi} \neq 0$.
By defining $\chi \equiv f_{,\varphi}=f_{,R}$, the Lagrangian 
of $f(R)$ gravity reduces to 
\be
L=\frac{M_{\rm pl}^2}{2} \chi R-V(\chi)\,,\quad {\rm where} 
\quad V(\chi)=\frac{M_{\rm pl}^2}{2} \left[ \chi \varphi (\chi)
-f(\varphi (\chi)) \right]\,.
\label{LfR}
\ee
Comparing this Lagrangian with Eq.~(\ref{lagbra2}), 
it follows that  $f(R)$ gravity corresponds to BD theory 
with $\omega_{\rm BD}=0$ \cite{Ohanlon,Chiba03}.  
Under the conformal transformation 
$\hat{g}_{\mu \nu}=\chi g_{\mu \nu}$, the Lagrangian 
(\ref{LfR}) is transformed to that 
in the Einstein frame with a canonical 
scalar field \cite{Maeda89,Bunn,fRreview0,fRreview}
\be
\phi=\sqrt{\frac{3}{2}}
M_{\rm pl} \ln f_{,R}\,,
\label{phifR}
\ee
which corresponds 
to the field $\phi$ appearing in the Jordan-frame Lagrangian 
(\ref{lagbra}) of BD theory. 
On using the correspondence $\chi=e^{-2Q \phi/M_{\rm pl}}=f_{,R}$, 
the constant $Q$ in metric $f(R)$ gravity, 
which plays the role of couplings between the field 
$\phi$ and matter in the Einstein frame \cite{AmenPRL,BD1}, 
is given by 
\be
Q=-\frac{1}{\sqrt{6}}\,.
\ee

In BD theory with an arbitrary BD parameter 
$\omega_{\rm BD}$, the modification of gravity 
is significant for $|Q|>{\cal O}(0.1)$. 
In the limit that $Q \to 0$, i.e., $\omega_{\rm BD} \to \infty$, 
the Lagrangian (\ref{lagbra}) recovers the minimally coupled quintessence in GR. 
If the scalar potential $V(\phi)$ is absent, gravitational experiments in the solar system 
constrain the BD parameter to be $\omega_{\rm BD}>40000$ \cite{Will}, 
which translates to $|Q|<2.5 \times 10^{-3}$. In the presence of $V(\phi)$, as in the case 
of $f(R)$ gravity, it is possible to suppress the fifth force 
induced by the scalar-matter interaction in the solar 
system under the chameleon mechanism even for 
$|Q|>{\cal O}(0.1)$  \cite{Bunn,Navarro,fR1,Capo07,Tamaki,Brax08}.
Indeed, the viable dark energy models in $f(R)$ 
gravity \cite{fR1,fR2,fR3,fR4,fR5}
and BD theory \cite{BD1,BD2} are constructed to have a large mass in regions of the high density, while the field mass is light on cosmological scales such that the effective potential of $\phi$ drives the late-time cosmic acceleration.

For the Lagrangian (\ref{lagbra}), the background Eqs.~(\ref{back1}) and (\ref{ddphi}) reduce, 
respectively, to  
\ba
& &
3M_{\rm pl}^2 H^2=
\frac{1}{2}(1-6Q^2) \dot{\phi}^2
+6Q M_{\rm pl}H \dot{\phi}+\left( \rho_m+V
\right)e^{2Q\phi/M_{\rm pl}}\,,\label{BDeq1}\\
& &
\ddot{\phi}+3H \dot{\phi}+V_{,\phi}e^{2Q\phi/M_{\rm pl}}
+\frac{Q}{M_{\rm pl}} \left[ \left( \rho_m-3P_m+4V 
\right)e^{2Q\phi/M_{\rm pl}}-2\dot{\phi}^2
\right]=0\,,\label{BDeq2}
\ea
where, for the derivation of Eq.~(\ref{BDeq2}), we used 
Eq.~(\ref{BDeq1}) to eliminate $H^2$. 
{}From Eq.~(\ref{BDeq2}), the coupling $Q$ can induce 
a local minimum even for a runaway potential $V(\phi)$. 
If the scalar field is nearly frozen around the local minimum 
during the matter era characterized by 
$\rho_m \gg \{P_m, V \}$, it follows that 
\be
M_{\rm pl}V_{,\phi}+Q \rho_m \simeq 0\,.
\label{mire1}
\ee
If $Q>0$ (or $Q<0$), then the solution to Eq.~(\ref{mire1}) 
exists for $V_{,\phi}<0$ (or $V_{,\phi}>0$). 
Since $\rho_m$ decreases in time, the field evolves slowly 
along the instantaneous minima determined 
by Eq.~(\ref{mire1}). 
The scalar potential $V$ is responsible for the late-time cosmic 
acceleration. After $V$ dominates over the matter density 
$\rho_m$, there exists a de Sitter solution characterized by 
\be
M_{\rm pl}V_{,\phi}+4QV=0\,.
\label{mire2}
\ee

{}From the above discussion, the field $\phi$ is almost 
frozen apart from the transient period from the matter era 
to the de Sitter solution. In other words, the deviation of 
$w_{\rm DE}$ from $-1$ starts to occur around today and 
$w_{\rm DE}$ finally approaches the asymptotic value $-1$. 
We note that, with decreasing $\rho_m$, the effective mass 
around the potential minimum induced by the coupling $Q$ 
gets smaller. In the early Universe, the field is nearly frozen 
due to the heavy mass associated with the large matter density 
$\rho_m$. In local regions of the Universe where the density 
$\rho_m$ is much larger than today's critical density $\rho_c$, 
the similar suppression of the propagation of fifth forces can 
occur under the operation of 
the chameleon mechanism \cite{chame1,chame2}.

We also compute quantities relevant to the evolution 
of linear cosmological perturbations for the Lagrangian (\ref{lagbra}).
Since $q_t=M_{\rm pl}^2 e^{-2Q\phi/M_{\rm pl}}$, 
the no-ghost condition of tensor perturbations 
is automatically satisfied. The quantities 
$\alpha_{\rm K}$, $\alpha_{\rm B}$, and 
$\alpha_{\rm M}$ are given by
\be
\alpha_{\rm K}=
\frac{(1 -6Q^2) \dot{\phi}^2}{M_{\rm pl}^2 H^2}\,,\qquad 
\alpha_{\rm B}=\frac{\alpha_{\rm M}}{2}
=-\frac{Q\dot{\phi}}{M_{\rm pl}H}\,.
\label{alBB}
\ee
{}From Eqs.~(\ref{qscon}) and (\ref{csge}), it follows that 
\be
q_s=2e^{-4Q\phi/M_{\rm pl}}M_{\rm pl}^2 \,,\qquad
c_s^2=1\,,
\ee
so there are neither ghost nor Laplacian instabilities 
of scalar perturbations in BD theories.
The dark energy EOS (\ref{wde}) 
reduces to 
\be
w_{\rm DE}=-1-\frac{4M_{\rm pl}^2 (1-e^{-2Q \phi/M_{\rm pl}}) \dot{H}
+4M_{\rm pl}Q e^{-2Q \phi/M_{\rm pl}} \ddot{\phi}
-2e^{-2Q \phi/M_{\rm pl}}\dot{\phi}[(1-2Q^2) \dot{\phi}+2M_{\rm pl}H Q]}
{(1-6Q^2)e^{-2Q \phi/M_{\rm pl}} \dot{\phi}^2+2V
+6M_{\rm pl}^2H^2(1-e^{-2Q \phi/M_{\rm pl}})
+12M_{\rm pl}H Qe^{-2Q \phi/M_{\rm pl}} \dot{\phi}}\,.
\label{wdeBD}
\ee
Existence of the nonvanishing coupling $Q$ allows 
the possibility for realizing 
$w_{\rm DE}<-1$ \cite{fR1,fR4,Moto}, without 
having ghost and Laplacian instabilities. 
Applying the bound (\ref{alpM}) to $\alpha_{\rm M}$ in 
Eq.~(\ref{alBB}), the field time derivative 
today (denoted as $\dot{\phi}_0$) is constrained to be 
\be
\left| \frac{Q \dot{\phi}_0}{M_{\rm pl}H_0} 
\right| <0.01\,,
\label{Qbound}
\ee
which limits the large deviation of $w_{\rm DE}$ from 
$-1$ at low redshifts.

{}From Eqs.~(\ref{Gefff2}) and (\ref{Sigmaf2}), 
we obtain
\be
\mu=e^{2Q\phi/M_{\rm pl}} \frac{1+2Q^2+{\cal F}_M (k)}
{1+{\cal F}_M(k)}\,,
\qquad 
\Sigma=e^{2Q\phi/M_{\rm pl}}\,, 
\label{SigmaBD}
\ee
where 
\be
{\cal F}_M (k) \equiv \frac{a^2M_{\phi}^2
e^{2Q \phi/M_{\rm pl}}}{k^2}\,.
\label{calFM}
\ee
As we will see below, the viable dark energy models 
in $f(R)$ gravity and BD theory have been 
constructed to have a growing mass $M_{\phi}$ 
in the asymptotic past with the scalar field nearly frozen around 
$\phi \simeq 0$. Since $M_{\phi}^2$ becomes  much 
larger than $H^2$, the quantity ${\cal F}_M(k)$
can exceed the order 1 even for perturbations deep 
inside the Hubble radius ($k^2 \gg a^2 H^2$).
Taking the limit ${\cal F}_M(k) \to \infty$ 
in Eq.~(\ref{SigmaBD}) in this massive regime, 
it follows that $\mu \simeq \Sigma=
e^{2Q\phi/M_{\rm pl}} \simeq 1$. 
Then, the evolution of linear perturbations is similar to that 
in GR for the same background expansion history.
 
Since $M_{\phi}^2$ gradually decreases in time, 
there is an instant at which ${\cal F}_M(k)$ crosses 1. 
For larger $k$, 
the entry to the regime ${\cal F}_M(k)<1$ occurs earlier.
Taking the limit ${\cal F}_M(k) \to 0$ in Eq.~(\ref{SigmaBD}), we obtain 
\be
\mu \simeq \left( 1+2Q^2 \right)\Sigma\,,\qquad 
\Sigma=e^{2Q\phi/M_{\rm pl}}\,.
\label{muBD}
\ee
This is the regime in which the modification of gravity 
manifests itself on scales relevant to the linear growth of 
large-scale structures. 
In particular the coupling $Q$ leads to the enhancement  
of $G_{\rm eff}$, so the growth rate of matter perturbations 
gets larger than that in quintessence and k-essence.
This allows the possibility for distinguishing $f(R)$ gravity 
and BD theories from dark energy models 
in the framework of GR.

\subsection{$f(R)$ gravity}
\label{fRsec}

In $f(R)$ gravity, the examples of models relevant to 
the late-time cosmic acceleration are 
given by \cite{fR1,fR2}
\ba
&  & {\rm (i)}~f(R)=R-\lambda R_{0}\frac{(R/R_{0})^{2n}}{(R/R_{0})^{2n}+1}\,,\label{imodel}\\
&  & {\rm (ii)}~f(R)=R-\lambda R_{0}\left[1-\left(1+\frac{R^{2}}{R_{0}^2}\right)^{-n}
 \right]\,,\label{iimodel}
\ea
where $\lambda$, $R_0$, and $n$ are positive constants. 
These models satisfy $f(R=0)=0$, so 
the cosmological constant disappears in the limit 
of flat spacetime. 
In the high-curvature regime satisfying $R \gg R_0$, 
they have the asymptotic behavior
\be
f(R) \simeq R-\lambda R_0 
\left[ 1-\left( \frac{R_0}{R} \right)^{2n} 
\right]\,,
\label{fRasy}
\ee
so that they approach the $\Lambda$CDM model 
($f(R)=R-\lambda R_0$). 
We note that the model $f(R)=R-\lambda R_0 \tanh (R/R_0)$ \cite{fR4}
also has the same asymptotic behavior.
The deviation from GR manifests itself after $R$ 
decreases to the order of $R_0$.

On using the asymptotic form (\ref{fRasy}), the scalar field $\phi$ 
defined by Eq.~(\ref{phifR}), which is called the scalaron \cite{Staro}, 
is expressed as 
\be
\phi \simeq \sqrt{\frac32}M_{\rm pl}\,\ln \left[1-2n \lambda 
\left( \frac{R}{R_0} \right)^{-(2n+1)} \right]<0\,,
\label{phifRre}
\ee
which approaches 0 as $R \to \infty$.
In the regime $R \gg R_0$,  
the scalar potential (\ref{LfR}), i.e., $V=
(M_{\rm pl}^2/2)(f_{,R}R-f)$, reduces to 
\be
V(\phi) \simeq \frac{\lambda R_0M_{\rm pl}^2}{2} \left[ 1-\frac{2n+1}
{(2n\lambda)^{2n/(2n+1)}} \left( 1-e^{\sqrt{2/3}\phi/M_{\rm pl}}
\right)^{2n/(2n+1)} \right]\,,
\label{Vphiap}
\ee
which approaches the constant $\lambda R_0M_{\rm pl}^2/2$ 
for $R \to \infty$. 
Since the potential (\ref{Vphiap}) satisfies $V_{,\phi}>0$ for $\phi<0$, 
the solutions to Eqs.~(\ref{mire1}) and (\ref{mire2}) exist for the 
above $f(R)$ models.
The field mass squared $M_{\phi}^2=V_{,\phi \phi}$ increases 
for larger $R$, with the divergence 
$M_{\phi}^2 \to \infty$ for $R \to \infty$. 
In terms of $f(R)$, the mass squared 
can be expressed as
\be
M_{\phi}^2=\frac{Rf_{,R}}{3m} \left( 1+m \right)\,,
\label{Mphi}
\ee
where the quantity $m \equiv Rf_{,RR}/f_{,R}$ 
characterizes the deviation from the $\Lambda$CDM 
model \cite{fRsta3}.
The increase of $M_{\phi}^2$ in the asymptotic past is 
attributed to the decrease of 
$f_{,RR}=2\lambda n (2n+1)R_0^{-1}(R/R_0)^{-2n-2}$ 
toward 0. Since $f_{,RR}$ is positive for 
the models (\ref{imodel}) and (\ref{iimodel}), 
the mass squared (\ref{Mphi}) is positive and hence 
there is no tachyonic instability.
Unlike the $\Lambda$CDM model in which $f_{,RR}$ is exactly 0, 
the above $f(R)$ models give rise to a large mass 
in the asymptotic past due to the small deviation of 
$f_{,RR}$ from $0$. 
In the regime $m \ll 1$ the mass squared (\ref{Mphi}) 
can be estimated as $M_{\phi}^2 \simeq 
Rf_{,R}/(3m) \simeq H^2/m \gg H^2$, where 
we used the properties $R=6(2H^2+\dot{H}) \simeq 3H^2$ 
and $f_{, R} \simeq 1$ in the matter era.
 
Unless the initial conditions of $\phi$ are carefully 
chosen to be close to 
the instantaneous minimum given by Eq.~(\ref{mire1}), 
the scalar field 
oscillates due to the heavy mass in the early 
cosmological epoch \cite{fR2,fR4}. 
Then, the system can even access the curvature singularity 
at $\phi=0$ \cite{Frolov}. 
This is the general problem of late-time $f(R)$ 
cosmic acceleration models. For relativistic stars 
in $f(R)$ gravity, the similar problem also arises as 
a fine-tuning of boundary 
conditions \cite{MaedafR,Tsuji09,Babi09,Upadhye}.
Provided that the initial conditions are fine-tuned such that 
the scalar field is very close to the position determined 
by Eq.~(\ref{mire1}), the field evolves slowly along the 
instantaneous minima with decreasing $\rho_m$.

In the regime $R \gg R_0$, the field stays in the region around $\phi=0$ 
and hence the potential $V(\phi)$ is nearly constant: 
$V(\phi) \simeq \lambda R_0 M_{\rm pl}^2/2$. 
In this region, $w_{\rm DE}$ is close to $-1$. 
After $R$ decreases to the order $R_0$, the field $\phi$ tends to 
be away from 0, so the deviation of $w_{\rm DE}$ from $-1$ 
starts to occur by today \cite{Moto}. 
Finally, the solution reaches the de Sitter 
fixed point satisfying Eq.~(\ref{mire2}). 

In $f(R)$ gravity, the condition (\ref{mire2}) is equivalent to \cite{fRsta3,fRreview}
\be
Rf_{,R}=2f\,.
\label{fRcon1}
\ee
The stability analysis around this fixed point shows that the 
de Sitter point satisfying (\ref{fRcon1}) is stable 
if the parameter $m=Rf_{,RR}/f_{,R}$ is in 
the range \cite{fRsta1,fRsta2,fRsta3}
\be
0<m \le 1\,.
\label{fRcon2}
\ee
In model (i), the conditions (\ref{fRcon1}) and (\ref{fRcon2}) 
translate, respectively, to 
\ba
& &
\lambda x_1^{2n-1}(2+2x_1^{2n}-2n)=(1+x_1^{2n})^2\,,
\label{imodellam}\\
& &
2x_1^{4n}-2(n+2)(2n-1)x_1^{2n}+
(2n-1)(2n-2) \ge 0\,,
\label{imodelcon}
\ea
where $x_1 \equiv R_1/R_0$, and $R_1$ is the value of $R$ 
at the de Sitter point.
For given $n$, the condition (\ref{imodelcon}) 
gives a lower bound on the parameter $\lambda$.
If $n=1$, we have $x_1 \geq \sqrt{3}$ and 
$\lambda \geq 8\sqrt{3}/9$.

The de Sitter fixed point corresponds to a stable spiral for 
$0<m<16/25$ \cite{fRsta3}, in which case the dark energy 
EOS approaches the asymptotic value $-1$ with oscillations 
around $w_{\rm DE}=-1$. 
The phantom EOS $w_{\rm DE}<-1$ can be 
realized by today without having ghost 
instabilities \cite{fR1,fR4,AT08,Moto}. 
This property is different from that in quintessence and 
k-essence where $w_{\rm DE}$ is always in the range 
$w_{\rm DE}>-1$. We note that, under the bound (\ref{Qbound}), 
the variation of $w_{\rm DE}$ given by Eq.~(\ref{wdeBD}) 
is limited around today, such that $|w_{\rm DE}+1|<{\cal O}(0.01)$. 
At the background level, the $f(R)$ models (\ref{imodel}) and (\ref{iimodel}) are 
consistent with the observational data \cite{fRob1,fRob2,fRob3}.

For $m \ll 1$, the mass squared (\ref{Mphi}) in the matter era 
can be estimated as $M_{\phi}^2 \simeq H^2/m$, so the function (\ref{calFM}) is approximately given by 
${\cal F}_M (k) \simeq a^2 H^2/(mk^2)$. 
This means that the entry from the massive 
regime (${\cal F}_M (k)>1$) to the massless regime 
(${\cal F}_M (k)<1$) occurs at 
\be
m \simeq \left( \frac{aH}{k} \right)^2\,.
\label{mre}
\ee
For the models (\ref{imodel}) and (\ref{iimodel}) the quantity 
$m$ grows in time during the matter era, while 
$(aH/k)^2$ decreases. Extrapolating the relation (\ref{mre}) 
up to the present epoch ($z=0$), the perturbations relevant to 
the linear growth of large-scale structures ($k \lesssim 300 H_0$) 
enter the massless regime by today 
under the criterion 
\be
m_0 \gtrsim 10^{-5}\,,
\ee
where $m_0$ is today's value of $m$.
Since $\mu$ and $\Sigma$ are close to 1 in the massive 
regime (${\cal F}_M (k)>1$), the evolution of 
perturbations is similar 
to that in GR, i.e., $\delta_m \propto t^{2/3}$ and 
$\psi_{\rm eff}=-2\Psi=2\Phi={\rm constant}$ 
during the matter dominance. 
After the entry to the regime ${\cal F}_M (k)<1$,  
the quantities $\mu$ and $\Sigma$ are 
given by Eq.~(\ref{muBD}) with $Q=-1/\sqrt{6}$.
Solving Eq.~(\ref{delmeqf}) with Eqs.~(\ref{Poi}) and (\ref{psieff}), 
the perturbations evolve as \cite{fR2,fR4}
\be
\delta_m \propto t^{(\sqrt{33}-1)/6}\,,\qquad 
\psi_{\rm eff}=-\frac{3}{2}\Psi=3\Phi \propto 
t^{(\sqrt{33}-5)/6}\,,
\label{delmfR}
\ee
during the matter dominance. 
The growth rates of matter perturbations and 
gravitational potentials are enhanced compared 
to those in GR. 
For $k \gtrsim 300 H_0$, the nonlinear effects on the evolution 
of linear perturbations can be important, see Refs.~\cite{non1,non2,non3,non4,non5,non6,non7} 
for $N$-body simulations in $f(R)$ gravity.

\begin{figure}[h]
\begin{center}
\includegraphics[height=2.5in,width=7.0in]{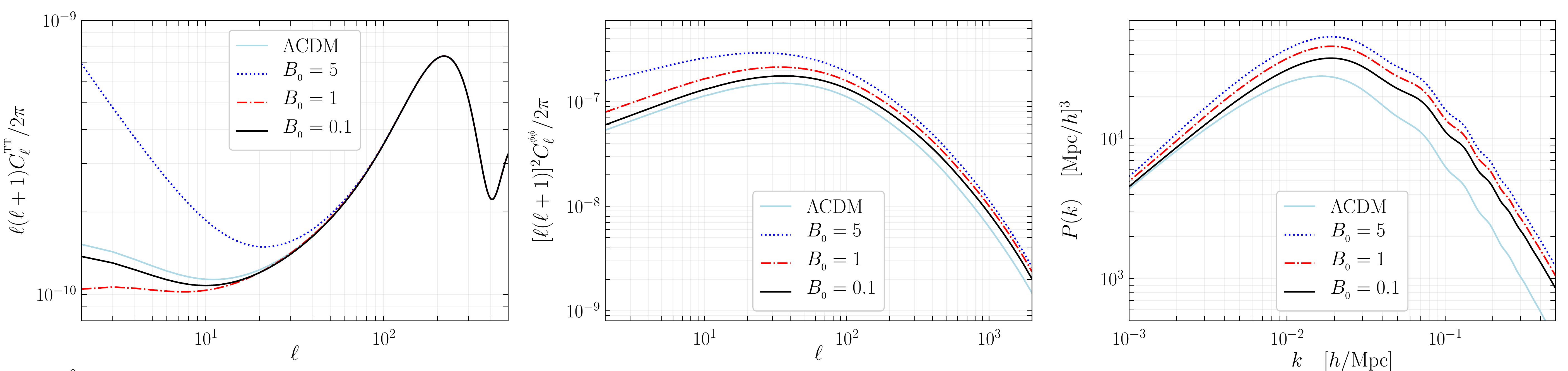}
\end{center}
\caption{\label{fig2}
The CMB angular temperature power spectrum (left), the lensing power spectrum 
(middle), and the linear matter power spectrum (right) in $f(R)$ models of 
the late-time cosmic acceleration with the  $\Lambda$CDM background.
We plot theoretical predictions of three $f(R)$ models 
($B_0=5, 1, 0.1$) besides that in the $\Lambda$CDM model. 
Taken from Ref.~\cite{Battye}. }
\end{figure}

The modified growth of perturbations affects the observables 
associated with the power spectra of galaxy clusterings, CMB, and 
weak lensing.
In the left panel of Fig.~\ref{fig2}, we show the CMB angular 
temperature power spectrum for three different $f(R)$ models 
and the $\Lambda$CDM model. 
The background expansion history  
is fixed to be the same as that in the $\Lambda$CDM model, 
but the evolution of perturbations is qualitatively similar to 
that in the $f(R)$ models (\ref{imodel}) and (\ref{iimodel}). 
We define the following quantity \cite{fR1}
\be
B \equiv m \frac{H}{\dot{H}} \frac{\dot{R}}{R}\,,
\label{Bdef}
\ee
whose today's value is denoted as $B_0$. 
We note that $B$ is of the same order as $m$. 
For increasing $B_0$, the CMB power spectrum exhibits stronger 
amplification relative to that in the $\Lambda$CDM model. 
This is attributed to the late-time ISW effect 
induced by the enhanced gravitational potential $\psi_{\rm eff}$. 
As we see in the middle panel of Fig.~\ref{fig2}, for increasing $B_0$, 
the weak lensing power spectrum is also subject to stronger enhancement. 
In the right panel, we can also confirm that the modified growth 
of $\delta_m$ leads to the amplification of the matter power spectrum.

For the $f(R)$ model (\ref{imodel}), the likelihood analysis based on the data 
of CMB, galaxy clusterings, weak gravitational lensing combined 
with the SN Ia, BAO, and $H_0$ 
data showed that the parameter $B_0$ is constrained to be \cite{Lomb}
\be
B_0<1.1 \times 10^{-3}\,,
\ee
at 95 \% CL. 
The similar bounds were also derived in Refs.~\cite{Sil13,Raveri14} 
and \cite{Battye} by taking the effective field theory approach and 
the designer approach to $f(R)$ gravity, respectively. 
In the designer approach, the time-dependence of $w_{\rm DE}$ is 
a priori assumed, under which the functional form of $f(R)$ is determined. 
For designer $f(R)$ models with constant $w_{\rm DE}$, the joint data analysis 
based on CMB, BAO, and lensing data placed the bounds $|w_{\rm DE}+1|<0.002$ 
and $B_0<0.006$ at 95 \% CL \cite{Battye}.

We also note that there are bounds on the value of $|f_{,R}-1|$ constrained 
from the suppression of fifth forces in dense environments. 
For the model (\ref{imodel}), the screening of fifth forces in the Milky Way 
and dwarf galaxies put the bounds $|f_{,R0}-1| \lesssim 10^{-5}$ \cite{fR1} and 
$|f_{,R0}-1| \lesssim 10^{-7}$ \cite{Jain12}, respectively, at 95 \% CL, 
where $f_{,R0}$ is today's value of $f_{,R}$.

These bounds show that there have been no observational signatures for the $f(R)$ 
modification of gravity. The crucial point is that the gravitational coupling 
with matter is large in the massless regime 
${\cal F}_M(k)<1$, such that $\mu=(4/3)e^{-2\phi/(\sqrt{6}M_{\rm pl})}$. 
To avoid the large enhancement of $\delta_m$, the perturbations 
need to be in the massive regime ${\cal F}_M(k)>1$ until recently. In this case, however, 
the scalar field hardly evolves by today, so the $f(R)$ models are not be distinguished from 
the $\Lambda$CDM model both at the levels of background and perturbations.

\subsection{BD theory with a scalar potential}

The BD theory with a light scalar (satisfying $M_{\phi}^2 \lesssim H^2$) can 
be consistent with local gravity constraints for the coupling 
$|Q|<2.5 \times 10^{-3}$.
In this case, however, it is difficult to distinguish BD theory from quintessence.
If  $|Q| \gtrsim 10^{-3}$, then the potential needs to be designed to have a large 
mass in regions of the high density for suppressing the propagation of fifth forces. 
The potential of this type is given by \cite{BD1}
\be
V(\phi)=V_0 \left[ 1-C (1-e^{-2Q \phi/M_{\rm pl}})^p \right]\,,
\label{Vphige}
\ee
where $V_0>0, 0<C<1, 0<p<1$ are constants. 
This is the generalization of the potential (\ref{Vphiap}) in 
$f(R)$ gravity to arbitrary couplings $Q$.

The background cosmological dynamics in BD theory with (\ref{Vphige}) 
is similar to that in $f(R)$ models explained in Sec.~\ref{fRsec}.
As long as $|\phi/M_{\rm pl}|\ll1$ in the matter era, the field slowly evolves along the instantaneous minimum 
at $\phi=\phi_m$ determined by the condition (\ref{mire1}), i.e., 
\be
\phi_m \simeq \frac{M_{\rm pl}}{2Q}
\left( \frac{2V_0p\,C}{\rho_m} \right)^{1/(1-p)}\,.
\label{phimexpre}
\ee
In the limit $\rho_m \to \infty$, $\phi_m$ approaches 0 with the 
divergent mass squared $M_{\phi}^2=V_{,\phi \phi}$. Hence 
the fine-tuning problem of initial conditions mentioned in 
Sec.~\ref{fRsec} also persists in this case.
During the early matter era, the variation of $\phi$ is suppressed due to 
the heavy mass, so $w_{\rm DE}$ is close to $-1$. 
The large deviation of $w_{\rm DE}$ from $-1$ starts to occur after 
$\rho_m$ drops to the order of $V(\phi)$.
Finally, the solution approaches the de Sitter fixed point 
given by Eq.~(\ref{mire2}). 
Under the bound (\ref{Qbound}), the deviation of $w_{\rm DE}$ from 
$-1$ around today should be also limited as in the case of $f(R)$ gravity.

Since $M_{\phi}^2$ is large in the early matter era, the perturbations 
relevant to the linear growth of large-scale structures are initially 
in the massive region ${\cal F}_M(k)>1$. 
After the entry to the massless regime ${\cal F}_M(k)<1$, 
the quantities $\mu$ and $\Sigma$ approach the values 
given by Eq.~(\ref{muBD}). 
Solving Eq.~(\ref{delmeqf}) with Eqs.~(\ref{Poi}) and (\ref{psieff})
in the massless regime, the evolution of perturbations 
during the matter-dominated epoch yields
\be
\delta_{m}\propto t^{(\sqrt{25+48Q^{2}}-1)/6}\,,\qquad 
\psi_{\rm eff}=-\frac{2}{1+2Q^2}\Psi
=\frac{2}{1-2Q^2}\Phi \propto 
t^{(\sqrt{25+48Q^{2}}-5)/6}\,.
\label{delmBD}
\ee
For the coupling $|Q|>1/\sqrt{6}$, the growth rate of $\delta_m$ is 
larger than that in $f(R)$ gravity. 
This is also the case for the gravitational potentials 
$\psi_{\rm eff}, \Psi, \Phi$. 
Then, the observational constraints on BD theory with $|Q|>1/\sqrt{6}$ should be more stringent than those on 
$f(R)$ models. As we explained in Sec.~\ref{fRsec} in $f(R)$ gravity, 
it is hard to observationally distinguish BD theory with the potential 
from the $\Lambda$CDM model.

If the chameleon mechanism works in local regions of the Universe, 
there is the bound arising from the violation of equivalence principle
in the solar system \cite{Will,chame2}. 
For the potential (\ref{Vphige}), this bound translates to \cite{BD1}
\be
p>1-\frac{5}{13.8-{\rm log}_{10}|Q|}\,.
\ee
For $|Q|=0.1$ and $|Q|=0.01$, we have $p>0.66$ and 
$p>0.68$, respectively. 
It may be of interest to study whether BD theory with 
$10^{-3} \lesssim |Q| \lesssim 0.1$ can leave some 
observational signatures different from those in the 
$\Lambda$CDM model 
(see Refs.~\cite{Fine1,Fine2,Fine3} for 
related works).

If we transform the action (\ref{lagbra}) to that in the Einstein frame
under the conformal transformation, the scalar field $\phi$ 
has a direct coupling $Q$ with nonrelativistic matter 
(dark matter and baryons) \cite{chame2,fRreview}. 
Instead, we may start from the  
Einstein frame action by assuming that the field $\phi$ 
is coupled to dark matter alone \cite{Amenco}.
Then, the solar-system constraints are evaded due 
to the absence of interactions between baryons and the scalar field.
In this coupled quintessence scenario with the
inverse power-law potential 
$V(\phi)=V_0\phi^{-p}$ ($p>0$), 
the Planck CMB data placed the upper bound 
$|Q|<0.066$ at 95 \% CL \cite{Planckdark}. 
Combined with the BAO and weak lensing data, there is 
a peak around $|Q|=0.04$ for the marginalized posterior 
distribution of $Q$ \cite{Petto,Planckdark}. 
The statistically strong observational evidence for supporting 
the nonvanishing coupling $Q$ has been still lacking, but the future 
high-precision observations may clarify whether the coupling 
of order $|Q|={\cal O}(0.01)$ is really favored from the data.

\section{Class (C): Kinetic braidings}
\label{classCsec}

The Lagrangian of class (C), which is known as 
kinetic braidings, is expressed in the form 
\be
L=G_2(\phi, X)+G_3(\phi, X) \square \phi
+\frac{M_{\rm pl}^2}{2}R\,.
\label{modelC}
\ee
Since $q_t=M_{\rm pl}^2$, the no-ghost 
condition of tensor perturbations is satisfied.
The parameter $\alpha_{\rm M}$ vanishes due 
to the absence of nonminimal couplings $G_4(\phi)$.
Then, the above theories automatically evade the 
bound (\ref{alpM}).

Unlike $f(R)$ gravity, we do not need to consider a heavy mass of $\phi$ in dense regions to suppress the propagation 
of fifth forces, so we will focus on the case in which the field 
mass squared $M_{\phi}^2=-G_{2,\phi \phi}$ does not 
exceed the order of $H^2$. 
Provided that the ratio $q_t/q_s$ is not much different from 1, 
the term $2a^2 M_{\phi}^2q_t/(c_s^2 k^2 q_s)$ in 
the expression of $\Delta_2$ of Eq.~(\ref{Delta2}) is 
much smaller than 1 for the modes deep inside the 
sound horizon. 
Substituting $\Delta_2 \simeq \tp^2 q_s c_s^2/(4\Mpl^2H^2)$
into Eq.~(\ref{muC}), it follows that 
\be
\mu=\Sigma \simeq 1+\frac{4M_{\rm pl}^4H^2\aB^2}
{\tp^2q_sc_s^2}\,,
\label{muIII}
\ee
which does not have the scale-dependence.
The evolution of linear perturbations is determined by 
the product $4M_{\rm pl}^4H^2\aB^2/(\tp^2q_sc_s^2)$. 
Due to the absence of nonminimal couplings in the 
Lagrangian (\ref{modelC}), the typical 
solution around a spherically symmetric and static body 
corresponds to $\phi={\rm constant}$ \cite{DKT12,Kase13}, 
so that the propagation of fifth forces is suppressed in local regions of the Universe.

\subsection{Concrete dark energy models}

We present models of the late-time cosmic acceleration which 
belong to the Lagrangian (\ref{modelC}).
The cubic covariant Galileon \cite{Galileons} corresponds to 
\be
{\rm Model~(C1)}:~~
L=\beta_1X-m^3 \phi+\beta_3X \square \phi
+\frac{M_{\rm pl}^2}{2}R\,,
\label{iii1}
\ee
where $\beta_1, \beta_3, m$ are constants. 
In the limit of Minkowski spacetime, the equations of motion following 
from the Lagrangian (\ref{iii1}) respects the Galilean symmetry 
$\partial_{\mu} \phi \to \partial_{\mu} \phi+b_{\mu}$.

In absence of the linear linear potential $V(\phi)=m^3 \phi$, 
it is known that there exists a tracker solution for the full 
Galileon Lagrangian containing quartic and quintic 
couplings \cite{DT10,DT10b}. 
The dark energy EOS along the tracker is 
given by $w_{\rm DE}=-2$ during the matter era. 
This is followed by the approach to the self-accelerating 
de Sitter solution characterized by $X={\rm constant}$.
The tracker solution is disfavored from the joint data analysis 
of SNIa, CMB, and BAO \cite{NDT10} because of the large 
deviation of $w_{\rm DE}$ from $-1$ in the matter era. 
Even in the case where the solutions approach the tracker 
at late times, the covariant Galileon without the linear 
potential is in strong tension with observational 
data \cite{AppleLin,Neveu,Barreira1,Barreira2,Renk,Peirone2} 
due to a very different structure formation pattern 
compared to that in the 
$\Lambda$CDM model \cite{Kase10}.

The likelihood analysis performed in 
Refs.~\cite{Barreira1,Barreira2,Renk,Peirone2} assumed that the 
Galileon potential is absent. In presence of the linear potential 
$V(\phi)=m^3 \phi$, we do not have the self-accelerating solution 
with $X={\rm constant}$, but the late-time cosmic acceleration 
can be driven by $V(\phi)$ (see also Ref.~\cite{Ali12} 
for other potentials). 
After the field $\phi$ enters the region $V(\phi)<0$, the 
Universe starts to enter the collapsing stage. 
Existence of the linear potential can modify the dynamics of 
background and perturbations, so it is worthy of studying 
the compatibility of model (\ref{iii1}) with observations.
In Sec.~\ref{mostsec}, we will discuss the cosmological dynamics 
of such a model by taking into account a more general nonminimal 
coupling $G_4(\phi)$.

Instead of the linear potential $V(\phi)=m^3 \phi$, we can consider the quadratic kinetic term $X^2$.
This theory is described by the Lagrangian \cite{Kase18}:
\be
{\rm Model~(C2)}:~~
L=\beta_1 X+\beta_2 X^2+\beta_3 X \square \phi
+\frac{M_{\rm pl}^2}{2}R\,,
\label{iii2}
\ee
where $\beta_1,\beta_2,\beta_3$ are constants. 
This can be regarded as the ghost condensate model 
with the cubic Galileon term. 
In absence of the term $\beta_2 X^2$, the solutions tend to 
approach the tracker solution mentioned above, which is 
observationally excluded. 
Moreover, the dominance of the cubic Galileon term  
$\beta_3 X \square \phi$ at low redshifts typically gives rise 
to the galaxy-ISW anti-correlation, which is also 
disfavored from the data \cite{Renk}.
For the model (\ref{iii2}), the new term $\beta_2 X^2$ prevents the 
approach to the tracker solution, so that the deviation 
of $w_{\rm DE}$ from $-1$ is smaller than that in the tracker case. 
Moreover, the cubic Galileon term can be subdominant to 
the contribution $G_2=\beta_1 X+\beta_2 X^2$, so it is expected that 
the galaxy-ISW anti-correlation does not necessarily occur.
In Sec.~\ref{cghostsec}, we study the cosmological dynamics 
and observational signatures for the model (\ref{iii2}).

\subsection{Cubic Galileon with ghost condensate}
\label{cghostsec}

As an example of class (C), we review the cosmology of 
model (C2) given by the Lagrangian (\ref{iii2}). 
In theories beyond Horndeski 
gravity \cite{Zuma,GLPV,Langlois1,Langlois2,Crisostomi16}, 
it is possible to introduce the $X$ dependence in the 
function $G_4$, while keeping the 
value $c_t^2=1$ due to an additional term outside the 
domain of Horndeski theories \cite{GWcon2,GWcon3}. 
Since this review focuses on Horndeski theories, we will not 
take into account such terms in the following.

For the matter sector, we take into account radiation 
(density $\rho_r$ and pressure $P_r=\rho_r/3$) and 
nonrelativistic matter (density $\rho_m$ and pressure $P_m=0$). 
We also introduce the following dimensionless quantities:
\be
\Omega_{\phi 1} \equiv 
\frac{\beta_1 \dot{\phi}^2}{6M_{\rm pl}^2 H^2}\,,
\qquad 
\Omega_{\phi 2} \equiv 
\frac{\beta_2 \dot{\phi}^4}{4M_{\rm pl}^2 H^2}\,,
\qquad 
\Omega_{\phi 3} \equiv 
-\frac{\beta_3\dot{\phi}^3}{M_{\rm pl}^2 H}\,,
\qquad 
\Omega_r \equiv \frac{\rho_r}{3 M_{\rm pl}^2 H^2}\,,
\ee
which correspond to the density parameters arising from 
$\beta_1 X$, $\beta_2 X^2$, $\beta_3 X \square \phi$, 
and radiation, respectively.
The Hamiltonian constraint (\ref{Frieq}) gives the relation
\be
\Omega_m \equiv \frac{\rho_m}{3M_{\rm pl}^2 H^2}
=1-\Omega_{\rm DE}-\Omega_r\,,\qquad 
\Omega_{\rm DE} \equiv 
\Omega_{\phi 1}+\Omega_{\phi 2}+\Omega_{\phi 3}\,.
\label{Omem}
\ee
The density parameters obey the differential equations:
\ba
\Omega_{\phi 1}' = 2\Omega_{\phi 1} 
\left( \epsilon_{\phi}-h \right)\,,\qquad
\Omega_{\phi 2}' = 2\Omega_{\phi 2} 
\left( 2\epsilon_{\phi}-h \right)\,,\qquad
\Omega_{\phi 3}' = \Omega_{\phi 3} 
\left( 3\epsilon_{\phi}-h \right)\,,\qquad
\Omega_r' = -2\Omega_r \left( 2+h \right)\,,
\label{Omereq}
\ea
where, from Eqs.~(\ref{dH}) and (\ref{ddphi}),  
the quantities $h \equiv \dot{H}/H^2$ and 
$\epsilon_{\phi} \equiv 
\ddot{\phi}/(H \dot{\phi})$ are given, respectively, by  
\ba
h &=&
-\frac{2(3\Omega_{\phi 1}+\Omega_{\phi 2}
+\Omega_r+3)(\Omega_{\phi 1}+2\Omega_{\phi 2})
+2\Omega_{\phi 3}(6\Omega_{\phi 1}+3\Omega_{\phi 2}
+\Omega_r+3)+3\Omega_{\phi 3}^2}
{4 \left( \Omega_{\phi 1}+2\Omega_{\phi 2}
+\Omega_{\phi 3} \right)+\Omega_{\phi 3}^2}\,,
\label{dh}\\
\epsilon_{\phi} 
&=&-\frac{4(3\Omega_{\phi 1}+2\Omega_{\phi 2})
-\Omega_{\phi 3}(3\Omega_{\phi 1}+\Omega_{\phi 2}
+\Omega_r-3)}{4 \left( \Omega_{\phi 1}+2\Omega_{\phi 2}
+\Omega_{\phi 3} \right)+\Omega_{\phi 3}^2}\,.
\label{epphi}
\ea
The dark energy EOS defined by Eq.~(\ref{wdedef}) yields
\be
w_{\rm DE}=\frac{3\Omega_{\phi 1}+\Omega_{\phi 2}
-\epsilon_{\phi}\Omega_{\phi 3}}
{3(\Omega_{\phi 1}+\Omega_{\phi 2}+\Omega_{\phi 3})}\,.
\label{wdecu}
\ee

For the dynamical system given by Eq.~(\ref{Omereq}),  
there exists a self-accelerating de Sitter solution 
satisfying $\dot{\phi}={\rm constant}$ and $H={\rm constant}$, 
in which case $\epsilon_{\phi}=0, h=0$, and $\Omega_r=0$. 
From Eqs.~(\ref{dh}) and (\ref{epphi}), we obtain the following 
two relations on the de Sitter solution:
\be
\Omega_{\phi 1}=-2+\frac{1}{2}\Omega_{\phi 3}\,,\qquad 
\Omega_{\phi 2}=3-\frac{3}{2}\Omega_{\phi 3}\,,
\label{Omephire}
\ee
under which the dark energy EOS (\ref{wdecu}) 
reduces to $w_{\rm DE}=-1$. 
As shown in Ref.~\cite{Kase18}, this de Sitter solution is 
always a stable attractor.
In the limit $\Omega_{\phi 3} \to 0$, we have 
$\Omega_{\phi 1}=-2$ and $\Omega_{\phi 2}=3$, 
so that $\beta_1=-\beta_2 \dot{\phi}^2$. 
Indeed, this is equivalent to the de Sitter solution 
in ghost condensate corresponding to $G_{2,X}=0$ 
in Eq.~(\ref{wdeke}). As in the case of ghost condensate, 
we consider the couplings $\beta_1<0$ and $\beta_2>0$ 
in the following discussion.

\begin{figure}[h]
\begin{center}
\includegraphics[height=3.2in,width=3.4in]{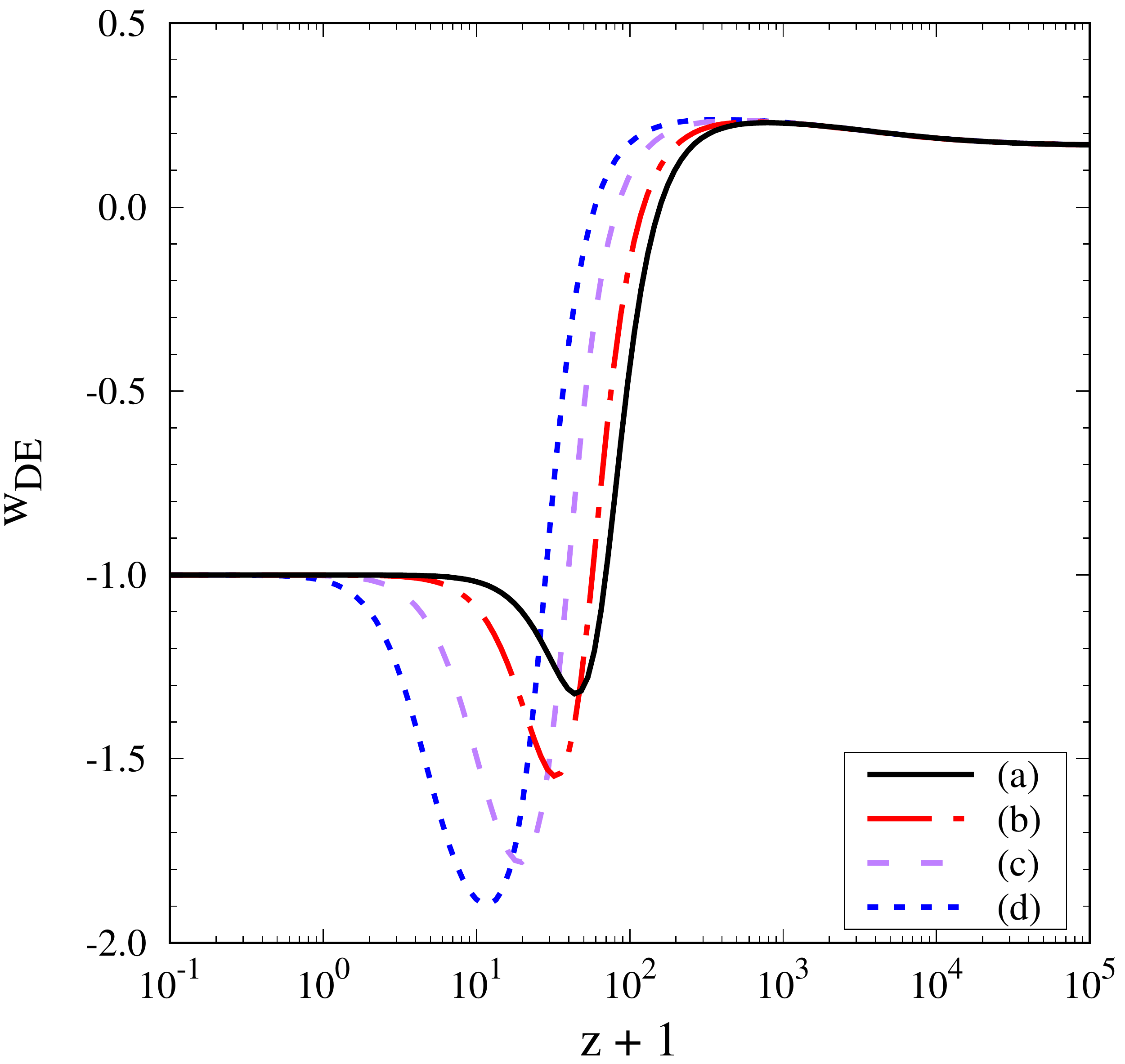}
\includegraphics[height=3.2in,width=3.4in]{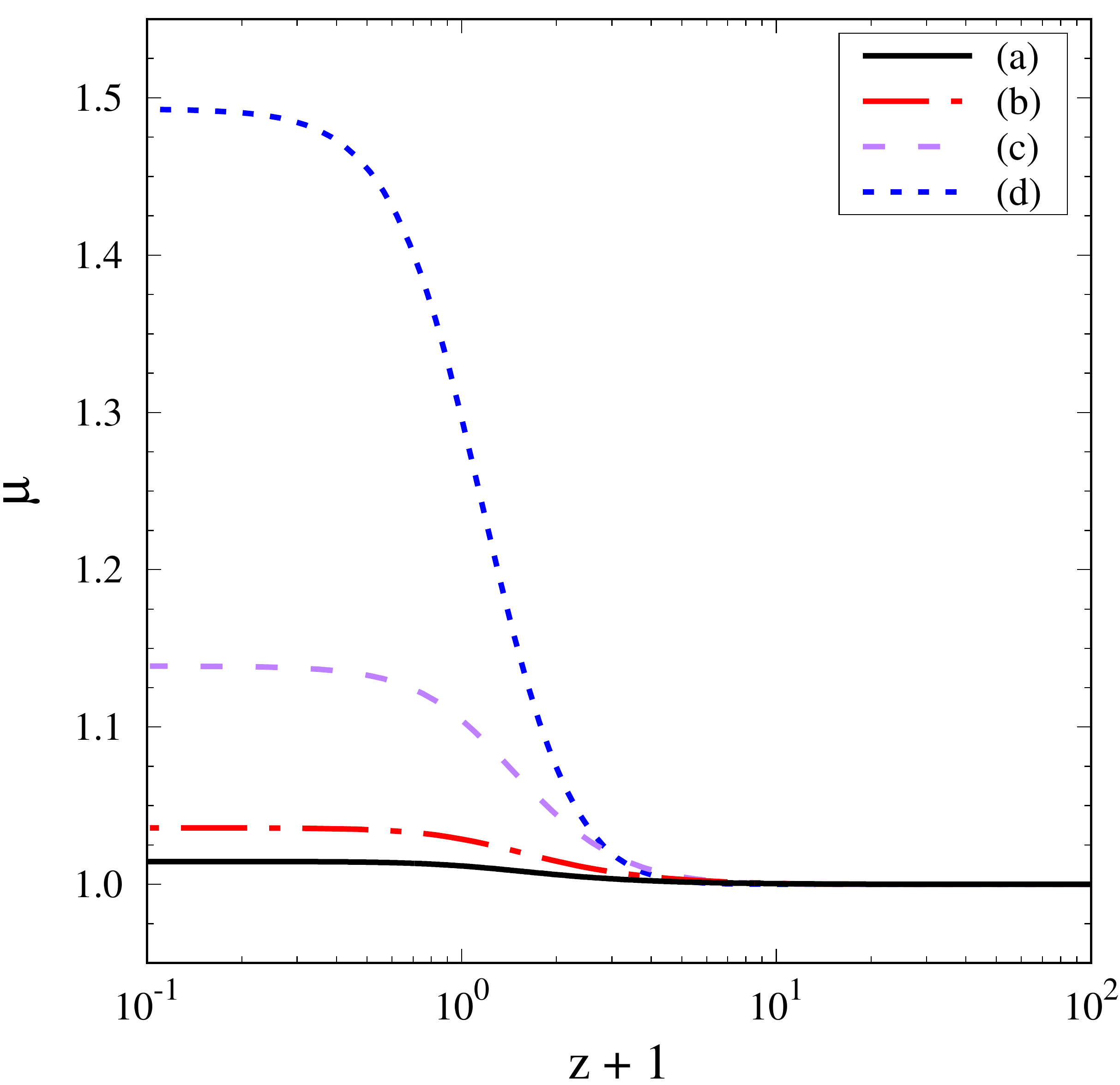}
\end{center}
\caption{\label{fig3}
(Left) Evolution of $w_{\rm DE}$ versus $z + 1$ 
for four different initial conditions: 
(a) $\Omega_{\phi 1}=-1.8 \times 10^{-14}$, 
$\Omega_{\phi 2}=1.0 \times 10^{-11}$, 
(b) $\Omega_{\phi 1}=-1.0 \times 10^{-14}$, 
$\Omega_{\phi 2}=3.0 \times 10^{-12}$, 
(c) $\Omega_{\phi 1}=-4.0 \times 10^{-15}$, 
$\Omega_{\phi 2}=5.0 \times 10^{-13}$, and 
(d) $\Omega_{\phi 1}=-1.8 \times 10^{-15}$, 
$\Omega_{\phi 2}=1.0 \times 10^{-13}$, 
at the redshift $z=1.3 \times 10^5$. 
The other initial conditions are chosen to be 
$\Omega_{\phi 3}=1.0 \times 10^{-6}$ and 
$\Omega_r=0.975$.
The present epoch ($z=0$) corresponds to 
$\Omega_{\rm DE}=0.68$ and $\Omega_r \simeq 10^{-4}$. 
(Right) Evolution of $\mu~(=\Sigma)$ corresponding to the cases 
(a), (b), (c), (d) plotted in the left panel.
}
\end{figure}

In the radiation and deep matter eras, we first consider the case 
in which $\Omega_{\phi 3}$ dominates over 
$|\Omega_{\phi 1}|$ and $\Omega_{\phi 2}$. 
In these epochs, the field density parameters are smaller than 
the order 1, so we obtain the approximate relations   
$h \simeq -(\Omega_r+3)/2$ and 
$\epsilon_{\phi} \simeq (\Omega_r-3)/4$
from Eqs.~(\ref{dh}) and (\ref{epphi}).
Substituting them into Eq.~(\ref{wdecu}), it follows that 
\be
w_{\rm DE} \simeq \frac{1}{4}-\frac{1}{12}\Omega_r\,,
\label{wdeG3}
\ee
and hence $w_{\rm DE} \simeq 1/6$ in the radiation era and 
$w_{\rm DE} \simeq 1/4$ in the matter era. 
This behavior can be confirmed in the numerical simulation 
of Fig.~\ref{fig3}.
Since $h \simeq -3/2$ and $\epsilon_{\phi} \simeq -3/4 $ 
during the matter dominance, we can integrate Eq.~(\ref{Omereq})
to give $|\Omega_{\phi 1}| \propto a^{3/2}, \Omega_{\phi 2} 
\propto a^0$, and $\Omega_{\phi 3} \propto a^{-3/4}$. 
Then, the density parameters $|\Omega_{\phi 1}|$ and 
$\Omega_{\phi 2}$ eventually catch up with 
$\Omega_{\phi 3}$. After this catch up, the evolution of 
$w_{\rm DE}$ is no longer described by Eq.~(\ref{wdeG3}). 
There is a transient period in which $w_{\rm DE}$ enters the 
region $w_{\rm DE}<-1$.
The solutions finally approach the de Sitter fixed point 
characterized by Eq.~(\ref{Omephire}).

For smaller initial conditions of $\Omega_{\phi 2}$, the minimum value 
of $w_{\rm DE}$ reached during the transient period tends to 
decrease toward $-2$, see the left panel of Fig.~\ref{fig3}. 
This can be understood as follows. Taking the limit 
$\Omega_{\phi 2} \to 0$ in Eq.~(\ref{Omereq}), the quantity 
$y \equiv 2\Omega_{\phi 1}+\Omega_{\phi 3}$ 
obeys the differential equation 
\be
y'=y \frac{12\Omega_{\phi 1}^2+\Omega_{\phi 1} (4\Omega_r-12+
27\Omega_{\phi 3})+\Omega_{\phi 3} (5\Omega_r-3
+3\Omega_{\phi 3})}
{4\Omega_{\phi 1}+4\Omega_{\phi 3}+\Omega_{\phi 3}^2}\,,
\ee
which admits the solution $y=0$, i.e., $\Omega_{\phi 3}=-2\Omega_{\phi 1}$. Substituting this relation and Eq.~(\ref{epphi}) 
into Eq.~(\ref{wdecu}), it follows that 
\be
w_{\rm DE}=-\frac{6+\Omega_r}{3(1-\Omega_{\phi 1})}\,.
\ee
This is the tracker solution \cite{DT10,DT10b} along 
which $w_{\rm DE}$ evolves as $-7/3$ (radiation era) 
$\to -2$ (matter era) $\to -1$ 
(de Sitter era satisfying $\Omega_{\phi 1}=-1$ and 
 $\Omega_{\phi 3}=2$). 
As we already mentioned, the tracker is in tension with 
the observational data \cite{NDT10} due to the large deviation of 
$w_{\rm DE}$ from $-1$. 
Since $\Omega_{\phi 2} \neq 0$ in the present model, 
the dominance of $\Omega_{\phi 2}$ over $\Omega_{\phi 3}$
prevents the approach to the tracker. 
For larger initial values of $\Omega_{\phi 2}$, the deviation 
of $w_{\rm DE}$ from $-1$ tends to be smaller. 
Moreover, the approach to $w_{\rm DE}=-1$ occurs 
at higher redshifts. 
Thus, the existence of the term $\beta_2 X^2$ allows an 
interesting possibility for realizing the phantom equation 
of state ($w_{\rm DE}<-1$) which can be compatible 
with observations. 

The quantities $q_s$ and $c_s^2$ can be expressed, 
respectively, as 
\be
q_s=\frac{3M_{\rm pl}^4H^2}{\tp^2} 
\left( 4 \Omega_{\phi 1}+8\Omega_{\phi 2}
+4\Omega_{\phi 3}
+\Omega_{\phi 3}^2 \right)\,,\qquad 
c_s^2=\frac{12\Omega_{\phi 1}+8\Omega_{\phi 2}
+4(\epsilon_{\phi}+2)\Omega_{\phi 3}-\Omega_{\phi 3}^2}
{3(4\Omega_{\phi 1}+8\Omega_{\phi 2}+4\Omega_{\phi 3}+
\Omega_{\phi 3}^2)}\,.
\label{stabC2}
\ee
The conditions $q_s>0$ and $c_s^2>0$ hold for all
the cases shown in Fig.~\ref{fig3}, so 
there are no ghost and Laplacian instabilities of scalar perturbations.
Substituting Eq.~(\ref{stabC2}) and $\aB=-\Omega_{\phi 3}/2$ into Eq.~(\ref{muIII}), we obtain
\be
\mu=\Sigma \simeq 1+\frac{\Omega_{\phi 3}^2}
{12\Omega_{\phi 1}+8\Omega_{\phi 2}
+4(\epsilon_{\phi}+2)\Omega_{\phi 3}-\Omega_{\phi 3}^2}\,, 
\label{muIII2}
\ee
so that the deviations of $\mu$ and $\Sigma$ arise from 
the cubic coupling. Using the relations (\ref{Omephire}) on the de Sitter 
fixed point, it follows that 
\be
\mu=\Sigma \simeq 1+\frac{\Omega_{\phi 3}}{2-\Omega_{\phi 3}}\,.
\label{mudS}
\ee
In the right panel of Fig.~\ref{fig3}, the evolution of $\mu$ is plotted 
for four different cases corresponding to those in the left panel. 
Solving the perturbation equations of motion numerically, we 
compute $\mu$ and $\Sigma$ according to their definitions given 
in  Eqs.~(\ref{Poi}) and (\ref{psieff}). We confirmed that the analytic 
estimation (\ref{muIII2}), which is derived under the quasi-static approximation, 
is sufficiently accurate for the modes deep inside the sound horizon.
In the deep matter era, $\mu$ is close to 1 in all these cases, 
but the difference arises at low redshifts depending on the values 
of $\Omega_{\phi 3}$. For the cases (a), (b), (c), (d), the numerical 
values of $\Omega_{\phi 3}$ today are given, respectively, by 
$2.3 \times 10^{-2}$, $5.6 \times 10^{-2}$, $2.0 \times 10^{-1}$, 
$5.1 \times 10^{-1}$, so the deviation of $\mu$ from 1 tends to be 
more significant for larger $\Omega_{\phi 3}$. 
This property is also consistent with the estimation (\ref{mudS}) 
on the de Sitter solution.

For the cases in which $\Omega_{\phi 3}$ gives the large contribution to the field density today 
(say $\Omega_{\phi 3}(z=0)>{\cal O}(0.1)$), 
the model may be tightly constrained from the galaxy-ISW correlation 
data. Moreover, in such cases, the deviation of $w_{\rm DE}$ 
from $-1$ at low redshifts tends to be significant. 
It will be of interest to place observational constraints on today's 
value of $\Omega_{\phi 3}$ by performing the joint likelihood analysis
including the galaxy-ISW correlation data. 

\section{Class (D): Nonminimally coupled scalar with cubic 
derivative interactions}
\label{mostsec}

The class (D) is the most general Horndeski theories given by 
the Lagrangian (\ref{lagcon}) in which $c_t^2$ is exactly equivalent to 1. 
In this case, the quantities $\alpha_{\rm B}$ and 
$\alpha_{\rm M}$ give rise to nonvanishing contributions 
to the second terms in Eqs.~(\ref{Gefff2}) and (\ref{Sigmaf2}). 
The difference between $\mu$ and 
$\Sigma$ is given by 
\be
\mu-\Sigma=-\frac{2M_{\rm pl}^2H^2q_t\alpha_{\rm M}
(\alpha_{\rm B}-\alpha_{\rm M})}
{\tp^2 q_s c_s^2}\,, 
\ee
which means that  the nonminimal coupling $G_4(\phi)R$ 
generally leads to $\mu \neq \Sigma$ due to the nonvanishing 
$\alpha_{\rm M}$. 
This property is different from that in the models of 
classes (A) and (C).

In Sec.~\ref{fRBDsec}, we showed that it is difficult to observationally 
distinguish $f(R)$ gravity and BD theory with a scalar potential
from the $\Lambda$CDM model. 
In such class (B) theories, the field potential needs to be designed to have 
a large mass in regions of the high density for the 
chameleon mechanism to work.
There is yet another mechanism dubbed the Vainshtein 
mechanism \cite{Vain} in which field nonlinear 
self-interactions (like the cubic Galileon) can suppress 
the propagation of fifth forces in local regions of the 
Universe. In such cases, the background cosmological 
dynamics and the evolution of perturbations are 
different from those in class (B). 

In class (D) the quantity $\Sigma$ can grow rapidly at 
low redshifts, while this is not generally the case for 
class (B) in which $\Sigma=1/(16\pi G G_4)$.
This allows the possibility for distinguishing the models 
between classes (B) and (D) from the observations of 
galaxy-ISW correlations further.
The class (D) is categorized as Horndeski theories containing 
a nonminimally coupled light scalar field with cubic derivative 
interactions. 

\subsection{Concrete dark energy models}

One of the theories in class (D) is known as generalized 
BD theories \cite{DTge} given by the Lagrangian 
\be
{\rm Model~(D1)}:~~L=\omega \left( \frac{\phi}{M_{\rm pl}} 
\right)^{1-n}X+\frac{\lambda}{\mu^3}  \left( \frac{\phi}{M_{\rm pl}} 
\right)^{-n}X \square \phi+\frac{M_{\rm pl}^2}{2} \left( \frac{\phi}{M_{\rm pl}} 
\right)^{3-n}R\,,
\label{LagGBD}
\ee
where $\mu~(>0)$ is a constant having a dimension of mass, 
and $n, \lambda, \omega$ are dimensionless constants. 
The original BD theory \cite{Brans} without the cubic coupling corresponds to the power $n=2$. 
In presence of the cubic derivative interaction, 
there is a de Sitter solution characterized by  
$\dot{\phi}/(H \phi)={\rm constant}$ even 
without the scalar potential. 
The case $n=2$ was first studied in Ref.~\cite{Silva}
(see also Refs.~\cite{Koba1,Koba2}) and then it was 
generalized to the more general case with $n \neq 2$ 
in Ref.~\cite{DTge}.
The power $n$ is constrained to be 
\be
2 \le n \le 3\,,
\ee
to satisfy theoretical consistent conditions throughout 
the cosmic expansion history \cite{DTge}. 
In model (D1), the cubic derivative coupling is the main 
source for the late-time cosmic acceleration. For $n=2$, it was 
shown in Ref.~\cite{Koba1} that the galaxy-ISW 
anti-correlation tends to be stronger for larger $|\omega|$. 
For $n \neq 2$, the galaxy-ISW correlation power spectrum 
was not computed yet, but it is expected that the anti-correlation can still persist for the case in which the cubic 
interaction provides the dominant contribution to the dark energy 
density. It remains to be seen to put observational  
constraints on model (D1) by using the galaxy-ISW 
correlation data.

There is also the nonminimally coupled cubic Galileon model 
given by the Lagrangian 
\be
{\rm Model~(D2)}:~~
L=\left( 1-6Q^2 \right) e^{-2Q (\phi-\phi_0)/M_{\rm pl}}X
-m^3 \phi+\beta_3 X \square \phi+\frac{M_{\rm pl}^2}{2} 
e^{-2Q (\phi-\phi_0)/M_{\rm pl}}R\,,
\label{modelD2}
\ee
where $Q, \phi_0, m, \beta_3$ are constants. 
In the limit $Q \to 0$, Eq.~(\ref{modelD2}) recovers 
the Lagrangian (\ref{iii1}) 
of model (C1), i.e., cubic Galileons with $\beta_1=1$.
In high-density regions, the cubic Galileon term can 
suppress the fifth force 
induced by the nonminimal coupling 
$e^{-2Q (\phi-\phi_0)/M_{\rm pl}}R$.
The cosmological dynamics of model (D2) was studied 
in Ref.~\cite{KTD15} in a more general context 
beyond the Horndeski domain. 
In this model, the main source for the late-time cosmic 
acceleration is the linear potential $V(\phi)=m^3 \phi$ 
rather than the cubic Galileon term $\beta_3 X \square \phi$. 
Then, unlike model (D1), the dominance of cubic interactions
over $V(\phi)$ does not typically occur at low 
redshifts. In Sec.~\ref{cosnonsec}, we discuss the cosmological 
dynamics and observational signatures of model (D2). 

We note that it is also possible to take into account the 
nonminimal coupling for model (C2) introduced 
in Sec.~\ref{classCsec}. 
It remains to be seen to explore whether such a model 
is consistent with cosmological and local gravity constraints.

\subsection{Cosmology in nonminimally 
coupled cubic Galileons}
\label{cosnonsec}

To study the cosmological dynamics in model (D2) given 
by the Lagrangian (\ref{modelD2}), we introduce the following dimensionless quantities:
\be
x_1 \equiv \frac{\dot{\phi}}{\sqrt{6}M_{\rm pl} H}\,,
\qquad 
\Omega_{\phi_2} \equiv \frac{m^3 \phi}
{3M_{\rm pl}^2 H^2 F}\,,
\qquad \Omega_{\phi_3} \equiv 
-\frac{\beta_3 \dot{\phi}^3}{M_{\rm pl}^2 H F}\,,
\qquad 
\Omega_r \equiv \frac{\rho_r}{3M_{\rm pl}^2 H^2 F}\,,
\qquad 
\lambda \equiv -\frac{M_{\rm pl}}{\phi}\,,
\label{moDdi}
\ee
where $F \equiv e^{-2Q (\phi-\phi_0)/M_{\rm pl}}$. 
The effective gravitational coupling in a screened 
environment is given by $G_{\rm eff}=G/F$, so we consider 
the case in which the field value $\phi$ today is 
equivalent to $\phi_0$, i.e., $F(z=0)=1$.  
Taking into account the radiation and nonrelativistic matter 
as in the analysis of Sec.~\ref{cghostsec}, the Hamiltonian constraint (\ref{Frieq}) leads to 
\be
\Omega_m \equiv \frac{\rho_m}{3M_{\rm pl}^2 H^2 F}
=1-\left( 1-6Q^2 \right)x_1^2 -2\sqrt{6} Q x_1 
-\Omega_{\phi 2}-\Omega_{\phi 3}-\Omega_r\,.
\label{Omem2}
\ee
The quantities defined by Eq.~(\ref{moDdi}) satisfy
the differential equations:
\ba
& &
x_1'=x_1 \left( \epsilon_{\phi}- h \right)\,,\qquad 
\Omega_{\phi 2}'= \Omega_{\phi 2} \left[ 
\sqrt{6} \left( 2Q-\lambda \right)x_1-2h \right]\,,\qquad 
\Omega_{\phi 3}'= \Omega_{\phi 3} \left( 
2\sqrt{6} Q x_1+3\epsilon_{\phi}-h \right)\,,\nonumber \\
& &
\Omega_r'=2 \Omega_r \left( \sqrt{6} Q x_1-2-h \right)\,,
\qquad 
\lambda'=\sqrt{6} \lambda^2 x_1\,,
\ea
where the quantities $h=\dot{H}/H^2$ and 
$\epsilon_{\phi}=\ddot{\phi}/(H \dot{\phi})$ are 
given, respectively, by 
\ba
h &=& -[  \Omega_{\phi 3}(6+2\Omega_r-6\Omega_{\phi 2}+3\Omega_{\phi 3})
+\sqrt{6}\Omega_{\phi 3} 
(2Q-\lambda \Omega_{\phi 2})x_1 
+2 \{ 3+\Omega_r-3\Omega_{\phi 2}+6\Omega_{\phi 3}
-6\lambda Q \Omega_{\phi 2} \nonumber \\
& &+6Q^2 (1-\Omega_r+3\Omega_{\phi 2}
-2\Omega_{\phi 3}) \}x_1^2+2\sqrt{6}Q (6Q^2-1) 
(\Omega_{\phi 3}-2)x_1^3+6(12Q^4-8Q^2+1)x_1^4]
/{\cal D}\,,\label{hD}\\
\epsilon_{\phi} &=& 
[ \sqrt{6} \Omega_{\phi 3} (\Omega_r-3\Omega_{\phi 2} 
-3)+12 \{Q (\Omega_r-1-3\Omega_{\phi 2}
-2\Omega_{\phi 3})+\lambda \Omega_{\phi 2} \} 
x_1 \nonumber \\
& &
+3\sqrt{6} \{ \Omega_{\phi 3}-4+2Q^2 
(4+\Omega_{\phi 3}) \}x_1^2 
+12Q (5-6Q^2)x_1^3]/(\sqrt{6}{\cal D})\,,
\label{epD}
\ea
with 
\be
{\cal D}=4x_1^2+4\Omega_{\phi 3}+4\sqrt{6} Qx_1 
\Omega_{\phi 3}+\Omega_{\phi 3}^2\,.
\ee
The dark energy EOS defined by Eq.~(\ref{wdedef}) 
can be expressed as\footnote{In Ref.~\cite{KTD15}, the dark energy 
equation of state was defined in a different way 
such that Eqs.~(\ref{Frieq}) and (\ref{Frieq2}) are expressed
to contain the terms $3M_{\rm pl}^2F_0H^2$ and $2M_{\rm pl}^2F_0 \dot{H}$ 
on the left hand sides of them, respectively, where $F_0$ is today's value of $F$. 
The definition of $w_{\rm DE}$ in this review corresponds to the choice $F_0=1$, 
which should be more suitable to constrain the model from observations.} 
\be
w_{\rm DE}=-\frac{3+2h-[3+2h+3(1+2Q^2)x_1^2-3\Omega_{\phi 2}
-\epsilon_{\phi}\Omega_{\phi 3}-2\sqrt{6}Q x_1(2+\epsilon_{\phi})]F}
{3-3[1-\Omega_{\phi 2}-\Omega_{\phi 3}+(6Q^2-1)x_1^2-2\sqrt{6}Qx_1]F}\,.
\label{wdeD}
\ee

Let us consider the case in which the cubic Galileon density parameter 
$\Omega_{\phi 3}$ (assumed to be positive) dominates over the other field density parameters
in the early cosmological epoch, except for $Qx_1$,  
such that $1 \gg \Omega_{\phi 3} \gg \{ x_1^2, \Omega_{\phi 2} \}$. 
Then, Eqs.~(\ref{hD}) and (\ref{epD}) approximately reduce to 
\ba
h \simeq -\frac{3}{2}-\frac{\Omega_r}{2}\,,\qquad 
\epsilon_{\phi} \simeq -\frac{3}{4}+\frac{\Omega_r}{4}
-\frac{\sqrt{6}}{2} 
\left( 1-\Omega_r \right) \frac{Qx_1}{\Omega_{\phi 3}}\,.
\label{hepes}
\ea
Depending on the initial ratio $Qx_1/\Omega_{\phi 3}$, 
there are two qualitatively different cases for the evolution of 
$x_1$, $\Omega_{\phi 2}$, and $\Omega_{\phi 3}$. 
If $|Qx_1| \lesssim \Omega_{\phi 3}$, then we have 
$h \simeq -2$ and $\epsilon_{\phi} \simeq -1/2$ during the 
radiation dominance, so we obtain the integrated solutions 
$x_1^2 \propto a^{3}$, $\Omega_{\phi 2} \propto a^4$, 
and $\Omega_{\phi 3} \propto a^{1/2}$.
In this case, even if $\Omega_{\phi 3}$ initially 
dominates over the other field densities, 
$\Omega_{\phi 2}$ quickly catches up with $\Omega_{\phi 3}$. 
After this catch up, both $x_1^2$ and $\Omega_{\phi 3}$ become
much smaller than $\Omega_{\phi 2}$, so the linear potential 
$V(\phi)=m^3 \phi$ gives the dominant contribution to 
the scalar-field dynamics.

\begin{figure}[h]
\begin{center}
\includegraphics[height=3.2in,width=3.3in]{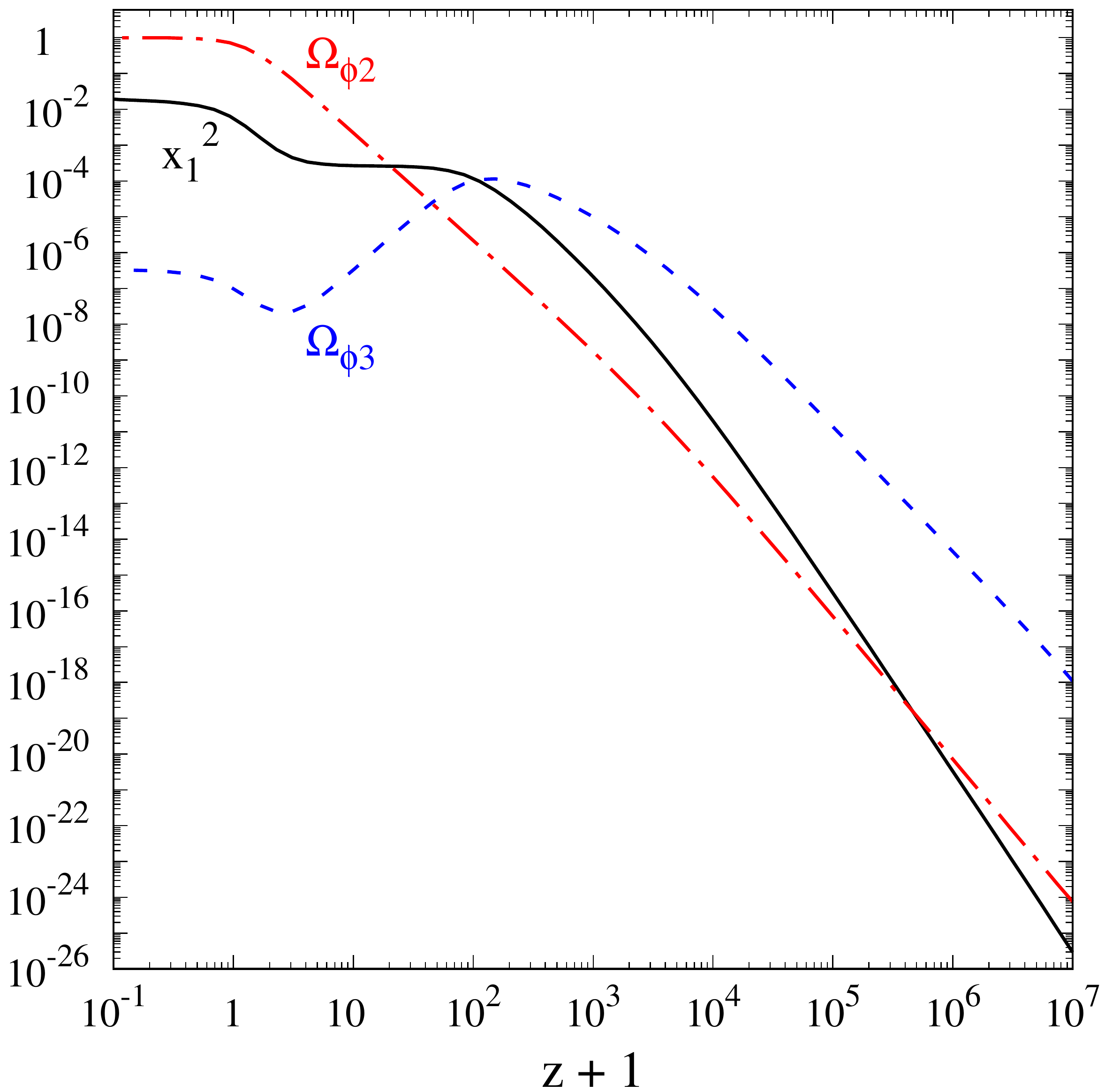}
\includegraphics[height=3.2in,width=3.3in]{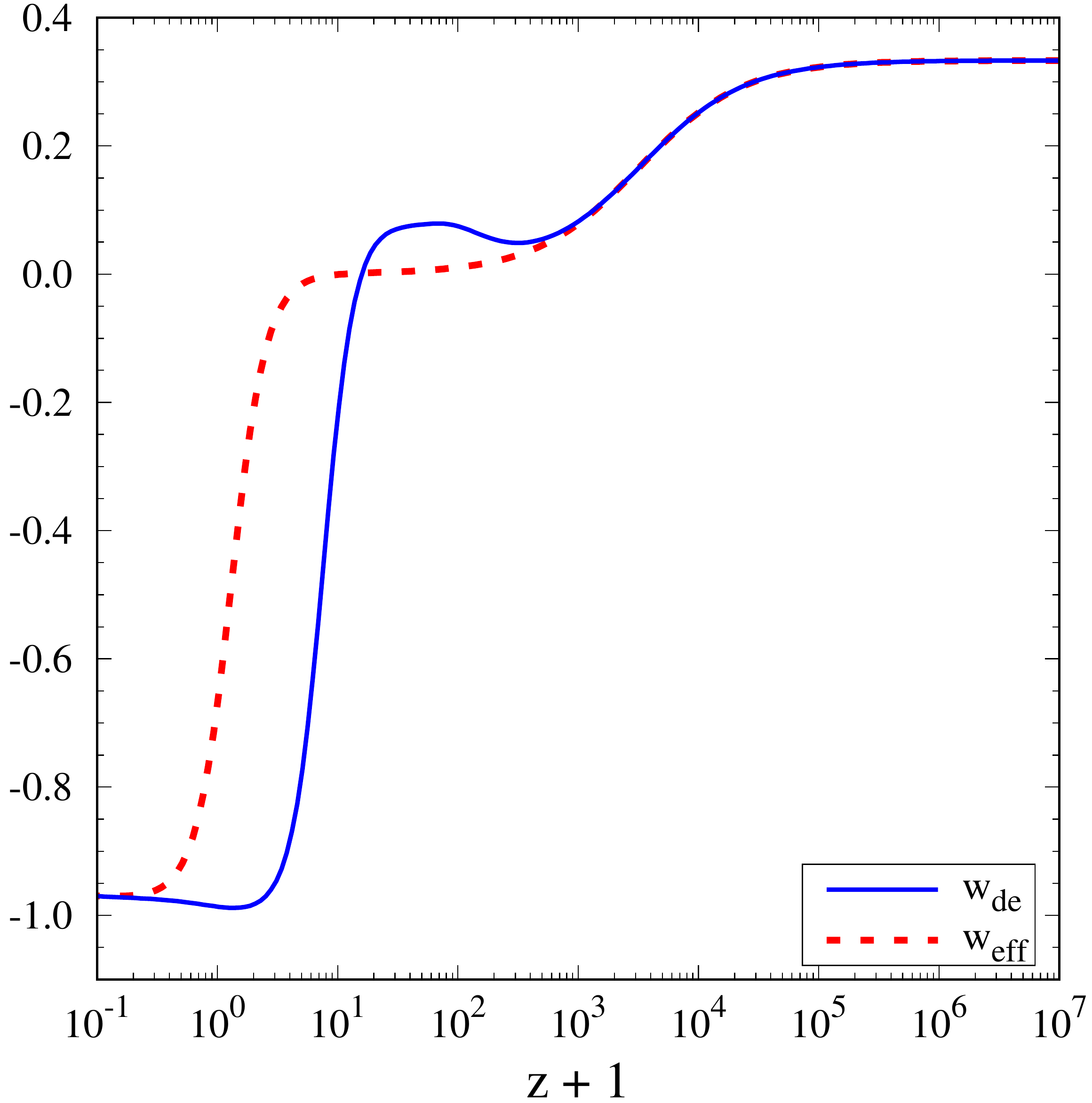}
\end{center}
\caption{\label{fig4}
(Left) Evolution of $x_1^2, \Omega_{\phi 2}, 
\Omega_{\phi 3}$ versus $z + 1$ 
for $Q=0.02$ with the initial conditions 
$x_1=-1.6 \times 10^{-13}$, $\Omega_{\phi 2}=7.0 \times 
10^{-25}$, $\Omega_{\phi 3}=1.0 \times 10^{-18}$, 
$\Omega_r=0.99968$, and $\lambda=-0.2$
at the redshift $z=1.01 \times 10^7$. 
The present epoch is identified by 
$\Omega_{m}=0.32$ and $\Omega_r \simeq 10^{-4}$. 
(Right) Evolution of $w_{\rm eff}$ and $w_{\rm DE}$ 
for the same value of $Q$ and initial conditions as 
those used in the left panel.}
\end{figure}

The other case corresponds to the initial conditions satisfying 
$|Qx_1| \gg \Omega_{\phi 3}$. Since there is the deviation of 
$\Omega_r$ from 1 even during the radiation-dominated epoch, 
it can happen that the last term in $\epsilon_{\phi}$ of Eq.~(\ref{hepes}) 
exceeds the order 1. For $Qx_1/\Omega_{\phi 3}<0$, 
$\epsilon_{\phi}$ can be positive during the radiation era.
In the simulation of Fig.~\ref{fig4}, for example, the numerical 
value of $\epsilon_{\phi}$ at the redshift $z \approx 10^6$ 
is around $0.5$, in which case the field density parameters evolve 
as $x_1^2 \propto a^{5}$, $\Omega_{\phi 2} \propto a^4$, 
and $\Omega_{\phi 3} \propto a^{3.5}$. 
Then, even if $x_1^2<\{ \Omega_{\phi 2},  \Omega_{\phi 3} \}$ 
initially, the Universe enters the stage in which $x_1^2$ 
dominates over $\Omega_{\phi 2}$ and $ \Omega_{\phi 3}$, 
see Fig.~\ref{fig4}. 
In this case, after the radiation dominance, the solutions temporally 
approach the so-called $\phi$-matter-dominated 
epoch ($\phi$MDE) \cite{Amenco}
characterized by the fixed point 
\be
\left( x_1,\Omega_{\phi 2}, \Omega_{\phi 3} 
\right) =\left( -\frac{\sqrt{6}Q}{3(1-2Q^2)},0,0 \right)\,,\quad 
{\rm with} \quad w_{\rm eff}=\frac{4Q^2}{3(1-2Q^2)}\,,\quad
{\rm and} \quad \Omega_m=\frac{3-2Q^2}{3(1-2Q^2)^2}\,.
\label{phiMDE}
\ee
For $Q>0$ satisfying the condition $Q^2<1/2$, the existence 
of $\phi$MDE requires that $x_1<0$, i.e., $\dot{\phi}<0$. 
In this case, the field rolls down along the linear potential 
$V(\phi)=m^3 \phi$ with $m>0$. 
The cosmic acceleration occurs in the region $\phi>0$, 
so the existence of $\phi$MDE requires that $\lambda=-M_{\rm pl}/\phi<0$.
Alternatively, as in Ref.~\cite{KTD15}, we can consider the cases $Q<0, x_1>0, \lambda>0$.

In the left panel of Fig.~\ref{fig4}, we can confirm that $x_1^2$ 
temporally approaches a constant determined by Eq.~(\ref{phiMDE}). 
As we observe in the right panel of Fig.~\ref{fig4}, the effective 
EOS $w_{\rm eff}$ starts to evolve from the value close to $1/3$
and temporally approaches the $\phi$MDE value $4Q^2/[3(1-2Q^2)]$.
After the linear potential $V(\phi)=m^3 \phi$ dominates over 
the other field densities, the Universe eventually enters the 
stage of cosmic acceleration ($w_{\rm eff}<-1/3$).
In Fig.~\ref{fig4}, we also show the evolution of 
$w_{\rm DE}$ 
given by Eq.~(\ref{wdeD}), whose value during the matter era 
is slightly larger than 0. Compared to $w_{\rm eff}$, 
the transition of $w_{\rm DE}$ to the value around $-1$ 
occurs earlier.

In comparison to model (C2) discussed in Sec.~\ref{cghostsec}, 
the contribution of $\Omega_{\phi 3}$ to the total field density 
tends to be negligible at an earlier cosmological epoch.
This is attributed to the fact that $\Omega_{\phi 2}$ in model (D2) 
evolves as $\Omega_{\phi 2} \propto a^3$ during the matter era, 
whereas, in model (C2), the density parameter 
$\Omega_{\phi 2}$ associated with the Lagrangian $\beta_2 X^2$ 
evolves much more slowly ($\Omega_{\phi 2} \propto a^0$). 
In the latter case, the cubic Galileon density can dominate 
over the other field densities at later cosmological epochs, 
so the dark energy 
EOS enters the region $w_{\rm DE}<-1$ by 
temporally approaching the tracker ($w_{\rm DE}=-2$). 
This behavior does not occur for model (D2), in which 
$w_{\rm DE}$ typically stays in the region $w_{\rm DE}>-1$, 
see the right panel of Fig.~\ref{fig4}.
Hence these two models can be observationally distinguished from 
each other even at the background level. 
In model (D2), $\Omega_{\phi 3}$ is usually much smaller than 1 today, 
so it is expected that the galaxy-ISW correlation data do not put tight 
constraints on the model parameters.

\begin{figure}[h]
\begin{center}
\includegraphics[height=3.2in,width=3.3in]{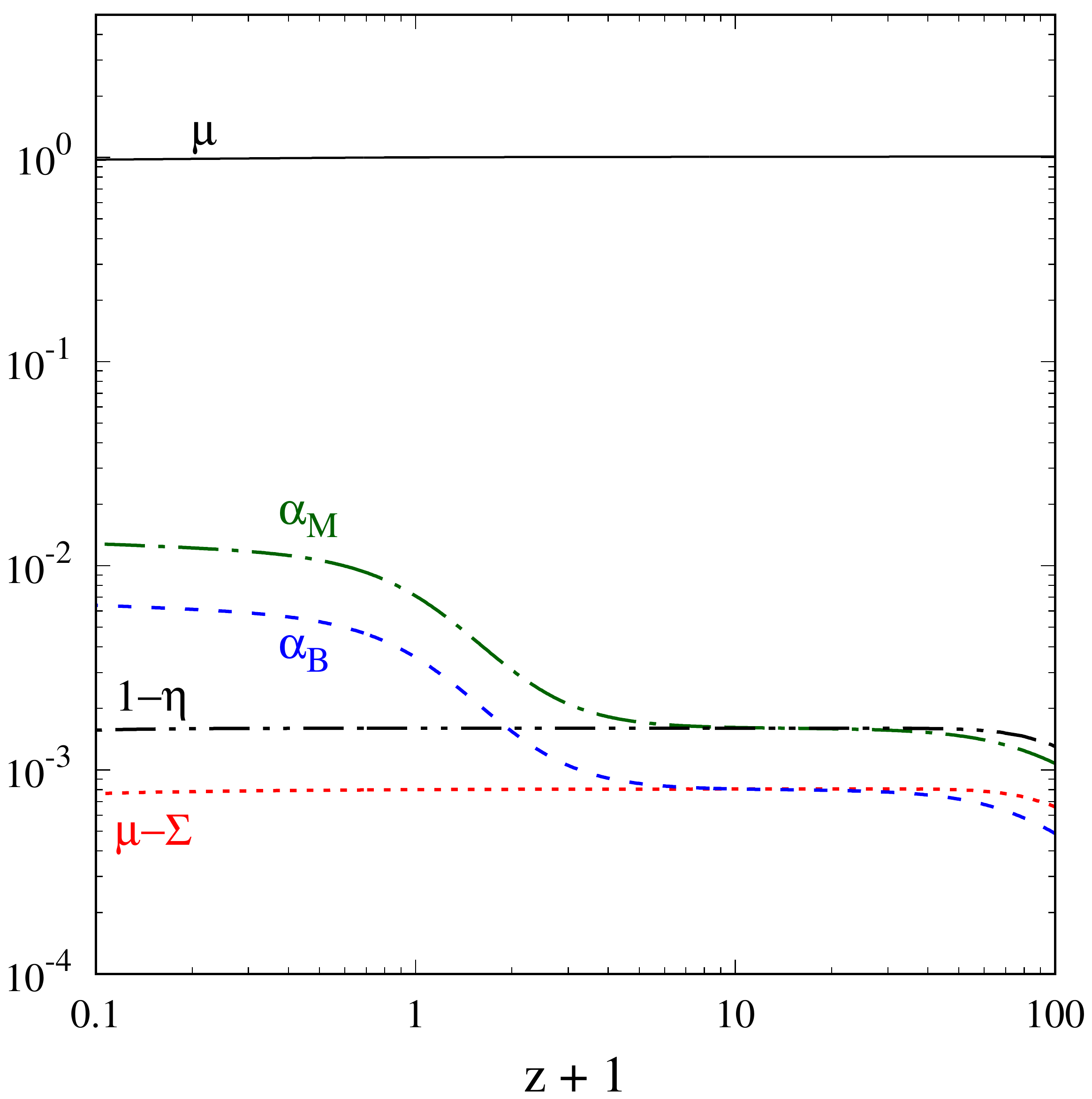}
\includegraphics[height=3.2in,width=3.4in]{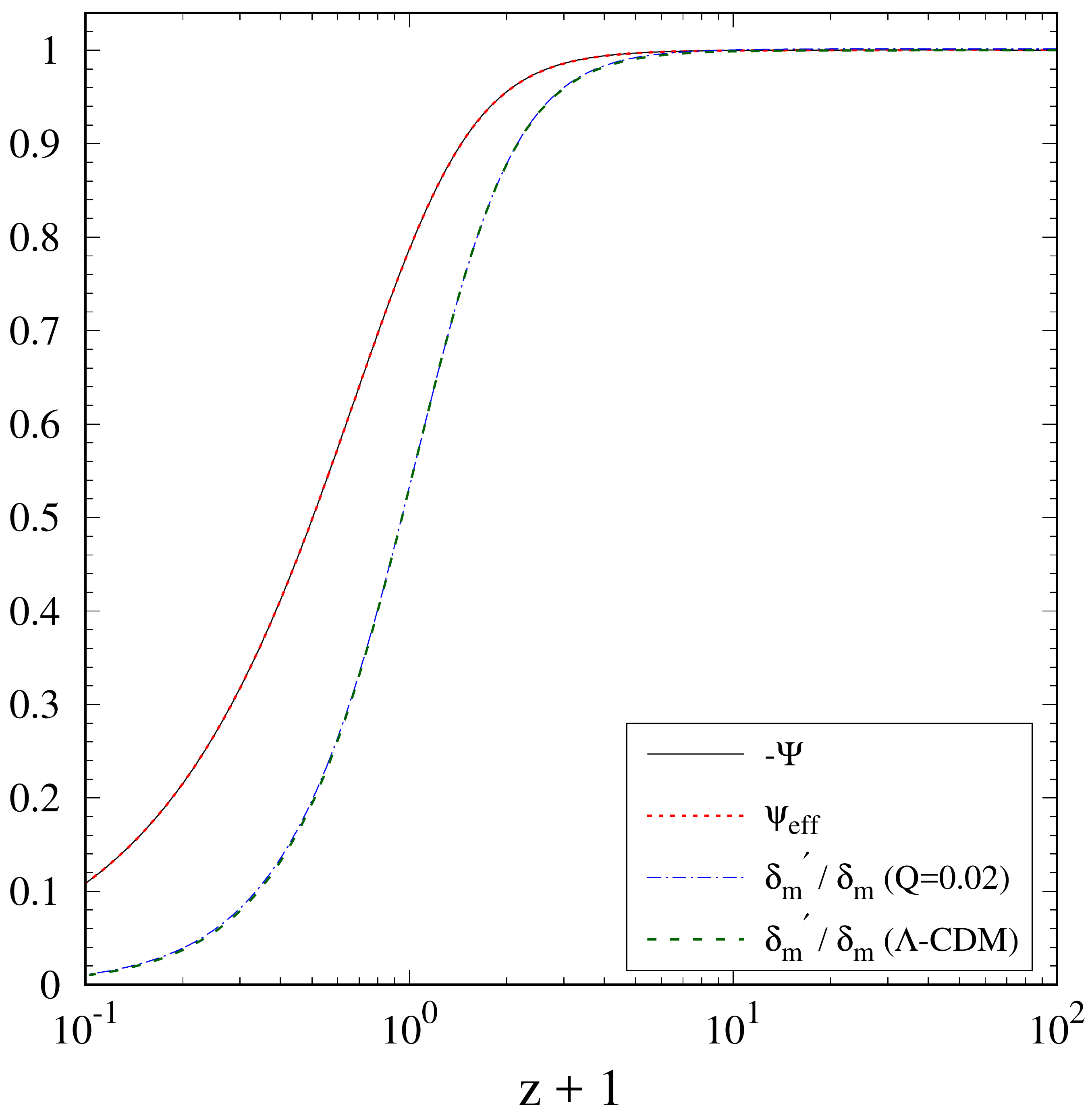}
\end{center}
\caption{\label{fig5}
Evolution of $\mu, 1-\eta, \mu-\Sigma, \alpha_{\rm M}, \alpha_{\rm B}$ 
(left) and $\delta_m'/\delta_m, -\Psi, \psi_{\rm eff}$ (right) 
for $Q=0.02$ and the initial conditions same 
as those used in Fig.~\ref{fig4}. 
The perturbations correspond to the wavenumber $k=200H_0$
with $\sigma_8(z=0)=0.81$. 
The gravitational potentials $-\Psi$ and $\psi_{\rm eff}$ are 
normalized by their initial values. 
In the right panel, we also show the evolution of $\delta_m'/\delta_m$ 
in the $\Lambda$CDM model as a dashed line.}
\end{figure}

The quantities $\alpha_{\rm M}$ and $\alpha_{\rm B}$, 
reduce, respectively, to 
\be
\alpha_{\rm M}=-2\sqrt{6} Q x_1\,,\qquad 
\alpha_{\rm B}=-\sqrt{6} Q x_1-\frac{\Omega_{\phi 3}}{2}\,.
\ee
After $\Omega_{\phi 3}$ becomes much smaller than $|Qx_1|$, 
the simple relation $\alpha_{\rm M} \simeq 2\alpha_{\rm B}$ holds.
Indeed, this property can be confirmed in the left panel 
of Fig.~\ref{fig5}. We recall that there is the bound (\ref{alpM}) 
on $\alpha_{\rm M}(z=0)$ arising from  
Lunar Laser Ranging experiments. 
In the numerical simulation on the left panel of Fig.~\ref{fig5}, 
which corresponds to $Q=0.02$, today's value of $\alpha_{\rm M}$ 
is $7.0 \times 10^{-3}$ and hence the bound (\ref{alpM}) is satisfied.
For the initial conditions similar to those used in Fig.~\ref{fig4}, 
the coupling needs to be in the range $|Q| \lesssim 0.04$ 
for the consistency with (\ref{alpM}).

The quantities associated with the stability of scalar perturbations are given by 
\be
q_s=\frac{M_{\rm pl}^2 F^2 (4x_1^2+4\Omega_{\phi 3}
+\Omega_{\phi 3}^2+4\sqrt{6}Q x_1 \Omega_{\phi 3})}
{2x_1^2}\,,\qquad
c_s^2=\frac{12x_1^2+4(2+\epsilon_{\phi})\Omega_{\phi 3}
-\Omega_{\phi 3}^2+4\sqrt{6}Q x_1 \Omega_{\phi 3}}
{3(4x_1^2+4\Omega_{\phi 3}+\Omega_{\phi 3}^2
+4\sqrt{6}Q x_1 \Omega_{\phi 3})}\,.
\label{qscsF}
\ee
At the late cosmological epoch in which the contribution of 
$\Omega_{\phi 3}$ is suppressed relative to the terms containing 
$x_1$ in Eq.~(\ref{qscsF}), it follows that 
$q_s \simeq 2\Mpl^2F^2$ and $c_s^2 \simeq 1$.
In the numerical simulations of Figs.~\ref{fig4} and \ref{fig5}, 
we confirmed that both $q_s$ and $c_s^2$ are 
positive from the radiation era to today. 
On using Eqs.~(\ref{Gefff2})-(\ref{Sigmaf2}) together 
with the property $M_{\phi}^2=0$ and ignoring the contribution 
$\Omega_{\phi 3}$ at late times, we obtain 
\be
\mu \simeq \frac{1}{F} \left( 1+2Q^2 \right)\,,\qquad 
\eta \simeq \frac{1-2Q^2}{1+2Q^2}\,,\qquad 
\Sigma \simeq  \frac{1}{F}\,.
\ee
For $Q>0$, the quantity $F=e^{-2Q(\phi-\phi_0)/M_{\rm pl}}$ is 
smaller than 1 in the past ($\phi>\phi_0$). 
Since $Q/\lambda<0$ for the existence of $\phi$MDE, 
both $\mu$ and $\Sigma$ are larger than 1. 
Since $Q^2=4 \times 10^{-4} \ll 1$ in the numerical simulation of Fig.~\ref{fig5}, 
the deviation of $\mu$ from 1 mostly arises from the term $1/F~(>1)$. 
This property is different from that in $f(R)$ gravity, in which the large 
coupling $Q^2=1/6$ gives rise to the strong gravitational interaction 
with matter. Since $\mu-\Sigma \simeq 2Q^2/F$, the difference between 
$\mu$ and $\Sigma$ is much smaller than 1, see Fig.~\ref{fig5}.

Let us estimate the evolution of $\delta_m$ during the $\phi$MDE.
Substituting $\mu \simeq (1+2Q^2)/F$ into Eq.~(\ref{delmeqf}), it 
follows that 
\be
\frac{d^2 \delta_m}{d{\cal N}^2}+\frac{1}{2} \left( 1-3w_{\rm eff} \right) 
\frac{d \delta_m}{d{\cal N}}-\frac{3}{2} \left(1 +2Q^2 \right) 
\Omega_m \delta_m \simeq 0\,.
\label{delmap2}
\ee
Using the values of $w_{\rm eff}$ and $\Omega_m$ given in Eq.~(\ref{phiMDE}), 
the growing-mode solution to Eq.~(\ref{delmap2}) for $Q^2<1/2$ is 
\be
\delta_m \propto a^{(1+2Q^2)/(1-2Q^2)}\,.
\label{delmQ}
\ee
In the limit $Q \to 0$, we recover the standard evolution $\delta_m \propto a$ 
during the matter era.  Although the coupling $Q$ leads to the faster growth 
of $\delta_m$, we recall that there exists the upper bound 
$|Q|<{\cal O}(0.01)$ from (\ref{alpM}).
In the right panel of Fig.~\ref{fig5}, we plot the evolution of $\delta_m'/\delta_m$ 
for $Q=0.02$ together with that in the $\Lambda$CDM model. 
{}From Eq.~(\ref{delmQ}), the value of $\delta_m'/\delta_m$ during the $\phi$MDE 
is estimated as $\delta_m'/\delta_m \simeq 1.0016$, so it is hardly distinguishable 
from that in the $\Lambda$CDM model. 
This property can be confirmed in the numerical simulation of Fig.~\ref{fig5} 
at high redshifts. Indeed, even at low redshifts, 
the evolution of $\delta_m'/\delta_m$ is almost the same as 
that in the $\Lambda$CDM model. 
As we observe in the left panel of Fig.~\ref{fig5}, 
the deviation of $\mu$ from 1 is at most of 
order $0.01$ and hence the growth of $\delta_m$ 
is hardly affected by the modification of gravity.

The gravitational potentials $-\Psi$ and $\psi_{\rm eff}$ also evolve in the similar 
way to those in the $\Lambda$CDM model with a small gravitational slip, i.e., 
$1-\eta \simeq 4Q^2/(1+2Q^2) \ll 1$ 
for $|Q|<{\cal O}(0.01)$.
Although the model (D2) does not leave distinguished observational 
signatures for the cosmic growth history, the background evolution of 
$w_{\rm DE}$ can be clearly distinguished from 
other (extended) Galileon models like (C2).

Finally, we should mention that, even in presence of the 
quartic Galileon 
term $G_4 \supset \beta_4 X^2$, there are cases in which the deviation of $c_t^2$ from 1 
at low redshifts is so small that the bound (\ref{ctbound}) can be consistently satisfied \cite{KTD15}. 
It remains to be seen whether the model with such a tiny value
of $|c_t^2-1|$ can survive in occurrence of further GW events 
with electromagnetic counterparts.

\section{Conclusions}
\label{concludesec} 

Two decades have passed after the first discovery of late-time cosmic acceleration, 
but we did not identify the source for this phenomenon yet. 
Although the cosmological constant $\Lambda$ is the simplest candidate 
for dark energy, it is still a challenging problem to explain the very small value 
of $\Lambda$ consistent with today's dark energy scale. 
Moreover, in the $\Lambda$CDM model, the values of $H_0$ and 
$\sigma_8 (z=0)$ constrained from the Planck CMB data have been in tension with 
those measured in low-redshift observations. 
If there are other possibilities to explain the cosmic acceleration, 
it is worthwhile to construct theoretically consistent dark energy models 
and explore their observational signatures to distinguish them from 
the $\Lambda$CDM model.

In dynamical dark energy models, there are usually additional propagating 
degrees of freedom to those appearing in standard model of particle physics. 
A scalar field $\phi$, which is compatible with the isotropic and homogenous cosmological background, is one of the most natural candidates 
for such a new degree of freedom.
If the scalar field has direct couplings to the gravity sector, Horndeski theories 
are known as the most general scalar-tensor theories with second-order equations 
of motion. As listed in Sec.~\ref{Hornsec}, Horndeski theories accommodate 
a wide variety of dark energy models proposed in the literature.

In Sec.~\ref{backsec}, we obtained the background equations of motion in Horndeski 
theories with a perfeclt fluid matter in the forms (\ref{back1}), (\ref{dH}), and (\ref{ddphi}), 
which can be directly applied to concrete models of dark energy with a given Lagrangian.
We also computed the second-order action of tensor perturbations and derived the 
speed of gravitational waves $c_t$ as Eq.~(\ref{ct}). 
The gravitational wave event GW170817 together with the gamma-ray burst 
GRB 170817A placed the tight observational bound on $c_t$.
If we strictly demand that $c_t^2=1$, 
the Lagrangian of Horndeski theories is constrained to be 
of the form (\ref{lagcon}). In extended versions of Horndeski gravity, such as 
Gleyzes-Langlois-Piazza-Vernizzi (GLPV) theories \cite{GLPV} and 
DHOST theories \cite{Langlois1,Langlois2,Crisostomi16}, 
there are models with $c_t^2=1$ even in presence of 
quartic-order derivative 
couplings \cite{GWcon1,GWcon2,GWcon3,GWcon4,GWcon5,GWcon6,Kase18}. 
In this review, we did not explore the dark energy cosmology 
beyond the domain of Horndeski theories. 

In Sec.~\ref{scalarsec}, we derived the linear scalar perturbation equations of motion 
in full Horndeski theories without fixing gauge conditions. 
This gauge-ready formulation is versatile in that any convenient gauges can be 
chosen depending on the problems at hand. 
The perturbation equations expressed in terms of gauge-invariant variables are 
directly applicable to the study of cosmic growth history (including initial 
conditions of perturbations). 
In Sec.~\ref{stasec}, we indentifid conditions for the absence of scalar ghost and 
Laplacian instabilities in the small-scale limit by choosing three different gauges 
and showed that they are given by Eqs.~(\ref{qscon}) and (\ref{csge}) 
independent of the gauge choices.

In Sec.~\ref{obsersec}, we introduced two quantities $\mu$ and $\Sigma$ associated 
with the linear growth of nonrelativistic matter perturbations and the weak lensing 
gravitational potential. 
We computed them under the quasi-static approximation for the modes 
deep inside the sound horizon, see Eqs.~(\ref{Gefff}) and (\ref{Sigmaf}). 
Under the stability conditions of tensor and scalar perturbations, 
the existence of scalar degree of freedom always enhances the gravitational 
coupling with matter, such that $\mu>c_t^2/( 8\pi G q_t)$. 
Taking the limit $c_t^2 \to 1$, we showed that $\mu$ and $\Sigma$ can be 
expressed as Eqs.~(\ref{Gefff2}) and (\ref{Sigmaf2}) in terms of two parameters 
$\alpha_{\rm M}$ and $\alpha_{\rm B}$ defined in Eq.~(\ref{nodim}). 
We classified surviving theories of the late-time cosmic acceleration 
into four classes according to these two parameters. 
In Sec.~\ref{screensec}, we also showed that today's value of $\alpha_{\rm M}$
is constrained as $|\alpha_{\rm M}(t_0)|<0.02$ from the Lunar Laser Ranging 
experiments measuring the variation of effective gravitational coupling 
in screened environments. In particular, this information can be used to constrain 
nonminimally coupled theories with the coupling $G_4(\phi)R$.

In Secs.~\ref{quinsec}-\ref{mostsec}, we reviewed observational signatures of four classes 
of Horndeski theories with the exact value $c_t^2=1$ by paying particular attention to 
the evolution of dark energy EOS and the cosmic growth 
history. We summarize the main points in the following.
\begin{itemize}
\item Class (A): This is categorized as the theories with 
$G_3=0$ and $G_4=M_{\rm pl}^2/2$. 
In this case, we have $\alpha_{\rm B}=0=\alpha_{\rm M}$ and 
hence $\mu=1=\Sigma$. Quintessence and k-essence belong to this class.

Depending on the evolution of $w_{\rm DE}$ (which is in the range $w_{\rm DE}>-1$ 
under stability conditions), we can classify quintessence into three subclasses: 
(a) thawing, (b) scaling, and (c) tracking, see Fig.~\ref{fig1}. 
The thawing quintessence is consistent with observations 
for $w_{\rm DE}(z=0) \lesssim -0.7$. 
The dark energy EOS in scaling quintessence realized by the potential 
(\ref{scaling}) can be well approximated by Eq.~(\ref{wdetran}) 
for $\lambda_2^2 \ll 1$. The transition scale factor is constrained to be 
$a_t<0.11$ (95 \%\,CL) to avoid the slow down of growth of structures 
induced by the scaling scalar field with $c_s^2=1$. 
In tracking quintessence with the inverse power-law potential (\ref{inpo}), 
the power $p$ is constrained to be $p<0.17$ (95 \% CL) and hence positive 
integers ($p \geq 1$) are excluded. 

In k-essence, the ghost condensate model (\ref{Lagghost}) predicts the 
evolution of $w_{\rm DE}$ consistent with observations for $\lambda \lesssim 0.36$. 
The k-essence Lagrangian (\ref{unilag}) gives rise to the unified description of dark energy 
and dark matter with the scalar sound speed $c_s^2$ much 
smaller than 1.

\item Class (B): Compared to class (A), a nonminimal coupling of the form 
$G_4(\phi)R$ is present in class (B). 
In this case, there is the particular relation 
$\alpha_{\rm B}=\alpha_{\rm M}/2 \neq 0$. The interaction between the scalar field and matter 
enhances the gravitational coupling $\mu=G_{\rm eff}/G$, while this is not the case for 
$\Sigma$. The $f(R)$ gravity and BD theory with the scalar potential belong to this class.
 
In $f(R)$ gravity, the models need to be constructed to suppress the propagation of fifth 
forces in regions of the high density, in which case the deviation of $w_{\rm DE}$ from $-1$ 
occurs at low redshifts. After the perturbations enter the massless regime, the large coupling 
$|Q|=1/\sqrt{6}$ leads to the strong amplification of matter density contrast and gravitational 
potentials. For the consistency with CMB, BAO, and weak lensing data, the dark energy 
EOS $w_{\rm DE}$ and the deviation parameter $B$ from the $\Lambda$CDM 
model need to be in the ranges $|w_{\rm DE}+1|<0.002$ and $B(z=0)<0.006$ at 95 \% CL. 
The similar tight constraints on the deviation from the $\Lambda$CDM 
model also persist in BD theories with the potential (\ref{Vphige}) 
and the coupling 
$|Q|>{\cal O}(0.1)$.

\item Class (C): In this class, the cubic coupling $G_3(\phi,X) \square \phi$ 
is added to the Lagrangian of class (A). 
Since $\alpha_{\rm M}=0$, there is no gravitational 
slip ($\eta=-\Phi/\Psi=1$) with $\mu=\Sigma>1$.
The bound (\ref{alpM}) from the Lunar Laser Ranging experiments is trivially satisfied.
In Sec.~\ref{classCsec}, we presented two models 
of this class, including the cubic Galileon with a linear potential [model (C1)]. 

In model (C2) given by the Lagrangian (\ref{iii2}), the existence of an additional term $\beta_2 X^2$ to $\beta_1 X$ 
gives rise to a self-accelerating de Sitter attractor ($X={\rm constant}$). 
If the dominance of this term over the cubic Galileon occurs 
at a later cosmological epoch, the deviations of $w_{\rm DE}$ from $-1$ 
and $\mu$ from 1 tend to be more significant, see Fig.~\ref{fig2}. 
The model (C2) allows an interesting possibility for realizing $w_{\rm DE}<-1$ 
without having ghosts, while modifying the large-scale structure growth at 
low redshifts. We expect that the galaxy-ISW correlation data can provide 
an upper bound on today's density parameter $\Omega_{\phi 3}$ 
of cubic Galileons.
\item Class (D): This class corresponds to the most general Lagrangian 
in Horndeski theories with $c_t^2$ exactly equivalent to 1. 
The difference from class (C) is that the nonminimal coupling 
$G_4(\phi)R$ is present. 
In model (D1) given by the Lagrangian (\ref{LagGBD}), the cubic 
coupling $G_3(\phi,X)$ is the main source for the late-time cosmic acceleration, 
but in such cases the model typically predicts 
the galaxy-ISW anti-correlation which can be incompatible with 
observational data. In model (D2), i.e., nonminimally coupled 
cubic Galileons, there exists the $\phi$MDE followed by the 
cosmic acceleration driven by the linear potential 
$V(\phi)=m^3 \phi$. In this case, the cubic Galileon density 
tends to be subdominant to the total field density at 
early cosmological epochs, so the dark energy density 
typically stays in the region $w_{\rm DE}>-1$.
{}From the bound (\ref{alpM}) the coupling $Q$ is constrained 
to be $|Q|<{\cal O}(0.01)$, in which case the cosmic 
growth pattern is not much different from that in GR.
However, it is possible to distinguish between model (D2) and 
uncoupled model (C2) due to 
different cosmic expansion histories and 
galaxy-ISW correlations.
\end{itemize}

We have thus shown that the surviving dark energy models 
after the GW170817 event predict different observational 
signatures which can be probed in current and future 
observational data. 
The dawn of GW astronomies opened up new possibilities 
for measuring today's Hubble parameter $H_0$ and 
the luminosity distance of GWs \cite{GW170817,GWlu2}.
In the concordance cosmological model there has been a tension 
of $H_0$ between CMB and low-redshift measurements, 
so new GW constraints on $H_0$ may give us a hint 
on whether we should go beyond the $\Lambda$CDM model 
or not.
We also note that the GW luminosity distance is modified 
by the nonminimal coupling $G_4(\phi)R$, 
so the accumulation of GW events will provide important 
information for the viability of models in classes (B) and (D).
The models in classes (A) and (C) can be distinguished further
by exploring whether the future observations favor 
the region $w_{\rm DE}>-1$ or $w_{\rm DE}<-1$.
We hope that upcoming observational data will allow us 
to distinguish between four classes of dark energy models 
and that the origin of late-time cosmic acceleration is 
eventually identified.

\section*{ACKNOWLEDGEMENTS} 
We thank M.~Sami for the invitation to write this review.
We are grateful to Antonio De Felice and 
Atsushi Naruko for useful discussions.
RK is supported by the Grant-in-Aid for Young Scientists B 
of the JSPS No.\,17K14297. 
ST is supported by the Grant-in-Aid for Scientific Research 
Fund of the JSPS No.~16K05359 and 
MEXT KAKENHI Grant-in-Aid for 
Scientific Research on Innovative Areas ``Cosmic Acceleration'' (No.\,15H05890).

\appendix

%
\section{Correspondence of notations and stability conditions with other papers}
\label{notation}
%

\begin{center}
\begin{table}[htb]
\begin{tabular}{|c|c|c|c|c|} \hline
& Cubic Lagrangian & Definition of $X$ 
& Tensor no-ghost condition & Scalar no-ghost condition\\ \hline
This paper & $+G_3(\phi,X)\Box\phi$ & $X=-\cfrac{1}{2}\nabla^{\mu}\phi\nabla_{\mu}\phi$ 
& $q_t$ & 
\begin{tabular}{lll}
Unitary gauge: & $\qsu$ &  \\
Flat gauge: & $q_s^{\rm (f)}$ & \\
\end{tabular}
\\ \hline
Kobayashi {\it et al.} \cite{KYY} & $-G_3(\phi,X)\Box\phi$ & $X=-\cfrac{1}{2}\nabla^{\mu}\phi\nabla_{\mu}\phi$ 
& ${\cal G}_T=q_t$ & ${\cal G}_S=\qsu$ 
\rule[-3mm]{0mm}{8.5mm}\\ \hline
De Felice and Tsujikawa \cite{DT12} & $-G_3(\phi,X)\Box\phi$ & $X=-\cfrac{1}{2}\nabla^{\mu}\phi\nabla_{\mu}\phi$ 
& $Q_T=\cfrac{q_t}{4}$ & $Q_S=\qsu$ 
\rule[-3mm]{0mm}{8.5mm}\\ \hline
Bellini and Sawicki \cite{Bellini} & $-G_3(\phi,X)\Box\phi$ & $X=-\cfrac{1}{2}\nabla^{\mu}\phi\nabla_{\mu}\phi$ 
& $M_*^2=q_t$ & $Q_{\rm S}=\qsu$ 
\rule[-3mm]{0mm}{8.5mm}\\ \hline
Gleyzes {\it et al.} \cite{Gleyzes14} & $+G_3(\phi,X)\Box\phi$ & $X=\nabla^{\mu}\phi\nabla_{\mu}\phi$ 
&  $M^2=q_t$ & $\cfrac{M^2 \left(\aK+6\aB^2 \right)}{(1+\aB)^2}=2\qsu$ 
\rule[-4mm]{0mm}{10.5mm} \\ \hline
\end{tabular}
\label{tab}
\caption{Notations and stability conditions in other papers.}
\end{table}
\end{center}

The cosmology in Horndeski theories has been extensively studied in the literature, 
but different notations were used depending on the papers. 
In Table I, we summarize the notation and sign convention 
for some quantities
adopted in four other papers \cite{KYY,DT12,Bellini,Gleyzes14}. 
The quantities $q_t$ and $\qsu$, $q_s^{\rm (f)}$ 
are defined, respectively, by Eqs.~(\ref{qt}) and 
(\ref{qsue}), (\ref{qsf}). 
In terms of the dimensionless parameters $\aM,\aB,\aK$ introduced in 
Eq.~(\ref{nodim}), we can express $\qsu, q_s^{\rm (f)}, q_s$ in the forms:
\be
\qsu=\frac{q_t \left(\aK+6\aB^2 \right)}{2(1+\aB)^2}\,,\qquad
q_s^{\rm (f)}=\frac{H^2 q_t \left(\aK+6\aB^2 \right)}{2\dot{\phi}^2 
(1+\aB)^2}\,,\qquad 
q_s=\frac{2H^2 q_t^2 \left(\aK+6\aB^2 \right)}{\dot{\phi}^2}\,.
\ee
As we mentioned in Eq.~(\ref{abre}), our definition of $\alpha_{\rm B}$ 
is different from that used by Bellini and Sawichi \cite{Bellini}, as 
$\alpha_{\rm B}=-\alpha_{\rm B}^{(\rm BS)}/2$. 

It may be also convenient to notice that the variables 
$w_1, w_2, w_3, w_4$ introduced in Ref.~\cite{DT12} 
are related to our
$q_t,c_t^2,D_1,D_6$, as 
\be
w_1=q_t\,,
\qquad 
w_2=2Hq_t-\tp D_6\,,
\qquad 
w_3=3\tp^2D_1-9H(Hq_t-\tp D_6)\,,
\qquad 
w_4=q_tc_t^2\,.
\ee
\section{Coefficients in the second-order action of scalar perturbations}
\label{coeff}

The coefficients $D_{1,\cdots,7}$ appearing in the background 
Eqs.~(\ref{back2}), (\ref{back3}) and the second-order action (\ref{Ss}) 
of scalar perturbations are given by
\ba
&&
D_1=H^3\tp\left(3 G_{5,X}+\frac72\tp^2G_{5,XX}+\frac12\tp^4G_{5,XXX}\right)
+3H^2\left[G_{4,X}-G_{5,\phi}+\tp^2\left(4G_{4,XX}-\frac52G_{5,X\phi}\right)\right.
\notag\\
&&\hspace{0.9cm}
\left.+\tp^4\left(G_{4,XXX}-\frac12G_{5,XX\phi}\right)\right]
-3H\tp\left[G_{3,X}+3G_{4,X\phi}+\tp^2\left(\frac12G_{3,XX}+G_{4,XX\phi}\right)\right]
\notag\\
&&\hspace{0.9cm}
+\frac{1}{2}\left[G_{2,X}+2G_{3,\phi}+\tp^2\left(G_{2,XX}+G_{3,X\phi}\right)
\right]
\,,
\notag\\
&&
D_2=-\left[2(G_{4,X}-G_{5,\phi})+\tp^2(2G_{4,XX}-G_{5,X\phi})
+H\tp(2G_{5,X}+\tp^2G_{5,XX})\right]\dot{H}
\notag\\
&&\hspace{0.9cm}
+\left[G_{3,X}+3G_{4,X\phi}+\tp^2\left(\frac{G_{3,XX}}{2}+G_{4,XX\phi}\right)
-2H\tp(3G_{4,XX}-2G_{5,X\phi})\right.
\notag\\
&&\hspace{0.9cm}
\left.-H\tp^3(2G_{4,XXX}-G_{5,XX\phi})
-H^2\left(G_{5,X}+\frac52\tp^2G_{5,XX}+\frac12\tp^4G_{5,XXX}
\right)
\right]\ddot{\phi}
\notag\\
&&\hspace{0.9cm}
-H^3\tp\left(2G_{5,X}+\tp^2G_{5,XX}\right)
-H^2\left[3(G_{4,X}-G_{5,\phi})+5\tp^2\left(G_{4,XX}-\frac12G_{5,X\phi}\right)
+\frac12\tp^4G_{5,XX\phi}\right]
\notag\\
&&\hspace{0.9cm}
+2H\tp(G_{3,X}+3G_{4,X\phi})-H\tp^3(2G_{4,XX\phi}-G_{5,X\phi\phi})
+\tp^2\left(\frac12G_{3,X\phi}+G_{4,X\phi\phi}\right)-G_{3,\phi}-\frac12 G_{2,X}\,,
\notag\\
&&
D_3=3\left[G_{4,\phi\phi}+\tp^2\left(\frac12G_{3,X\phi}+G_{4,X\phi\phi}\right)
-2H\tp(G_{4,X\phi}-G_{5,\phi\phi})\right.
\notag\\
&&\hspace{0.9cm}
\left.-H\tp^3(2G_{4,XX\phi}-G_{5,X\phi\phi})
-\frac{H^2\tp^2}{2}\left(3G_{5,X\phi}+\tp^2G_{5,XX\phi}\right)
\right]\dot{H}
\notag\\
&&\hspace{0.9cm}
-\left[\frac12G_{2,X\phi}+G_{3,\phi\phi}
+\frac12\tp^2(G_{2,XX\phi}+G_{3,X\phi\phi})
-3H\tp(G_{3,X\phi}+3G_{4,X\phi\phi})\right.
\notag\\
&&\hspace{0.9cm}
\left.
-3H\tp^3\left(\frac12G_{3,XX\phi}+G_{4,XX\phi\phi}\right)
+3H^2(G_{4,X\phi}-G_{5,\phi\phi})
+3H^2\tp^2\left( 4G_{4,XX\phi}-\frac52G_{5,X\phi\phi} \right)\right.
\notag\\
&&\hspace{0.9cm}
\left.
+3H^2\tp^4\left(G_{4,XXX\phi}-\frac12G_{5,XX\phi\phi}\right)
+H^3\tp\left(3G_{5,X\phi}+\frac72\tp^2G_{5,XX\phi}
+\frac12\tp^4G_{5,XXX\phi}\right)
\right]\ddot{\phi}
\notag\\
&&\hspace{0.9cm}
-\frac32H^4\tp^2(3G_{5,X\phi}+\tp^2G_{5,XX\phi})
-H^3\tp\left[9(G_{4,X\phi}-G_{5,\phi\phi})
+\tp^2\left(9G_{4,XX\phi}-\frac72G_{5,X\phi\phi}\right)
+\frac12\tp^4G_{5,XX\phi\phi}\right]
\notag\\
&&\hspace{0.9cm}
+3H^2\left[2G_{4,\phi\phi}
+\tp^2\left(\frac32G_{3,X\phi}+3G_{4,X\phi\phi}
+\frac12G_{5,\phi\phi\phi}\right)
-\tp^4\left(G_{4,XX\phi\phi}-\frac12G_{5,X\phi\phi\phi}\right)\right]
\notag\\
&&\hspace{0.9cm}
-3H\tp\left[\frac{1}{2}G_{2,X\phi}+G_{3,\phi\phi}
-\tp^2\left(\frac12G_{3,X\phi\phi}+G_{4,X\phi\phi\phi}\right)\right]
-\frac12\tp^2(G_{2,X\phi\phi}+G_{3,\phi\phi\phi})
+\frac{G_{2,\phi\phi}}{2}\,,
\notag\\
&&
D_4=-H^3\tp^2\left(15G_{5,X}+10\tp^2G_{5,XX}+\tp^4G_{5,XXX}\right)
-3H^2\tp \bigl[6(G_{4,X}-G_{5,\phi})
\notag\\
&&\hspace{0.9cm}
+\tp^2(12G_{4,XX}-7G_{5,X\phi})+\tp^4(2G_{4,XXX}-G_{5,XX\phi}) \bigr]
+3H\left[2G_{4,\phi}+\tp^2(3G_{3,X}+8G_{4,X\phi})\right.
\notag\\
&&\hspace{0.9cm}
\left.+\tp^4(G_{3,XX}+2G_{4,XX\phi})\right]
-\tp^3(G_{2,XX}+G_{3,X\phi})-\tp (G_{2,X}+2G_{3,\phi})\,,
\notag\\
&&
D_5=-H^3\tp^3(5G_{5,X\phi}+\tp^2G_{5,XX\phi})
+3H^2\left[2G_{4,\phi}-\tp^2(4G_{4,X\phi}-3G_{5,\phi\phi})
-\tp^4(2G_{4,XX\phi}-G_{5,X\phi\phi})\right]
\notag\\
&&\hspace{0.9cm}
+3H\tp\left[
2 G_{4,\phi\phi}
+\tp^2(G_{3,X\phi}+2G_{4,X\phi\phi})\right]
-\tp^2(G_{2,X\phi}+G_{3,\phi\phi})+G_{2,\phi}
\,,
\notag\\
&&
D_6=H^2\tp^2(3G_{5,X}+\tp^2G_{5,XX})
+2H\tp\left[2(G_{4,X}-G_{5,\phi})+\tp^2(2G_{4,XX}-G_{5,X\phi})\right]
-\tp^2(G_{3,X}+2G_{4,X\phi})-2G_{4,\phi}
\,,
\notag\\
&&
D_7=H^3\tp^2(3G_{5,X}+\tp^2G_{5,XX})
+2H^2\tp\left[
3(G_{4,X}-G_{5,\phi})+\tp^2(3G_{4,XX}-2G_{5,X\phi})\right]
\notag\\
&&\hspace{0.9cm}
-H\left[2G_{4,\phi}
+\tp^2(3G_{3,X}+10G_{4,X\phi}-2G_{5,\phi\phi})\right]
+\tp(G_{2,X}+2G_{3,\phi}+2G_{4,\phi\phi})
\,.
\ea
%


\end{document}